\documentclass[12pt]{article}

\usepackage{amsfonts,amsmath}

\usepackage{tikz}
\usetikzlibrary{decorations.markings}

\numberwithin{equation}{section}

\newcommand\encadremath[1]{\vbox{\hrule\hbox{\vrule\kern8pt
\vbox{\kern8pt \hbox{$\displaystyle #1$}\kern8pt}
\kern8pt\vrule}\hrule}}
\def\enca#1{\vbox{\hrule\hbox{
\vrule\kern8pt\vbox{\kern8pt \hbox{$\displaystyle #1$}
\kern8pt} \kern8pt\vrule}\hrule}}

\newcommand\figureframex[3]{
\begin{figure}[bth]
\hrule\hbox{\vrule\kern8pt
\vbox{\kern8pt \vbox{
\begin{center}
{\mbox{\epsfxsize=#1.truecm\epsfbox{#2}}}
\end{center}
\caption{#3}
}\kern8pt}
\kern8pt\vrule}\hrule
\end{figure}
}
\newcommand\figureframey[3]{
\begin{figure}[bth]
\hrule\hbox{\vrule\kern8pt
\vbox{\kern8pt \vbox{
\begin{center}
{\mbox{\epsfysize=#1.truecm\epsfbox{#2}}}
\end{center}
\caption{#3}
}\kern8pt}
\kern8pt\vrule}\hrule
\end{figure}
}

\newtheorem{theorem}{Theorem}[section]

\newtheorem{remark}[theorem]{Remark}
\newtheorem{proposition}[theorem]{Proposition}
\newtheorem{definition}[theorem]{Definition}
\newtheorem{lemma}[theorem]{Lemma}
\newtheorem{corollary}[theorem]{Corollary}
\def\br{\begin{remark}\rm\small}
\def\er{\end{remark}}
\def\bt{\begin{theorem}}
\def\et{\end{theorem}}
\def\bd{\begin{definition}}
\def\ed{\end{definition}}
\def\bp{\begin{proposition}}
\def\ep{\end{proposition}}
\def\bl{\begin{lemma}}
\def\el{\end{lemma}}
\def\bc{\begin{corollary}}
\def\ec{\end{corollary}}
\def\beaq{\begin{eqnarray}}
\def\eeaq{\end{eqnarray}}
\newcommand{\proof}{{\noindent \bf Proof:}\par\smallskip}
\newcommand{\eproof}{$\square$\par}

\newcommand{\boxtimes}{{\square\!\!\!\!\!\times}}

\newcommand{\td}{\tilde}

\newcommand{\beq}{\begin{equation}}
\newcommand{\eeq}{\end{equation}}
\newcommand{\bea}{\begin{eqnarray}}
\newcommand{\eea}{\end{eqnarray}}

\newcommand{\Tr}{\operatorname{Tr}}

\newcommand{\diff}{\operatorname{d}}
\newcommand{\Ho}{\operatorname{H}}

\newcommand{\Lieg}{{\mathfrak g}}
\newcommand{\Lieh}{{\mathfrak h}}

\newcommand{\Weyl}{{\mathbf w}}

\newcommand{\Ad}{\operatorname{Ad}}
\newcommand{\Ker}{\operatorname{Ker}}
\newcommand{\Img}{\operatorname{Im}}

\newcommand{\CC}{\mathbb C}
\newcommand{\RR}{\mathbb R}

\newcommand{\modsp}{{\mathcal M}}

\newcommand{\DD}{\mathcal D}
\newcommand{\dd}{{\partial}}

\newcommand{\Tau}{{\mathfrak T}}

\newcommand{\curve}{{\Sigma}}
\newcommand{\curverond}{\overset{\circ}{\curve}}
\newcommand{\genus}{{\mathbf g}}
\newcommand{\genusrond}{\overset{\circ}{\genus}}
\newcommand{\acycle}{{\cal A}}
\newcommand{\bcycle}{{\cal B}}
\newcommand{\acyclerond}{\overset{\circ}{\acycle}}
\newcommand{\bcyclerond}{\overset{\circ}{\bcycle}}
\newcommand{\omegarond}{\overset{\circ}{\omega}}

\newcommand{\Brond}{\overset{\circ}{B}}
\newcommand{\Erond}{\overset{\circ}{\mathcal E}}

\newcommand{\curveuniv}{\widetilde{\Sigma}}

\newcommand{\Res}{\mathop{\,\rm Res\,}}
\usepackage[pdftex]{hyperref}
\hypersetup{colorlinks,urlcolor=magenta,citecolor=red,linkcolor=blue,filecolor=black}

\begin{document}

\sloppy

\addtolength{\baselineskip}{0.20\baselineskip}
\begin{center}
\vspace{1cm}

{\Large \bf {From the quantum geometry of Fuchsian systems
\\ to conformal blocks of W--algebras
}}

\vspace{1cm}

{Rapha\"el Belliard}$^{1,2}$ and
{Bertrand Eynard}$^{3,4}$

\vspace{5mm}
$^1$\ DESY Theory,
\\
Notkestrasse 85, 22607 Hamburg, Germany
\\
$^2$\ Department of Mathematics, University of Hamburg
\\
Bundesstrasse 55, 20146 Hamburg, Germany
\\
$^3$\ Institut de physique th\'eorique, Universit\'e Paris Saclay, 
\\
CEA, CNRS, F-91191 Gif-sur-Yvette, France
\\
$^4$\ Centre de recherches math\'ematiques, Universit\'e de Montr\'eal,
\\
2920, Chemin de la tour, 5357 Montr\'eal, Canada.
\vspace{5mm}
\\ 
\end{center}

\begin{center}
{\bf Abstract}
\end{center}
\begin{quote}

We consider the moduli space of holomorphic principal  bundles for reductive Lie groups over Riemann surfaces (possibly with boundaries) and equipped with meromorphic connections. We associate to this space a point--wise notion of \textit{quantum spectral curve} whose generalized periods define a new set of moduli. 
We define homology cycles and differential forms of the quantum spectral curve, allowing to derive quantum analogs of the form--cycle duality and Riemann bilinear identities of classical geometry. 
A tau--function is introduced for this system in the form of a theta--series and in such a way that the variations of its coefficients with respect to moduli, isomonodromic or not, can be computed as quantum period integrals.
This lays new grounds to relate our study to that of integrable hierarchies, isomonodromic deformation of meromorphic connections and non--perturbative topological string theory. In turn, we define amplitudes on the quantum spectral curve which have an interpretation in conformal field theory when the Lie algebra is of simply--laced type. They are moreover related by W--constraints, so--called loop equations, allowing one to compute recursively a certain asymptotic expansion of the tau--function, namely the one corresponding both to the heavy--charge regime of conformal field theory and to the weak--coupling regime of topological string theory.
\end{quote}

\pagebreak

\tableofcontents 

\section{Introduction}

In the last decades, the boundary interaction between fundamental theoretical Physics and Mathematics has been the source of tremendous progress on both sides. In particular, methods arising from the huge symmetry content exhibited by (possibly quantum) integrable systems such as semi--classical integrable hierarchies, integrable spin chains or conformal field theories have been transported through physical dualities to gain an ever growing scope of mathematical applications ranging from the computation of enumerative geometry quantities to getting a better understanding of the geometric Langlands correspondence. A central notion appearing throughout this theoretical landscape is that of tau--functions \cite{JMU1981,SW1985}. They come up as functions on the moduli (phase) spaces under considerations that contain the information needed to solve the problem at stake. They can be defined as satisfying bilinear identities and are candidates to quantize theta--functions on Riemann surfaces. Their theta--series expansions in the context of Riemann--Hilbert problems and conformal field theory of free fermions were for instance studied in \cite{CPT2018}. 

Building up on a conjecture of the authors and collaborators in \cite{BER2018} we give the general definition of a tau--function associated to the moduli space of meromorphic connections in a holomorphic principal bundle over a compact Riemann surface. We define it as a theta--series expansion whose coefficients satisfy special geometry relation (related to hyper--K\"ahler structures). The choices of corresponding complex structure defining the Riemann surface, reductive complex structure group and lagrangian submanifold appear as parameters of the construction and make contact with its variety of possible applications.

Let us therefore denote by $G$ a reductive complex Lie group, by $\Lieg$ the associated Lie algebra with universal enveloping algebra $\mathcal U(\Lieg)$,  $\Lieh\subset\Lieg$ a given Cartan subalgebra, $\Weyl$ its Weyl group and $\mathfrak R\subset\Lieh^*$ a chosen root system.

Recall that a reductive Lie algebra $\Lieg$ admits a root decomposition $\Lieg=\Lieh \oplus_{\mathfrak r\in \mathfrak R} \Lieg_{\mathfrak r}$, where $\Lieg_{\mathfrak r}\underset{def}{=}E_{\mathfrak r}\otimes \CC $ is a  one--dimensional subspace. We shall often abuse notation by writing $E_0\underset{def}{=}\Lieh$ for the Cartan subalgebra whose dimension is by definition the rank of $\Lieg$. Such a Cartan subalgebra contains in particular the center of $\Lieg$, inclusion denoted $\mathcal Z\Lieg\subset\Lieh$ and not to be confused with the center of the enveloping algebra, itself denoted $\mathcal Z\,\mathcal U(\Lieg)\subset\mathcal U(\Lieg)$.

Let us fix once and for all an Adjoint--invariant multilinear map $\left\langle\, \bullet\, \right\rangle:\mathcal U(\Lieg)\longrightarrow\mathbb C$ restricting to a non--degenerate symmetric bilinear form on $\Lieg$ denoted by $\left\langle a,b\right\rangle$ for generic elements $a,b\in\Lieg$. Since we are consider reductive Lie algebras only, this symmetric bilinear form is necessarily proportional to the Killing form wherever this latter is non--zero. For any given faithful representation $\rho$ of $\Lieg$, such a multilinear map is provided by the trace $\underset{\rho}{\Tr}$ in that representation.

Let us fix $\curverond$ to be a compact Riemann surface of genus $\genusrond\geq 0$ (we consider the case of a base--curve with boundaries in \ref{boundaries}). We shall now study the moduli space $\modsp=\big\{(\mathcal E,\nabla)\big\}\big/\mathcal G$ of pairs of a principal holomorphic $G$--bundle $\mathcal E$ over $\curverond$ endowed with a meromorphic connection $\nabla$ and considered up to holomorphic gauge equivalence embodied in the action of the gauge group $\mathcal G$. We will restrict ourselves to Fuchsian connections, i.e. having simple poles at prescribed points $p\underset{def}{=}\{p_1,\dots,p_M\}$, $M\in\mathbb N^*$, on $\curverond$, and extend to wild connections in section \ref{sec:wildconnection}. This prescription foliates the moduli space and we shall consider the leaf subspaces $\modsp_p$ (and $\modsp_{p,[\alpha]}$) of pairs with fixed positions of poles $p_j$ (and spectra of corresponding residues, so--called charges, the right invariant quantities). We will introduce a tau--function on $\modsp_p$ as well as a conformal block on the leaf moduli space $\modsp_{p,[\alpha]}$.

\section{Principal bundle and connection}

Let $\mathcal P\longrightarrow\curverond$ be a stable principal $G$--bundle over a compact base Riemann surface $\curverond$, with $\nabla$ a meromorphic connection, with a finite number of poles $p_1,\dots,p_M\in\curverond$.
Here we first consider Fuchsian connections, i.e. those having only simple poles, and postpone higher order poles to section \ref{sec:wildconnection}.

In any local chart $U\subset\mathbb C$ of the bundle, one has a trivialization $\mathcal P_{|\varphi(U)}\sim_\varphi U\times G$ and the connection takes the form 
\beq
\nabla \simeq_\varphi \diff-\Phi_U
\eeq
with $\Phi_U$ a $\Lieg$--valued 1--form on the chart $U$, with simple poles at the intersection $\{p_1,\dots,p_M\}\cap \varphi(U)$.

\pagebreak

\bd[Gauge equivalence]
$(\mathcal P,\nabla)$ and $(\mathcal P',\nabla')$ are said to be gauge equivalent if and only if there exists a bi--holomorphic map $g:\mathcal P\longrightarrow \mathcal P'$, acting by fiberwise left--multiplication,
such that in each chart:
\beq
\Phi'_U(x) = \Ad_{g_U(x)} \Phi_U(x)  + \diff g_U(x)\cdot g_U(x)^{-1}.
\eeq
\ed
\bea
\mathcal P & \longrightarrow & \mathcal P'  \cr
\downarrow & & \downarrow \cr
\curverond & \longrightarrow & \curverond
\eea

In particular, the group of gauge automorphisms is the set of holomorphic sections of $\mathcal P$ denoted $\Ho^0(\curverond,\mathcal P)$. Since the base curve $\curverond$ is compact, 
 we have an isomorphism
\beq
\Ho^0(\curverond,\mathcal P) \sim G.
\eeq

Back to local behavior, if a chart $U\subset\mathbb C$ contains a singularity $z_j\underset{def}{=}\varphi^{-1}(p_j)$, we have
\beq
\Phi_U(x) \underset{x\sim p_j}{\simeq} \frac{\Phi_j^U}{x-z_j}\diff x
\eeq
with generic residue $\Phi^U_j\in \Lieg$ having a  Cartan decomposition of the form
\beq
\Phi_j^U= \Ad_{V^U_j}( \alpha_j^U) \qquad V^U_j\in G, \ \alpha_j^U\in \Lieh.
\eeq
This decomposition is not unique as $\alpha_j^U$ is defined modulo the action of the Weyl group $\Weyl$ and $V^U_j$ is defined modulo $\Weyl$ and modulo right--multiplication by elements of the torus $\exp\Lieh$. 
The equivalence class $[\alpha^U_j]\in\Lieh/\Weyl$ is well defined and independent of the chart $U\subset\mathbb C$. It is furthermore invariant under holomorphic gauge transformations as the corresponding affine terms are locally derivatives of holomorphic quantities and have no contribution to the residue. We shall therefore drop the chart in the notation and denote
\beq
\Phi^{(j)}\underset{def}{=} -\underset{p_j}{\Res}\nabla = \Ad_{V_j}\alpha_j
\eeq
\bd[Weights, charges]
The equivalence class $[\alpha_j]\underset{def}{=}[\alpha^U_j]\in \Lieh/\Weyl$ is called the \textbf{weight} of the connection $\nabla$ at the singularity $p_j$. We collectively denote them as the array $[\alpha]\underset{def}{=}\big\{[\alpha_1],\dots,[\alpha_M]\big\}$.
In the context of conformal field theory, these are called the external \textbf{charges} or \textbf{momenta}.
\ed

\subsection{Adjoint bundle and quantum spectral curve}

The adjoint bundle of $\mathcal P$ denoted $\Ad\mathcal P$ is the bundle over $\curverond$ associated to $\mathcal P$ and the Adjoint representation of $G$. Its generic fiber is isomorphic to the Lie algebra $\Lieg$ and the connection $\nabla$ acts on sections of $\Ad\mathcal P$ accordingly, namely in any given chart $U$ and local section $\sigma_U$ by
\beq
\nabla \sigma_U = \diff\sigma_U - [\Phi_U,\sigma_U].
\eeq

The adjoint bundle will play a key role in our study and we will in particular describe it in terms of the space of analytic continuations of solutions to the flat section equation
\beq
\nabla\Psi=0
\eeq

By analogy with the definition of a classical spectral curve (corresponding to the total space of a covering of Riemann surfaces), we shall call \textit{quantum spectral curve} the total space of the adjoint bundle as follows.

\bd[Quantum spectral curve]
For any Fuchsian connection in a principal bundle $\mathfrak m=(\mathcal P,\nabla)$ over $\curverond$, the associated quantum spectral curve  denoted by $\widehat\curve_{\mathfrak m}$ is the total space of the bundle $(\overset{\circ}{\pi})^*\Big(\Ad\mathcal P_{\big |\curverond_p}\Big)$, the adjoint bundle restricted to the punctured Riemann surface $\curverond_p\underset{def}{=}\curverond-\{p_1,\dots,p_M\}$ and then pulled-back by its universal covering map $\overset{\circ}{\pi}$. Elements of $\widehat\curve_{\mathfrak m}$ are of the form $\widetilde x\cdot\sigma$ where $\widetilde x\in(\overset\circ\pi)^*\curverond_p$ is a point on the universal covering and $\sigma$ is a multivalued $\nabla$-flat section of the adjoint bundle.
\ed

\br
Recall that a classical spectral curve is defined from a branched cover over $\curverond_p$ with as many sheets as the rank of (a representation of) the Lie algebra $\Lieg$. Intuitively, its quantization should consist in arbitrary linear combinations of the branches, thus reconstructing a bundle of Cartan subalgebras isomorphic to the previously introduced subalgebra $\Lieh\subset\Lieg$. 
However, quantization should replace the deck transformations of the classical spectral curve, by the monodromies associated to the flat connection $\nabla$ and hence have no reason in general to leave $\Lieh$ invariant. In other words, the quantum spectral curve will be the vector bundle with fiber $\Lieg$ whose transition functions are implied by the monodromy of $\nabla$, namely the adjoint bundle parameterized by $\nabla$--flat sections.
\er

\subsection{Flat sections and behavior near singularities}

Recall the following facts:

\begin{itemize}
\item For any local $\nabla$--flat section $\Psi_U$ defined on a chart $U$ of $\mathcal P$,
\beq
\diff\Psi_U = \Phi_U\cdot \Psi_U.
\eeq
and as such $\Psi_U$ will generically have singularities if the open set contains one or more of the $p_j$'s.
\item Any pair $\Psi_U$, $\td\Psi_U$ of local flat sections are related by right--multiplication by a constant group element $C\in G$, 
\beq
\td\Psi_U = \Psi_U\cdot C
\qquad C\in G.
\eeq
\item Local flat sections $\sigma_U$ of the adjoint bundle can be obtained from that of $\mathcal P$ by Adjoint action 
\beq
\sigma_U= \Ad_{\Psi_U}(E)
\eeq
on an element $E\in\Lieg$ viewed as a constant section on a given chart $U$.

\item Any flat section has a local behavior near a singularity $p_j$ of the form
\beq\label{Psinearzj}
\Psi_U(x) \underset{x\sim p_j}{=}  V_j \cdot\big(\bold 1_G+\mathcal O(x-z_j)\big) \cdot (x-z_j)^{\alpha_j}\cdot C_j
\eeq
where $O(x-z_j)$ stands for analytic expressions of the local variable $x$ in a neighborhood of $p_j$. Similarly, 
a flat local section of the adjoint bundle behaves as
\beq
\sigma_U(x) \underset{x\sim p_j}{\simeq} \Ad_{V_j\cdot(x-z_j)^{\alpha_j}}\cdot\Ad_{C_j} E,
\eeq
and if we decompose $\Ad_{C_j}E\in\Lieg$ with respect to the root system $\mathfrak R$,
\beq
\Ad_{C_j}E = E_0 + \sum_{\mathfrak r\in \mathfrak R} E_{\mathfrak r}
\eeq
with $E_0\in \Lieh$ and $E_{\mathfrak r}\in \Lieg_{\mathfrak r}$, we then have
\beq
\sigma_U(x) \underset{x\sim p_j}{=} \Big( \Ad_{V_j} E_0  +  \sum_{\mathfrak r\in \mathfrak R} (x-z)^{\mathfrak r(\alpha_j)} \Ad_{V_j} E_{\mathfrak r}  \Big) \big(1+O(x-z_j)\big).
\eeq

\end{itemize}

\subsection{Fuchsian moduli spaces}

In this paper we consider several moduli spaces that are related to each other either by being subspaces or cosets of one another.

\bd[Moduli space]
Let $\modsp'$ be the moduli space of holomorphic principal $G$--bundles $\mathcal P\longrightarrow\curverond$ over the base curve together with a meromorphic connection $\nabla$ having only simple poles. The holomorphic gauge group $\mathcal G\underset{def}{=}\Ho^0(\curverond,\mathcal P)\sim G$ acts faithfully on $\modsp'$ in such a way that one can consider the moduli space of orbits under gauge transformations denoted
\beq
\modsp\underset{def}{=}\big\{(\mathcal P,\nabla)\big\}\big/\mathcal G.
\eeq

As mentioned in the introduction, we restrict ourselves to connections with prescribed locus $p\underset{def}{=}\{p_1,\dots,p_M\}$ of simple poles, defining the moduli spaces with fixed positions of singularities and that with fixed weights as well. We denote them as 
\beq
\modsp_{p,[\alpha]}\subset \modsp_p\subset\modsp
\eeq
and similarly before reducing to the orbits under the holomorphic gauge group as 
\beq
\modsp'_{p,[\alpha]}\subset \modsp'_p\subset\modsp'
\eeq
\ed

We recall the well-known dimension of that space (another proof of which can be found below in this article \ref{genusbasis})
\bt
\beq
\dim \modsp_{p,[\alpha]}=2\genus.
\eeq
where the genus $\genus$ is defined by the expression
\beq\label{defgenus}
\genus\underset{def}{=}(\genusrond-1)\dim\Lieg +M\frac{\dim\Lieg-\dim\Lieh}{2}.
\eeq
\et
Remark: $\genus$ is actually the same as the genus of the classical spectral curve.

\subsection{Monodromy representation and Betti space}

Let $\mathfrak m=(\mathcal P,\nabla)\in\modsp'$ be a Fuchsian connection in  a principal bundle and let $\curverond_p$ be the  corresponding base curve with singularities of $\nabla$ removed. Any given $\nabla$--flat local section of $\mathcal P$ can be analytically continued to a flat global section of the bundle obtained by pulling back to the universal covering of $\curverond_p$.  Let $\Psi$ denote such a global section (generically multivalued on $\curverond_p$), its image by any element of the holonomy group of the connection, that is its image after parallel transport along any closed loop $\gamma\in \pi_1(\curverond_p,o)$ starting from a generic reference point $o\in\curverond_p$, is also such a global section. Recall that $\Psi$ is defined on the universal covering of $\curverond_p$, each point $\widetilde x$ of which -- a homotopy class of paths on $\curverond_p$ from $o$ to a given $x\in\curverond_p$ -- yields an isomorphism
\bea
\widetilde x : &\pi_1(\curverond_p,o)&\overset{\sim}{\longrightarrow}\, \pi_1(\curverond_p,x)\nonumber\\
&\gamma&\longmapsto\quad \gamma_x
\eea
such that $\Psi_\gamma(\widetilde x)\underset{def}{=}\Psi(\widetilde x + \gamma_x)$ and $\Psi$ must be related by right multiplication by a group element, constant by flatness of $\nabla$, called the monodromy of $\Psi$  along $\gamma$. Namely
\beq
\Psi_\gamma= \Psi \cdot S_\gamma
\eeq
where $S_\gamma\in G$ is independent of $x$. This yields the group morphism
\bea
\pi_1(\curverond_p) & \overset{S}{\longrightarrow} & G \cr
\gamma & \longmapsto & S_\gamma
\eea
called the monodromy representation of $\nabla$, $S\in\operatorname{Hom}\big(\pi_1(\curverond_p),G\big)$, and representing the action of $\pi_1(\curverond_p)$ on the flat sections of $\nabla$. Had we chosen a different reference point, this would have changed the monodromy data by conjugating it by a constant group element, this is a particular case of change of flat section $\Psi$. Furthermore, the monodromy map gets conjugated by gauge transformations and as such yields a class $[S]\in\modsp_{\operatorname{Betti}}$ with the following definition,

\bd[Betti moduli space]
\beq
\modsp_{\operatorname{Betti}}\underset{def}{=}\operatorname{Hom}\big(\pi_1(\curverond_{\mathfrak m}),G\big)\big/G
\eeq

\ed

\br Had we considered meromorphic gauge transformations -- instead of holomorphic ones -- on $\curverond$, with singularities located at $p_1,\dots,p_M$, this would imply a diffeomorphism between $\modsp_{\operatorname{Betti}}$ and $\modsp_p$ that is \textit{not} algebraic by the Riemann-Hilbert correspondence. However, since we are restricting ourselves to holomorphic gauge transformations, considering two meromorphic connections related by shifts of the weights of the form $\alpha_j\longrightarrow \alpha_j+n_j$ where $n_j\in\Lieg$ satisfies $\exp(2\pi\mathbf i\, n_j)=\bold 1_G$ for all $j\in\{1,\dots,M\}$, defines two distinct classes in $\modsp_p$ even though the two connections will have identical monodromy data. The lattice of solutions $n\in\Lieg$ of the equation $\exp(2\pi\mathbf i\, n)=\bold 1_G$ is therefore identified with the generic fiber of a covering $\modsp_p\longrightarrow\modsp_{\operatorname{Betti}}$.
\er

\section{Quantum geometry of the adjoint bundle}

\subsection{Forms and cycles}
\label{formsandcycles}

Had we considered holomorphic connections, corresponding to a number $M=0$ of singularities, a natural definition for the space of holomorphic one--forms on the adjoint bundle would have simply been to consider the homology space $\Ho^0\big(\curverond, \text{K}_{\curverond} \otimes \Ad\mathcal P^*\big)$. However, since we are mainly interested in the case where $\nabla$ has Fuchsian singularities, we shall add corresponding constraints on the behavior the differential forms have to exhibit near the poles of the connection.

\bd[Differential forms]
Generalized differential forms are defined as flat meromorphic sections of the bundle $\operatorname K_{\curverond_p}\otimes\Ad\mathcal P^*$, where $\Ad\mathcal P^*$ is the dual adjoint bundle. We distinguish three kinds of generalized differential forms.
\begin{itemize}
\item A generalized differential form $\omega$ is of the first kind (or holomorphic), if and only if it is holomorphic on $\curverond_p$ and in any chart containing a pole $p_j$, we have that for any root $\mathfrak r\in\mathfrak R$ and any Lie algebra element $E_{\mathfrak r}\in \Lieg_{\mathfrak r}$,
\beq
(x-z_j)^{\mathfrak r(\alpha_j)}\, \omega\left(x\cdot\Ad_{V_j} E_{\mathfrak r}\right)
\eeq
 is holomorphic at $x=z_j$. The corresponding space of holomorphic forms is denoted as $(\widehat\Ho{}^1_{\mathfrak m})'$.

\item A generalized differential form $\omega$ is of the third kind if  and only if it is meromorphic on $\curverond_p$ and in any chart containing a pole $p_j$, we have that for any root $\mathfrak r\in\mathfrak R$ and any Lie algebra element $E_{\mathfrak r}\in\Lieg_{\mathfrak r}$,
\beq
(x-z_j)^{\mathfrak r(\alpha_j)} \, \omega\left(x\cdot\Ad_{V_j} E_{\mathfrak r}\right)
\eeq
has at most a simple pole at $x=z_j$. The corresponding space of third kind differentials is denoted as $(\widehat\Ho{}^1_{\mathfrak m})'''$.

\item A generalized differential form $\omega$  meromorphic on $\curverond_p$ with poles of arbitrary degree is called second kind.
The corresponding space of second kind differentials is denoted as $(\widehat\Ho{}^1_{\mathfrak m})''$.

\end{itemize}

Note furthermore that 
\beq
(\widehat\Ho{}^1_{\mathfrak m})'\subset(\widehat\Ho{}^1_{\mathfrak m})'''\subset(\widehat\Ho{}^1_{\mathfrak m})''
\eeq
and that if we restrict ourselves to the leaf moduli space with fixed spectra of residues $\mathfrak m\in\modsp_{p,[\alpha]}$, then these spaces become independent of the choice of connection and hence define trivial bundles $(\widehat\Ho{}^1)'\subset(\widehat\Ho{}^1)'''\subset(\widehat\Ho{}^1)''\longrightarrow\modsp'_{p,[\alpha]}$. We then extend $(\widehat\Ho{}^1_{\mathfrak m})''$ to the space $\widehat{\mathfrak M}^1_{\mathfrak m}$ of generalized meromorphic differential forms by allowing unconstrained additional meromorphic singularities away from $p_1,\dots,p_M$.
\ed


Let as before $\mathfrak m=(\mathcal P,\nabla)\in\modsp'$  be a pair composed of a principal bundle and Fuchsian connection. In 1984 \cite{Goldman84} (following \cite{Dold72,Steenrod51}), Goldman introduced homology with local coefficients in the $\Lieg$--isomorphic fibers of $\Ad\mathcal P$ and associated to $\nabla$, in the case when it is a holomorphic connection. Let us denote this homology as $\Ho_*(\curverond,\Ad\mathcal P)$. It is the complex of smooth singular chain simplices $\gamma\otimes\sigma$, with $\gamma$ a smooth singular chain simplex from the closed interval $[0,1]$ to $\curverond$ and $\sigma$ a $\nabla$--flat section of $ \Ad\mathcal P$ defined on $\gamma$. The boundary operator is simply $\widehat\dd\underset{def}{=}\dd\otimes \text{ev}$, acting on a chain simplex $\gamma\otimes\sigma$ through
\beq
\widehat\dd\left(\gamma\otimes\sigma\right)\underset{def}{=}\gamma(1)\otimes\sigma\big(\gamma(1)\big)-\gamma(0)\otimes\sigma\big(\gamma(0)\big)
\eeq
Cycles are then defined as linear combinations of chain complexes whose total boundary vanishes (linear combinations use the vector space structure of $\Lieg$ in the fibers over each point). This yields a gauge equivariant chain complex \cite{Goldman84} that moreover defines the homology of $\widehat\curve$.
However, since $\nabla$ has poles, we shall slightly amend the construction of Goldman.

\medskip

%
%
%
%
%
%
%

\bd
\beq
\widehat\Ho{}'_k(\mathfrak m)\underset{def}{=} \Ho_k(\curverond_p,\Ad\mathcal P)
\eeq
\ed
Since $\curverond$ is a compact Riemann surface, we have $\widehat \Ho{}'_2(\mathfrak m)=0$, and the only relevant space of generalized cycles is
\beq
\widehat \Ho{}'_1(\mathfrak m) = \Ho_1(\curverond_p,\Ad\mathcal P)
\eeq

Let us observe at this point that there exist some rather trivial elements in the space of cycles $\widehat \Ho{}'_1(\mathfrak m)$, as follows. Consider a small Jordan loop $\gamma_{p_j}$ around $p_j$, with associated monodromy $S_{\gamma_{p_j}}$.
The chain $\gamma_{p_i}\otimes\sigma$ is closed, i.e. $\widehat\dd \left(\gamma_{p_i}\otimes\sigma\right)=0$ if and only if $T^{\parallel}_{\gamma_{p_j}}\sigma=\sigma$, namely the flat section $\sigma$ is left invariant under parallel transport by $\nabla$ along $\gamma_{p_j}$ (with holonomy denoted $T^{\parallel}_{\gamma_{p_j}}$) . This is equivalent to the requirement that the value of $\sigma$ at the starting point of the loop $\gamma_{p_j}$ lies in the centralizer of the monodromy $S_{\gamma_{p_j}}$.
Generically (i.e. considering the action of $S_{\gamma_{p_j}}$ to be non--degenerate) this centralizer is isomorphic to a Cartan sub--algebra, which we will denote $\Lieh_j$ in the rest of this paper. Note that it contains in particular the center $\mathcal Z\Lieg$ of $\Lieg$.

\bd
Define the reduced homology space by quotienting out these cycles localized at the singularities of the connection, that is
\beq
\widehat \Ho_1(\mathfrak m) \underset{def}{=} \widehat \Ho{}'_1(\mathfrak m)  \Big/   \bigoplus_{j=1}^M \gamma_{p_j}\otimes\Lieh_j
\eeq
\ed

In the construction that follows, we will need to integrate some generalized differential forms on cycles whose end points are singularities of $\nabla$. We shall therefore need to enlarge our space of generalized cycles by allowing the presence of open arcs ending at the $p_j$'s. However, since flat sections of the connection are not defined over the locus of poles, we define that any cycle ending at a given $p_j$ has vanishing boundary contribution from  this point.

\bd[Third kind cycles]
Define the space of generalized cycles of the third kind to be
\beq
\widehat \Ho{}'''_1(\mathfrak m) \underset{def}{=} \Ho_1(\curverond,\Ad \mathcal P)
\eeq
where we defined the boundary of a chain ending at some pole $p_j$ to vanish. 
\ed

\bt
All these spaces of generalized cycles are gauge equivariant and have finite dimensions equal to (recall $\genus$ from \eqref{defgenus})
\beq
\dim \widehat \Ho{}'_1(\mathfrak m) = 2\genus+M \dim\Lieh .
\eeq
\beq
\dim \widehat \Ho_1(\mathfrak m) = 2\genus .
\eeq
\beq
\dim \widehat \Ho{}'''_1(\mathfrak m) = 2\genus +2M\dim\Lieh\underset{def}{=}2\genus'''.
\eeq

\et
Notice that the spaces $\widehat \Ho_1(\mathfrak m)$ and $\widehat \Ho{}'''_1(\mathfrak m)$ are of even dimension and they moreover carry a symplectic structure that we shall describe in the next paragraphs.

\proof We postpone the proof to section \ref{basiscycles}, where we will explicitly construct a basis for each of these homology spaces. They are built using a choice of fundamental domain of $\curverond_p$, together with a choice of basis of flat sections over the homology classes defined by smooth singular one--simplices.
\eproof

\subsection{Intersection}

Goldman \cite{Goldman84} defined an intersection product on $\Ho_k(\curverond_{\mathfrak m},\Ad\mathcal P)$ by the oriented algebraic sum of the Killing bracket of the section components of the cycles over their base intersection points.
Goldman showed that this is a non degenerate symplectic form on $\widehat \Ho_1(\mathfrak m)$.
We now extend it to a non--degenerate symplectic form on $\widehat \Ho{}'''_1(\mathfrak m)$.

\bd[Intersection of generalized cycles]
Define the intersection product of two generalized chains $\Gamma_1$ and $\Gamma_2$ to be 
\beq
\Gamma_1\bigcap\Gamma_2\underset{def}{=} \left(\gamma_1\cap\gamma_2\right)\left\langle E_1\, E_2\right\rangle
\eeq
when $\Gamma_i=\gamma_i\otimes \sigma_i$ and $E_i=\sigma_i(p)$ if $\gamma_1$ and $\gamma_2$ intersect at a unique point $p$, then extended the definition linearly to the whole space $\widehat \Ho{}'''_1(\mathfrak m)$.
\ed

\subsection{Integration Poincar\'e--pairing}

\bd
The generalized Poincar\'e--pairing between $\widehat\Ho_1(\mathfrak m)$ and generalized differential forms is defined by the integration of forms on cycles and evaluation of Lie algebra elements on their duals
\beq\label{defPoincarepairing}
\left\langle\Gamma,\omega\right\rangle \underset{def}{=} \underset\Gamma\oint \omega = \sum_i c_i \underset{t=0}\int^1 \big\langle\omega\big(\gamma_i(t)\big) , \sigma_i\big(\gamma_i(t)\big)\big\rangle
\eeq
This is independent of the representative of $\Gamma=\sum_i c_i \left(\gamma_i\otimes \sigma_i\right)$, and independent of a choice of chart, and moreover this is gauge invariant.
\ed
\br \eqref{defPoincarepairing} is defined only if $\omega$ has no poles at $p_j$,
 and we need to define $\left\langle\Gamma,\omega\right\rangle$ also for all 3rd kind cycles and forms, but in order to make the presentation concise, we extend the definition of "regularized" integration pairing for 3rd kind cycles in appendix \ref{appintegral3rdkind}. Extension that we now use.
 \er

\bd[From forms to cycles]
Recalling that the dimension of the space of third kind generalized cycles was denoted by $2\genus'''=\dim\widehat\Ho{}'''_1(\mathfrak m)$, define the map

\bea
\widehat C\underset{def}{:} &(\widehat\Ho{}_{\mathfrak m}^1)''' &\longrightarrow\ \  \widehat \Ho{}'''_1(\mathfrak m)\cr
&\omega&\longmapsto\ \ \widehat C(\omega) = \sum_{k,l=1}^{2\genus'''} \acycle_k \ (I^{-1})_{k,l} \ \underset{\acycle_l}\oint \omega
\eea
 for any given basis $\{\acycle_i\}_{i=1}^{2\genus'''}$ with intersection matrix  $I=(I_{k,l})_{k,l=1}^{2\genus'''}$ with $I_{k,l}\underset{def}{=}\acycle_k\bigcap\acycle_l$.
\ed

\br This definition is manifestly invariant under changes of basis of $\widehat\Ho{}'''_1(\mathfrak m)$
\er

\subsubsection{Riemann bilinear identity}

Let $\Upsilon$ be a one--face graph with oriented edges such that $\curve_\Upsilon\underset{def}{=}\curverond-\Upsilon$ defines a fundamental domain of $\curverond_p$, see appendix \ref{appfunddom}. Its boundary is
\beq
\dd \curve_\Upsilon = \sum_{e\, :\, \text{edge of }\Upsilon} \quad e_+ - e_-
\eeq
where $e_+$ (resp. $e_-$) denotes the right (resp. left) side of the oriented edge $e$ in $\curve_\Upsilon$. This fundamental domain is by definition homeomorphic to a disc.

\bl
\label{lemmaprimitivetubularbnd}
Let $\omega$ be a generalized meromorphic one--form and consider a tubular neighborhood $U$ of $\dd \curve_\Upsilon$ that avoids all poles of $\omega$ which are not the poles of $\nabla$. Then there exists a local section $f\in \Ho^0(U, \Ad \mathcal P^*) $ such that
\beq
\diff f=\omega.
\eeq
\el
\proof
The tubular neighborhood $U$ of the boundary of a disc has the topology of an annulus and given a generic reference point $o'\in U$, there exist two independent homology classes of paths from $o'$ to any other point $x\in U$. Their difference is  the boundary homology class $\partial\Sigma_\Upsilon$ which actually vanishes on $\curverond$ by definition since it is the sum of the differences of edges that are identified. We therefore have $\oint_{\partial\Sigma_\Upsilon}\omega=0$ which exactly implies that the point-wise expression $f(x)\underset{def}{=}\int_{o'}^x\omega$ defines unambiguously the wanted function.
\eproof

\bt[Riemann bilinear identity]
\label{RBId}
Let $\omega_1,\omega_2\in\widehat{\mathfrak M}^1_{\mathfrak m}$ be two generalized meromorphic one--forms on $\widehat\curve_{\mathfrak m}$ and let $f_1$ (resp. $f_2$) be a primitive of $\omega_1$ (resp. $\omega_2$) in the tubular neighborhood $U$ as in lemma \ref{lemmaprimitivetubularbnd}. Consider a basis $\{\acycle_k\}_{k=1}^{2\genus'''}$ of the space of third--kind cycles $\widehat \Ho{}'''_1(\mathfrak m)$, with intersection matrix $I_{k,l}=\acycle_k\bigcap\acycle_l$. Then we have
\beq
\underset{\dd \curve_\Upsilon} \oint \langle\,\omega_1,\, f_2\,\rangle = -\underset{\dd \curve_\Upsilon} \oint  \langle\,\omega_2, \,  f_1\,\rangle
= \sum_{k,l=1}^{2\genus'''}\, \big(\underset{\acycle_k}\oint \omega_1 \big) \, (I^{-1})_{k,l}\,  \big(\underset{\acycle_l}\oint \omega_2\big)
\eeq
\et

\proof
Let $e_1,e_2,\dots,e_{\dim\Lieg}$ be an arbitrary basis of $\Lieg$, and consider its dual basis $e^1,\dots,e^{\dim\Lieg}$. It satisfies by definition $\langle e_r,e^s\rangle=\delta_{r,s}$.
We have
\bea
 \underset{\dd \curve_\Upsilon} \oint \langle\,\omega_1,\, f_2\,\rangle
 &=& \sum_{e\,:\, \text{edge of }\Upsilon} \underset{e_+-e_-}\int \langle\,\omega_1,\, f_2\,\rangle \cr
 &=& \sum_{e\,:\, \text{edge of }\Upsilon} \sum_{r=1}^{\dim\Lieg} \underset{e_+-e_-}\int \langle\,\omega_1,\, e_r\,\rangle\,\langle \,e^r ,\, f_2\,\rangle \cr
 &=& \sum_{e\,:\, \text{edge of }\Upsilon} \sum_{r=1}^{\dim\Lieg} \underset{e_+}\int \langle\,\omega_1,\, e_r\,\rangle\, \underset{\gamma_e} \oint \langle \, e^r ,\, \omega_2\,\rangle  \cr
 &=& \underset{\Gamma(\omega_2)}\oint \omega_1
\eea
where we defined a sum of $\Lieg$--weighted arcs by
\beq
\Gamma(\omega_2)\underset{def}{=}\sum_{e\,:\, \text{edge of }\Upsilon} \sum_{r=1}^{\dim\Lieg} \left(e_+\otimes e_r\right)\cdot \big( \underset{\gamma_e}\oint \langle\, e^r ,\,\omega_2\,\rangle \big)
\eeq
In order to show that $\Gamma(\omega_2)\in \widehat \Ho{}'''_1(\mathfrak m)$, we have to check that its boundary vanishes.
It is a sum of $\Lieg$ weighted vertices of $\Upsilon$:
\beq
\widehat\partial\, \Gamma(\omega_2) = \sum_{v\,:\, \text{vertex of } \Upsilon} v\otimes \Big(\sum_{e \mapsto v}\  \underset{\gamma_e}\oint \ \omega_2 \Big)
\eeq
which indeed vanishes because $\sum_{e\mapsto v} \gamma_e=0$ at each vertex of $\Upsilon$. 



We can thus generically decompose it on the basis of $\widehat \Ho{}'''_1(\mathfrak m)$ as
\beq
\Gamma(\omega_2)\underset{def}{=} \sum_{k=1}^{2\genus'''}
c_k(\omega_2)  \acycle_k.
\eeq
and the skew--symmetry $\underset{\Gamma(\omega_2)}\oint \omega_1 = -\underset{\Gamma(\omega_1)}\oint \omega_2$ implies furthermore that
\beq
c_k =\sum_{l=1}^{2\genus'''} c_{k,l}\, \acycle_l 
\eeq
where $c_{k,l}=-c_{l,k}$ denotes the coefficient of a skew--symmetric matrix. Define therefore
\beq
\Gamma \underset{def}{=} \sum_{k,l=1}^{2\genus'''} c_{k,l} \acycle_k \otimes \acycle_l = \sum_{e\,:\,\text{edge of }\Upsilon} \sum_{r=1}^{\dim\Lieg} (e_+\otimes e_r) \otimes (\gamma_e\otimes e^r)
\eeq
Each cycle $\acycle_k$ can be homotopically deformed either to the boundary of $\Upsilon$, and be written as
\beq
\acycle_k = \sum_{e\,:\, \text{edge of }\Upsilon} e_+\otimes E_{e,k},
\eeq
or equivalently by deforming it to the dual graph
\beq
\acycle_k = \sum_{e\,:\, \text{edge of }\Upsilon} \gamma_e\otimes\widetilde E_{e,k}.
\eeq
which implies the following formula for the intersection matrix
\beq
I_{k,l} = \acycle_k \bigcap \acycle_l = \sum_{e\,:\, \text{edge of }\Upsilon} \langle E_{e,k} , \widetilde E_{e,l}\rangle
\eeq
The intersection product $\bigcap$ is here viewed as a bilinear form on $\widehat\Ho{}'''_1(\mathfrak m)^{\otimes 2}$ and it therefore defines a bilinear form $\underset\otimes\bigcap$ on $\widehat\Ho{}'''_1(\mathfrak m)^{\otimes 2}\otimes\widehat\Ho{}'''_1(\mathfrak m)^{\otimes 2}$ by intersecting the first and third, respectively second and fourth, tensor factors of pure tensors together, multiplying the two obtained results and then extending linearly to the whole space. In particular 
\beq
\left(\acycle_k\otimes \acycle_l\right)\, \underset\otimes\bigcap\ \Gamma =  \sum_{k',l'=1}^{2\genus'''} I_{k,k'}  c_{k',l'} I_{l,l'}
\eeq
and using the expressions obtained by deformation implies
\bea
\left(\acycle_k\otimes \acycle_l\right)\, \underset\otimes\bigcap\ \Gamma 
&=& \sum_{e\,:\, \text{edge of }\Upsilon} \sum_{r=1}^{\dim\Lieg} \left(\acycle_k\bigcap e_+\otimes e_r\right)\cdot \left(\acycle_l\bigcap \gamma_e\otimes e^r\right) \cr
&=& - \sum_{e\,:\, \text{edge of }\Upsilon} \sum_{r=1}^{\dim\Lieg} \langle\widetilde E_{e,k},e_r\rangle \cdot \langle E_{e,j},e^r\rangle  \cr
&=& - \sum_{e\,:\, \text{edge of }\Upsilon} \langle \widetilde E_{e,k},E_{e,l}\rangle  \cr
&=& - I_{k,l} 
\eea
In other words we showed that $I \cdot c\cdot  I^T = -I = I^T$ which implies $c= I^{-1}$ and in turn the wanted formula.
\eproof

\subsubsection{Riemann bilinear inequality}

Recall that a real Lie algebra $\Lieg_{\mathbb R}$ is a real--form of the reductive Lie algebra $\Lieg$ if it satisfies $\Lieg\sim\Lieg_{\mathbb R} \otimes\mathbb C$. As such it defines a complex structure on $\Lieg$.
 
\bt[Riemann bilinear inequality]
\label{RBIn}
For any basis $\{\mathcal A_k\}_{k=1}^{2\genus'''}$ of third--kind generalized cycles with intersection matrix $I=(I_{k,l})_{k,l=1}^{2\genus'''}$ and any non--zero generalized meromorphic differential $\omega\in\widehat{\mathfrak M}^1_{\mathfrak m}$,
\beq
0\,<\,\frac{\mathbf i}2 \sum_{k,l=1}^{2\genus'''}\, \big(\underset{\acycle_k}\oint \omega \big) \, (I^{-1})_{k,l}\,  \big(\underset{\overline{\acycle_l}}\oint \overline\omega\big)
\eeq
\et
\proof 
Introduce a L$^2$--type norm
\bea
0<||\omega||^2&\underset{def}{=}& \frac{\mathbf i}2 \underset{\curve_\Upsilon}\iint \langle\,\omega\wedge\overline\omega\,\rangle \\
&=& \frac{\mathbf i}2 \underset{\partial\curve_\Upsilon}\oint \langle\,\omega,\,\overline f\,\rangle
\eea
where we used the bilinear bracket $\langle\,\bullet\,\rangle$, introduced a primitive $\omega=\diff f$ and applied Stoke's theorem to push the integration to the boundary. The Riemann bilinear identity then yields the wanted inequality.
\eproof

The Riemann bilinear inequality takes the form
\beq
0\,<\mathbf i\underset{\widehat C(\overline\omega)}\oint \omega=\,-\mathbf i\underset{\widehat C(\omega)}\oint \overline\omega
\eeq
which implies in particular that $\widehat C(\omega)\neq 0$ for any non--zero first--kind generalized differential form $\omega$. The map $\widehat C$ is therefore injective and as a consequence, its image in $\widehat\Ho{}'''_1(\mathfrak m)$ is isomorphic to $(\widehat\Ho{}^1_{\mathfrak m})'''$. We shall prove in the following that it defines a Lagrangian subspace of $\widehat\Ho{}'''_1(\mathfrak m)$ with respect to the previously introduced intersection form. For this purpose, let us start by introducing a left-inverse $\widehat B$ of $\widehat C$. It is a map $\widehat B :  \widehat\Ho{}'''_1(\mathfrak m)  \longrightarrow (\widehat\Ho{}^1_{\mathfrak m})'''$ that satisfies by definition
\bea
\widehat B\,\circ\,\widehat C & =& 2\pi\mathbf i \operatorname{Id}_{(\widehat\Ho{}^1_{\mathfrak m})'''}\\
\widehat\Ho{}'''_1(\mathfrak m) & =& \Img\widehat C\oplus \Ker\widehat B
\eea
and we will express it in terms of deformations of the quantum Liouville form that we now define.

\subsection{Quantum Liouville form and universal cycle}
\label{qLiouville}

Recall that $\Upsilon\subset\curverond$ is a one--face graph with oriented edges such that $\curve_\Upsilon=\curverond-\Upsilon$ defines a fundamental domain of $\curverond$ over which all the considered fiber bundles are trivial. We now work in this setup and identify all fibers of $\mathcal P$ ($\Ad\mathcal P$ respectively) over $\curve_\Upsilon$. We will moreover show in the next section that such a choice of graph allows to define useful sets of 
coordinates over $\modsp'_p$.

\bd[Self-reproducing kernel]
\label{selfrepker}
Associate to any multivalued global $\nabla$--flat section $\Psi$ a kernel defined by
\beq
\mathcal K_\Psi(\widetilde x,\widetilde y) = \underset{z=y}{\Res}\left( \frac{\Psi(\widetilde x)^{-1}\cdot\Psi(\widetilde z)}{\mathcal E(x,z)\,\mathcal E(z,y)}\right)
\eeq
where $\mathcal E$ is Fay's twisted prime--form that we define in appendix \ref{primeform}. $\mathcal K_\Psi$ is a multivalued section of the twisted bundle $(\operatorname K_{\curverond_p}^{\frac 12}\otimes \mathcal P)^{\otimes 2}\longrightarrow (\curverond_\Upsilon)^2$ which is discontinuous on the diagonal $\Delta_2\subset(\curve_\Upsilon)^2$.
\ed

It solves the double Riemann-Hilbert problem defined by the property that for any points $\widetilde x,\widetilde y\in \overset{\circ}{\widetilde\Sigma}_p$ and any loop $\gamma\in\pi_1(\curverond_p,o)$, with corresponding representatives $\gamma_x\in\pi_1(\curverond_p,x)$ and $\gamma_y\in\pi_1(\curverond_p,y)$, 
\bea
\mathcal K_\Psi(\widetilde x+\gamma_x,\widetilde y) &=& S_\gamma^{-1}\cdot\mathcal K_\Psi(\widetilde x,\widetilde y)\\
\mathcal K_\Psi(\widetilde x,\widetilde y+\gamma_y) &=& \mathcal K_\Psi(\widetilde x,\widetilde y)\cdot S_\gamma
\eea
where $S_\gamma$ is the monodromy of $\Psi$ around $\gamma$. It moreover has the diagonal asymptotics
\bea
\mathcal K_\Psi(\widetilde x,\widetilde y)  &\underset{\widetilde x\sim \widetilde y+\gamma_y}{=}&  \frac {S_\gamma^{-1}}{x-y}\sqrt{\diff x\, \diff y} +\, \mathcal O(1)
\eea

We now define a non-perturbative analogue of the Seiberg--Witten differential, or dispersive tautological one-form. In order to do so, let us first introduce the connection potential $\Phi$, an adjoint--valued meromorphic one-form $\Phi\in\Ho^0(\curve_\Upsilon,\operatorname{K}_{\curve_\Upsilon}\otimes \Ad \mathcal P)$ satisfying $\nabla_\Upsilon = \diff - \Phi$, $\nabla_\Upsilon$ denoting the restriction of $\nabla$ to $\curve_\Upsilon$.

\bd[Quantum Liouville form]
Let $W_1$ be the multivalued third--kind quantum differential on $\widehat\curve_{\mathfrak m}$ defined on the quantum spectral curve by 
\bea
W_1:&\widehat\curve_{\mathfrak m} &\longrightarrow (\overset\circ\pi)^*\operatorname K_{\curverond_p}\\
& \widetilde x\cdot\sigma & \longmapsto \left\langle \Phi(x),\sigma(\widetilde x)\right\rangle
\eea
and if $\Psi$ is a multivalued $\nabla$-flat section, we equivalently rewrite this as 
\beq
W_1\left(\widetilde x\cdot E\right)\underset{def}{=}\big\langle\Phi( x),\sigma_\Psi(\widetilde x\cdot E)\big\rangle
\eeq 
\beq
\text{with}\quad \Phi( x)=\Ad_{\Psi(\widetilde x)}\,\mathcal K_\Psi(\widetilde x,\widetilde x),
\eeq
$\mathcal K_\Psi$ being the self-reproducing kernel associated to $\Psi$. $\sigma_\Psi(\bullet,\cdot E)\in\widehat\curve_{\mathfrak m}$ denotes the unique multivalued holomorphic section of $\Ad\mathcal P$ over $\curve_\Upsilon$ with prescribed initial condition $\sigma_\Psi(\{o\}\cdot E) = E$ and such that for any constant Lie algebra element $E\in\Lieg$, it has monodromy with respect to the base curve given by
\beq
\sigma_\Psi(\widetilde x+\gamma_x\cdot E)=\sigma_\Psi(\widetilde x\cdot\Ad_{S_\gamma}E)
\eeq
where $\gamma_x\in\pi_1(\curverond_p,x)$ is associated to the loop $\gamma\in\pi_1(\curverond_p,o)$ from the corresponding base-point $x\in\curverond_p$. $W_1$ depends on the choice of $\nabla$--flat section $\Psi$ only through composition with the conjugation by a constant group element here denoted $C\in G$
\beq
\Psi\longmapsto \Psi\cdot C\qquad \Longrightarrow\qquad W_1(\widetilde x\cdot E)\longmapsto W_1(\widetilde x\cdot \Ad_CE)\qquad\qquad\\
\eeq
and similarly, the monodromy of the section $\Psi$ implies
\bea
W_1(\widetilde x+\gamma_x\cdot E) & = & W_1(\widetilde x\cdot \Ad_{S_\gamma}E)
\eea

$W_1$ depends on $\mathfrak m$ (we didn't write it explicitly) but only in a gauge invariant way. Using the isomorphism $\widehat\curve_{\mathfrak m}\underset{\Ad_\Psi^{-1}}{\simeq}\Lieg$, the quantum Liouville form $W_1:\widehat\curve_{\mathfrak m}\longrightarrow (\overset{\circ}{\pi})^*\operatorname K_{\curverond_p}$ is interpreted as a multivalued element $W_1\in(\widehat\Ho{}^1_{\mathfrak m})'''$ but defines in general a gauge invariant holomorphic bundle map called the quantum Liouville form
\beq
W_1\in\underset{\modsp_p}{\operatorname{Bun}}\left(\widehat\curve\big/\mathcal G,(\overset{\circ}{\pi})^*\operatorname{K}_{\curverond_p}\right)
\eeq
from the bundle $\widehat\curve\big/\mathcal G\longrightarrow\modsp_p$ of quantum spectral curves up to gauge transformations to $(\overset{\circ}{\pi})^*\operatorname{K}_{\curverond_p}\longrightarrow\modsp_p$, the trivial bundle whose fiber is the space of $\Lieg^*$-valued holomorphic one--forms over the universal covering $\overset{\circ}{\widetilde\curve}_p$ of $\curverond_p$.
\ed

\proof
Compactness of the base curve $\curverond$ implies that any holomorphic gauge transformation $g\in\mathcal G=\Ho^0(\curverond,\mathcal P)$ is a constant. The second term of the right-hand side of
\beq
g\cdot W_1=W_1+\left\langle \diff g\cdot g^{-1},\,\bullet\,\right\rangle
\eeq
therefore vanishes and $W_1$ is gauge invariant.
\eproof

Let $o\in\curverond_\Upsilon$ be a smooth reference point in $\curverond_\Upsilon$. $\Psi$ behaves near a singularity $p_j$ as \eqref{Psinearzj}:
\beq
\Psi(\widetilde x) \underset{x\sim p_j}= V_j \cdot\big(\bold 1_G+\mathcal O(x-z_j)\big) \cdot (x-z_j)^{\alpha_j}\cdot C_j
\eeq
where $C_j\in G$ is a constant in the group and $\Phi_j=\Ad_{V_j}(\alpha_j)=\underset{p_j}{\Res}\nabla$. We immediately get the asymptotics 
\bea
\Phi(x)&\underset{x\sim p_j}{=}& \frac{\Phi_j}{x-z_j}\diff x\,+\,\mathcal O(1)\\
W_1(\widetilde x\cdot E) &\underset{x\sim p_j}{=}& \frac{\left\langle \alpha_j, \Ad_{C_j}E\right\rangle}{x-z_j}\diff x\, + \, \mathcal O(1)
\eea

\bd[Universal cycle]
\beq
\widehat\Gamma\underset{def}{=}\widehat C(W_1)\in \Ho^0(\modsp_p,\widehat\Ho{}'''_1)
\eeq
\ed

\section{Deformations}

\subsection{Form-cycle duality}

Goldman \cite{Goldman84} showed that the tangent space $T_{\mathfrak m}\modsp$ carries a symplectic structure, isomorphic to that of $\widehat \Ho_1(\mathfrak m)$.  We now extend this construction to the Fuchsian case where the connection $\nabla$ has simple poles.

\smallskip

Let us therefore consider a fundamental domain $\curve_\Upsilon=\curverond-\Upsilon $ of $\curverond_p$ (see appendix \ref{appfunddom}), with $\Upsilon$ a one--face graph on $\curverond$. We still denote by $\Psi$ the restriction of a multivalued $\nabla$--flat section of $\mathcal P$ to $\curve_\Upsilon$ and furthermore, for each edge of $\Upsilon$, we denote $S_e$ the monodromy of $\Psi$ across $e$, i.e. the constant group element defined by the relation $\Psi(x+e^\perp) = \Psi(x)\cdot S_e$, where we introduced $e^\perp$ as the unique oriented homology class of $\curverond$ crossing $\Upsilon$ only once, along $e$, and oriented such that it crosses from $e_-$ to $e_+$.

\bt\label{thTangenttocycles}
The fiberwise map from tangent vectors to generalized cycles defined by 
\bea
T_{\mathfrak m} \modsp_p & \longrightarrow & \widehat \Ho{}'''_1(\mathfrak m)\Big/   \bigoplus_{j=1}^M \gamma_{p_j}\otimes\Lieh_j \cr
\delta & \longmapsto & \Gamma_\delta \underset{def}{=} \frac 1{2\pi\mathbf i} \sum_{e\, : \,\operatorname{edge\, of }\Upsilon} e\otimes (\delta S_e  \cdot S_e^{-1})
\eea
is well defined and is invertible with inverse given by

\pagebreak 

\bea
\widehat \Ho{}'''_1(\mathfrak m)\Big/   \bigoplus_{j=1}^M \gamma_{p_j}\otimes\Lieh_j & \longrightarrow & T_{\mathfrak m} \modsp_p  \cr
\Gamma \quad & \longmapsto & \partial_\Gamma 
\eea
where the tangent vector $\partial_\Gamma$ appearing in the last map is defined by
\bea
\partial_\Gamma \Phi(x) &\underset{def}{=}& \diff F_\Gamma(x) + [F_\Gamma(x),\Phi(x)]\\
\text{with}\quad F_\Gamma(x)& \underset{def}{=} & \underset{X'\in\Gamma}\oint \omega'''_{x,o}(x') \sigma_\Psi(X')
\eea
where $\omega'''_{x,o}\in\Ho^0\big(\curverond,\operatorname K_{\curverond}(x+o)\big) $ is any meromorphic (third--kind) differential on $\curverond$ with simple poles at the point $x$ and at a chosen generic reference point $o$ with corresponding residues $\underset x\Res\, \omega'''_{x,o}=1=-\underset o\Res\, \omega'''_{x,o}$.
\et

\proof
Proof in appendix \ref{apptangenttocycles}
\eproof.

\bc \label{holcor}
For any holomorphic differential $\omega$ on $\curverond$ and any $\Gamma\in\widehat\Ho{}'''_1(\mathfrak m)$,
\beq
\underset{X\in\Gamma}\oint \omega(x)\sigma_\Psi(X)=0
\eeq
\ec

\proof
Since the previous construction does not depend on the choice of third--kind differential $\omega_{x,o}'''$, we get the result for any $\Gamma\in\widehat\Ho{}'''_1(\mathfrak m)\Big/   \bigoplus_{j=1}^M \gamma_{p_j}\otimes\Lieh_j$. For the remaining cycles, observe that $\sigma_\Psi(x\cdot\,\bullet\,)$ restricted to $\Lieh_j$ (for any $j\in\{1,\dots,M\}$) is holomorphic at $x=p_j$. This implies in particular that for any $H\in\Lieh_j$,
\beq
\underset{x\cdot H\in \gamma_{p_j}\otimes H}\oint\,\omega(x)\sigma_\Psi(x\cdot H)=\underset{x=p_j}\Res\, \omega(x)\sigma_\Psi(x\cdot H)=0 
\eeq
which is what was left to be shown.
\eproof

\bt[Goldman symplectic form]
The intersection pairing on the coset $\widehat \Ho{}'''_1(\mathfrak m)\Big/   \bigoplus_{j=1}^M \gamma_{p_j}\otimes\Lieh_j$ pulls back to the Goldman symplectic form on 
$T_{\mathfrak m} \modsp_{p}$.
\et

\br
We can study tangent vectors that keep the charges $[\alpha]$ fixed, $\Gamma_{\delta}$ would then in fact lie in $\widehat\Ho_1(\mathfrak m)$.
\beq
 T_{\mathfrak m} \modsp_{p,[\alpha]} \sim  \widehat\Ho_1(\mathfrak m) .
\eeq
\er

\bp
\label{defrepker}
For any generic deformation $\delta\in T_{\mathfrak m}\modsp_p$,
\beq
\delta\mathcal K_\Psi( x, y) = -\underset{z\cdot E\in\Gamma_\delta}\oint \mathcal K_\Psi( x, z)\cdot E\cdot \mathcal K_\Psi( z, y)
\eeq
\ep

\proof
The deformation $\delta\Psi \underset{def}= F_\delta\cdot\Psi$ and the resulting equation $\delta\Phi  =  \diff F_\delta+[F_\delta,\Phi]$ allow to explicit the deformation for generic $\widetilde x$ and $\widetilde y$ as
\bea
\delta \mathcal K_\Psi (\widetilde x,\widetilde y) &=& \frac{\delta\Psi(\widetilde x)^{-1}\cdot\Psi(\widetilde y)}{\mathcal E( x, y)}+\frac{\Psi(\widetilde x)^{-1}\cdot\delta\Psi(\widetilde y)}{\mathcal E( x, y)}\\
&=& \Ad_{\Psi(\widetilde x)}^{-1}\big(-F_\delta(\widetilde x)+F_\delta(\widetilde y)\big) \cdot \mathcal K_\Psi(\widetilde x,\widetilde y)\\
&=&-\underset{Z\in\Gamma_\delta}\oint \frac{\omega'''_{x,y}(z)}{\mathcal E( x, y)} \Psi(\widetilde x)^{-1}\cdot \sigma_\Psi(Z)\cdot \Psi(\widetilde y)
\eea
where the expressed the difference of values of $F_\delta$ like in the proof of theorem \ref{thTangenttocycles}.  $\frac{\omega'''_{x,y}(z)}{\mathcal E( x, y)} - \frac 1{\mathcal E(x,z)\mathcal E(z,y)}$ is furthermore holomorphic in $z$ thus yielding the wanted formula by corollary \ref{holcor}. To extend the formula to the coinciding base points we use the general definition of the self--reproducing kernel
\bea
\delta\mathcal K_\Psi( x, y) &=& \delta \underset{t=y}\Res \left(\frac{\mathcal K_\Psi( x, t)}{\mathcal E(x,t)}\right)\\
&=&\underset{t=y}\Res\left(\frac{\delta\mathcal K_\Psi( x, t)}{\mathcal E(x,t)}\right)\\
&=& - \underset{z\cdot E\in\Gamma_\delta}\oint \underset{t=y}\Res \left(\frac{\mathcal K_\Psi( x, z)\cdot E\cdot \mathcal K_\Psi( z, t)}{\mathcal E(x,t)}\right)\\
&=&-\underset{z\cdot E\in\Gamma_\delta}\oint \mathcal K_\Psi( x, z)\cdot E\cdot \mathcal K_\Psi( z, y)
\eea
\eproof

We compute similarly the deformation properties of the quantum Liouville form.

\bt[Bertola--Malgrange form]
\bea
\delta W_1 &=&  \widehat B (\Gamma_\delta)\\
\text{with}\qquad \widehat B (\Gamma)(X) & \underset{def}{=} & \underset{X'\in\Gamma}\oint\diff_x\omega'''_{x,o}(x')\,\big\langle\sigma_\Psi(X'),\sigma_\Psi(X)\big\rangle\\
&&-\underset{X'\in\Gamma}\oint\,\omega'''_{x,o}(x')\,\big\langle\big[\Phi(x),\sigma_\Psi(X')\big],\,\sigma_\Psi(X)\big\rangle
\eea
where we explicited $\widehat B:\widehat\Ho{}'''_1(\mathfrak m)\longrightarrow (\widehat\Ho{}^1_{\mathfrak m})'''$ (left--inverse of the map $\widehat C$) as yielding the curvature of the Bertola--Malgrange one--form \cite{M2004,B2010,B2016}.
\et

\pagebreak

\proof
The equation $\delta\sigma_\Psi=[F_\delta,\sigma_\Psi]$ is used to compensate the variation of $\Psi$ in the definition of the quantum Liouville form that therefore deforms as
\bea
\delta W_1 &=& \delta\left\langle\,\Phi\,,\,\sigma_\Psi\,\right\rangle\,-\,\left\langle\,\Phi\,,\,\delta\sigma_\Psi\,\right\rangle\\
&=&\left\langle\,\diff F_\delta\,,\,\sigma_\Psi\,\right\rangle\,+\,\big\langle[F_\delta,\Phi],\,\sigma_\Psi\big\rangle
\eea
We therefore conclude that
\beq
\delta W_1(X)=\underset{X'\in\Gamma_\delta}\oint\,\diff_x\omega'''_{x,o}(x')\,\big\langle\sigma_\Psi(X'),\,\sigma_\Psi(X)\big\rangle\,-\underset{X'\in\Gamma_\delta}\oint\,\omega'''_{x,o}(x')\big\langle\big[\Phi(x),\sigma_\Psi(X')],\,\sigma_\Psi(X)\big\rangle
\eeq
To prove that $\widehat B$ is indeed the left--inverse of $\widehat C$, we use the Cauchy residue formula and the Riemann bilinear identity for any $\omega=\diff f\in (\widehat\Ho{}^1_{\mathfrak m})'''$ and generic values of the arguments
\bea
\omega(x\cdot E) &=& \frac 1{2\pi\mathbf i}\underset{x'\in\mathcal C_x}\oint\, f(x'\cdot E)\,\diff_x\omega'''_{x,o}(x')\\
&=&\frac 1{2\pi\mathbf i}\underset{x'\in\mathcal C_x}\oint\sum_{r=1}^{\dim\Lieg}\, f\left(x'\cdot e_r\right) \nonumber\\
&&\qquad\qquad\qquad\ \times\ \diff_x\omega'''_{x,o}(x')\,\big\langle\, \sigma_\Psi(x'\cdot e^r)\,,\, \sigma_\Psi(x'\cdot E) \,\big\rangle
\eea
where we used the $\nabla$--flat basis $\{\Ad_{\Psi(\widetilde x')}e_r\}_{r=1}^{\dim\Lieg}$ and its dual $\{\Ad_{\Psi(\widetilde x')}e^r\}_{r=1}^{\dim\Lieg}$ to decompose Lie algebra elements. Note that the fundamental domain abelianizes the base--curve. Now replacing $\sigma_\Psi(x'\cdot E)$ by it's Taylor series expansion 
\beq
\sigma_\Psi(x'\cdot E)\underset{x'\sim x}= \sigma_\Psi(x\cdot E)+(x'-x)\big[\Phi(x),\sigma_\Psi(x\cdot E)\big]+\mathcal O(x'-x)^2
\eeq
around $x'=x$, using the invariance of the bracket $\left\langle\bullet,\bullet\right\rangle$ and then using the Riemann bilinear identity gives
\bea
\omega(x\cdot E) &=& \frac 1{2\pi\mathbf i}\,\Big( \underset{X'\in\widehat C(\omega)}\oint\,\diff_x\omega'''_{x,o}(x')\,\big\langle \sigma_\Psi(X'),\, \sigma_\Psi(x\cdot E) \big\rangle\nonumber\\
&&\qquad\qquad\qquad - \underset{X'\in\widehat C(\omega)}\oint\,(x'-x)\diff_x\omega'''_{x,o}(x')\,\big\langle\big[\Phi(x), \sigma_\Psi(X')\big],\sigma_\Psi(x\cdot E)\big\rangle\Big)\nonumber\\
\eea
Furthermore, $(x'-x)\diff_x\omega'''_{x,o}(x')-\omega'''_{x,o}(x')$ is holomorphic in $x'$ such that the last equality reads $\omega=\frac 1{2\pi\mathbf i}\, \widehat B\big(\widehat C(\omega)\big)$. In other words, $\widehat B \circ \widehat C = 2\pi\mathbf i\,\operatorname{Id}_{(\widehat \Ho{}^1_{\mathfrak m})'''}$.
\eproof

\bt
\label{berginter}
For any pair of third--kind generalized cycles $\Gamma,\Gamma'\in\widehat\Ho{}'''_1(\mathfrak m)$,
\beq
\underset\Gamma\oint\, \widehat B(\Gamma')-\underset{\Gamma'}\oint\,\widehat B(\Gamma) = 2\pi\mathbf i\  \Gamma\bigcap\Gamma'
\eeq
\et

\proof
Both terms appearing in the difference are double integrals which both have non--zero contributions coming from the intersection of the generalized cycles. There, the difference has vanishing residue and the defining integrand of $\widehat B$ has a double pole (since its two arguments coincide) which counts the algebraic intersection of the base homology classes underlying $\Gamma$ and $\Gamma'$. This linearly extends to the wanted result.
\eproof

\br 
\label{kerbergint}
If $\Gamma\in\Ker\widehat B$,
\beq
\underset\Gamma\oint\widehat B(\Gamma') = 2\pi\mathbf i\, \Gamma\bigcap\Gamma'
\eeq
\er

\bc
The intersection form $\bigcap$ vanishes on $\Ker\widehat B\subset\widehat\Ho{}'''_1(\mathfrak m)$.
\ec

\subsection{Rigid frames for $\widehat\Ho{}'''_1$ and $T\modsp$}
\label{basiscycles}


Given a choice of basis of the fundamental group $\pi_1(\curverond_p,o)$ and using the Chevalley basis of $\Lieg$, one can define a canonical rigid basis of $\widehat\Ho{}'''_1(\mathfrak m)$. In turn, it can be used as a local frame in a neighborhood of any generic point in the moduli space. Let us emphasize that this basis depends only on discrete data and as such, admits a trivial connection making this rigid basis flat. Let us therefore start by using a fundamental domain of $\curverond_p$ to introduce a canonical basis of $\widehat\Ho{}'''_1(\mathfrak m)$ from which we compute its dimension and extract the rigid frame.

\bt\label{thdimmaxH1}

\beq
\dim \widehat\Ho_1(\mathfrak m) \leq \dim  {\mathcal M}_{p,[\alpha]} = 2\genus
\eeq
where the genus $\genus$ of $\widehat\curve_{\mathfrak m}$ is given by
\bea
\genus 
&\underset{def}{=}& (M-2+2\genusrond)\frac{\dim\Lieg-3\dim\Lieh}{2} + (M-3+3\genusrond) \dim\Lieh\cr
&=& \frac12 \big( (M-2+2\genusrond) \dim\Lieg - M \dim\Lieh\big).
\eea
\et

\proof
$\pi_1(\curverond_p,o)$ is generated by $M+2\genusrond$ closed loops starting and ending at the base point $o$, but only $M+2\genusrond-1$ are independent.
We thus write any $\Gamma\in\widehat \Ho{}'_1(\mathfrak m)$ as
\beq
\Gamma= \sum_{k=1}^{M+2\genusrond-1} \gamma_k \otimes E_k
\eeq
with $\{\gamma_k\}_{k=1}^{M+2\genusrond-1}$ a generating family of loops starting and ending at $o$.

The condition $\widehat\partial\,\Gamma=0$ then defines a submanifold of $\Lieg^{M+2\genusrond-1}$, of codimension $\geq 1$, namely $\dim \widehat \Ho{}'_1(\mathfrak m)\leq (M+2\genusrond-2)\dim\Lieg$ and to get $\widehat\Ho_1(\mathfrak m)$ we subtract the $M\dim\Lieh$ so-called trivial cycles.
\eproof

Let us recall for completeness how the dimension of $ \modsp_{p,[\alpha]}$ can be computed. Choose a fundamental domain as a polygon with $4\genusrond+2M$ edges and write as before $\nabla_\Upsilon=\diff-\Phi$ in this trivializing chart. The $2\genusrond+M$ gluings of pairs of edges amount to $2\genusrond+M$ gauge transformations, $2\genusrond+M-1$ of which are independent. Up to global gauge transformations we subtract another $\dim\Lieg$ to get
\beq
\dim {\mathcal M_p}= (M+2\genusrond-2)\dim\Lieg,
\eeq
Fixing the charges then yields
\beq
\dim {\mathcal M}_{p,[\alpha]} =  (M+2\genusrond-2)\dim\Lieg -M\dim\Lieh = 2\genus.
\eeq

Let us now construct an explicit basis of $\widehat\Ho_1(\mathfrak m)$ and show that equality holds.
\bt
The dimension of $\widehat\Ho_1(\mathfrak m)$ is finite and coincides with the dimension of the leaf moduli space $\modsp_{p,[\alpha]}$, namely
\beq
\dim \widehat\Ho_1(\mathfrak m) =  2\genus = (M-2+2\genusrond) \dim\Lieg - M \dim\Lieh
\eeq
where $\genus$ is called the genus of the quantum spectral curve $\widehat\curve_{\mathfrak m}$.
\label{genusbasis}
This implies moreover the dimension
\beq
\dim \widehat\Ho{}'''_1(\mathfrak m) = 2\genus + 2M\dim\Lieh =(M-2+2\genusrond) \dim\Lieg+M\dim\Lieh=2\genus'''
\eeq
\et

\proof
Let us build an independent family of $\widehat\Ho_1(\mathfrak m)$ consisting of $2\genus$ elements, this time by introducing a pair of pants--decomposition of $\curverond-\bigcup_{j=1}^M \mathcal{D}_{p_j}$, where $ \mathcal{D}_{p_j}$ is a small disc around $p_j$. We also introduce a graph that is dual to that pants decomposition:
\begin{equation} \label{graphpants}
\includegraphics[scale=0.5]{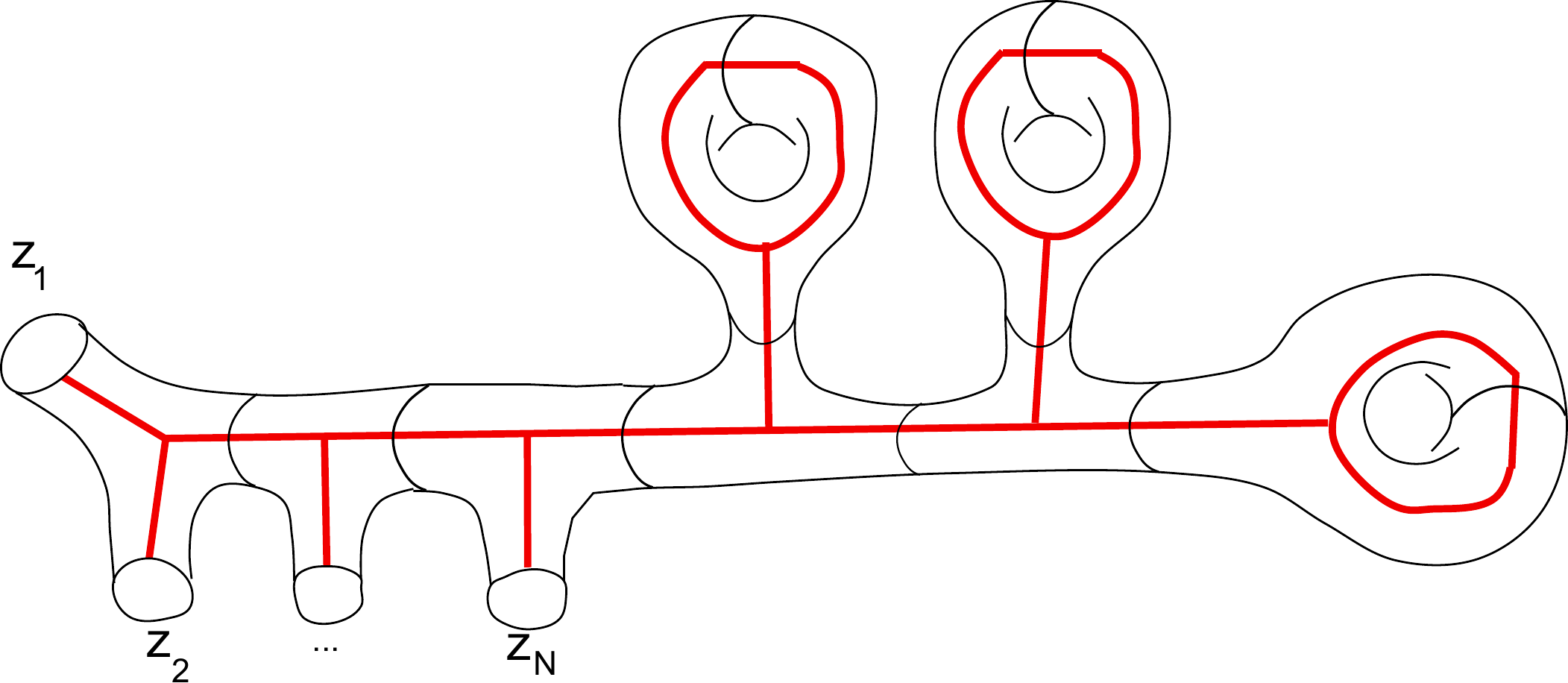}
\end{equation}

\pagebreak

This graph is made of 
\begin{itemize}
\item[*] $M-2+2\genusrond$ trivalent vertices,
\item[*] $M-3+3\genusrond$ internal edges,
\item[*] $M$ external edges, ending at the punctures.
\item[*] $\genusrond$ internal edges, each forming a loop.
\end{itemize}

Now let us introduce the basis elements of $\widehat\Ho_1(\mathfrak m)$:

$\bullet$ Each edge $e$ is dual to a pants boundary, and therefore to a cycle $\gamma_e$ in $\curverond_p$:
\[
\includegraphics[scale=0.5]{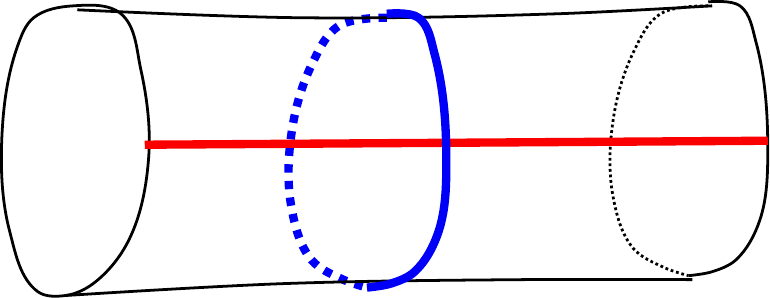}
\]

$\bullet$ If the edge $e$ is an external one, the cycle $\gamma_e$ is actually $\gamma_{p_j}$ for some $j\in\{1,\dots,M\}$, and the associated $\dim\Lieh$ cycles $\gamma_{p_j}\otimes\Lieh_j$ belong to $\widehat\Ho{}'_1(\mathfrak m)$ but are by definition vanishing in the coset $\widehat\Ho_1(\mathfrak m)$. 

$\bullet$ If the edge $e$ is an internal one however, we also obtain $\dim\Lieh$ cycles of the form $\acycle_e\underset{def}{=}\gamma_{e}\otimes E$ with $E$ in the centralizer $\Lieh_{\gamma_e}$ of the monodromy  $S_{\gamma_e}$.  These cycles belong to $\widehat\Ho_1(\mathfrak m)$, and are the first $(M-3+3\genusrond) \dim \Lieh$ elements of our basis.

$\bullet$ To each trivalent vertex $v$ of $\Upsilon$ are now associated cycles as well. Let indeed $e_1,e_2,e_3$ be the three edges ending at $v$, in the direct order around $v$ defined by the orientation of the surface. Let $\gamma_{e_i}$, $i\in\{1,2,3\}$, be the cycle of $\curverond_p$ that is dual to $e_i$ in a pair of pant neighborhood of $v$:
$$
\includegraphics[scale=0.5]{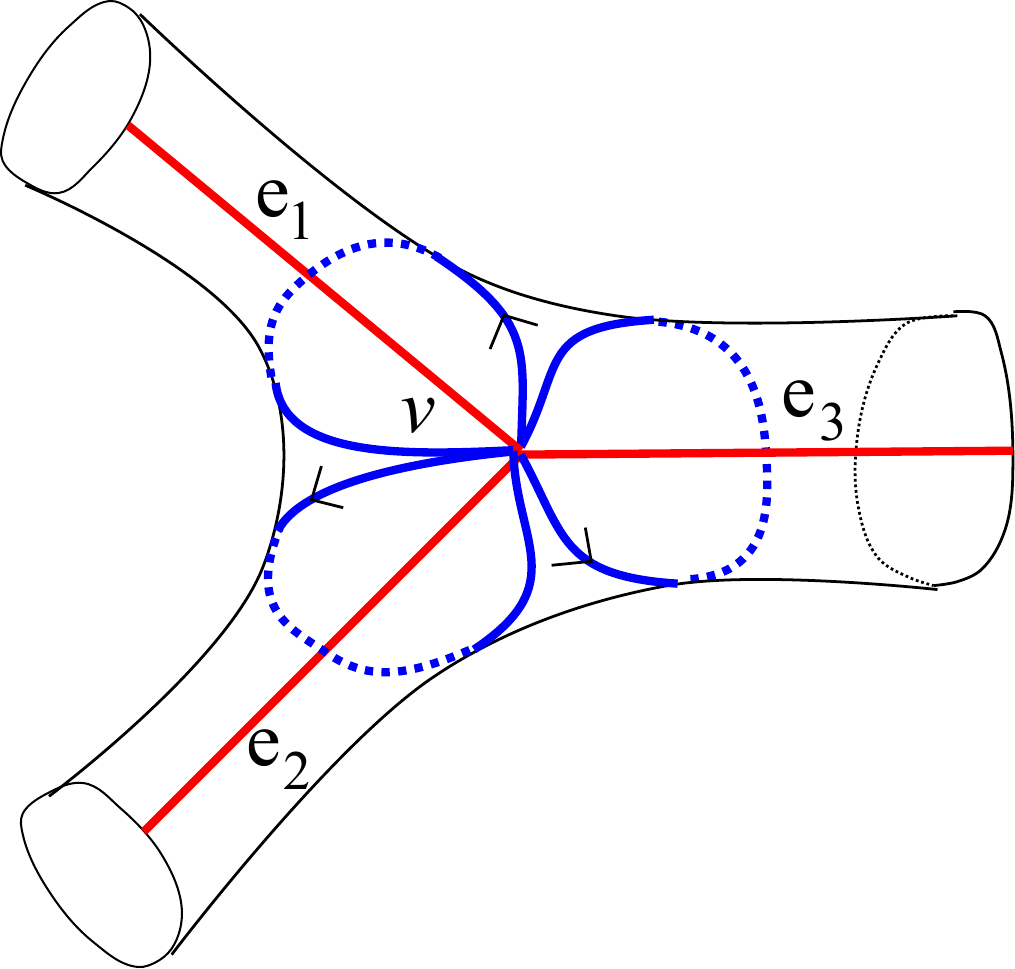}
$$
Let us now consider cycles of the form
\beq
\Gamma_v 
= \gamma_{e_1}\otimes E_1+\gamma_{e_2}\otimes E_2+\gamma_{e_3}\otimes E_3
\label{gv}
\eeq
The dimension of the space of these cycles is apparently $3\dim\Lieg$. 
However, the condition $\widehat\partial\, \Gamma_v=0$ reduces it to $2\dim\Lieg$, and the relation $\gamma_{e_1}+\gamma_{e_2}+\gamma_{e_3}=0$ in $\pi_1(\curverond_p,o)$ implies that only $\dim\Lieg$ such cycles are homologically independent. 
We should furthermore subtract the $3\dim\Lieh$ cycles of the form $\gamma_{e_i}\otimes E_i$ with $E_i\in\Lieh_{\gamma_{e_i}}$, associated to the meeting edges and already accounted for. 
The number of independent cycles associated to our  vertex is therefore 
\beq
\dim \Big(\big\{\Gamma_v\, \big|\, \widehat\partial \,\Gamma_v=0\big\} \Big/ \bigoplus_{i=1}^3 \gamma_{e_i}\otimes\Lieh_{\gamma_{e_i}}\Big) = \dim\Lieg - 3\dim \Lieh.
\eeq

$\bullet$ Finally, let us associate more cycles to each internal edge. If that edge $e$ is a loop that starts and ends at the same vertex, then it can be identified with a cycle that surrounds a hole:
$$
\includegraphics[scale=0.5]{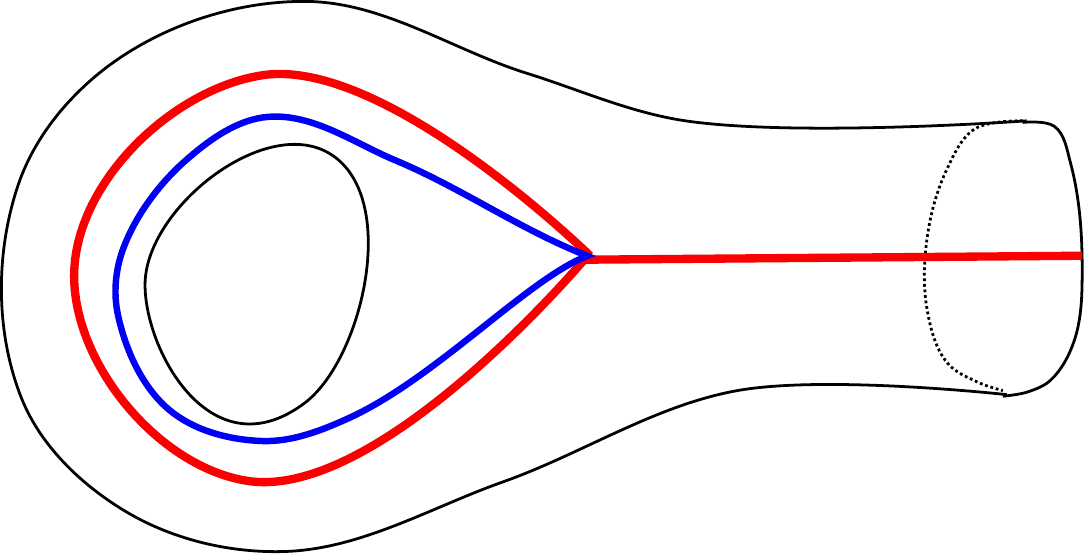}
$$
We then have $\dim\Lieh$ cycles of the form $\bcycle_{e\cdot E}\underset{def}{=}e\otimes E$ with $E\in \Lieh_{e}$.

$\bullet$ If however our internal edge joins two different vertices, let us introduce six arcs $\gamma_1,\dots,\gamma_6$ of $\curverond_p$ that end on these vertices:
$$
\includegraphics[scale=0.5]{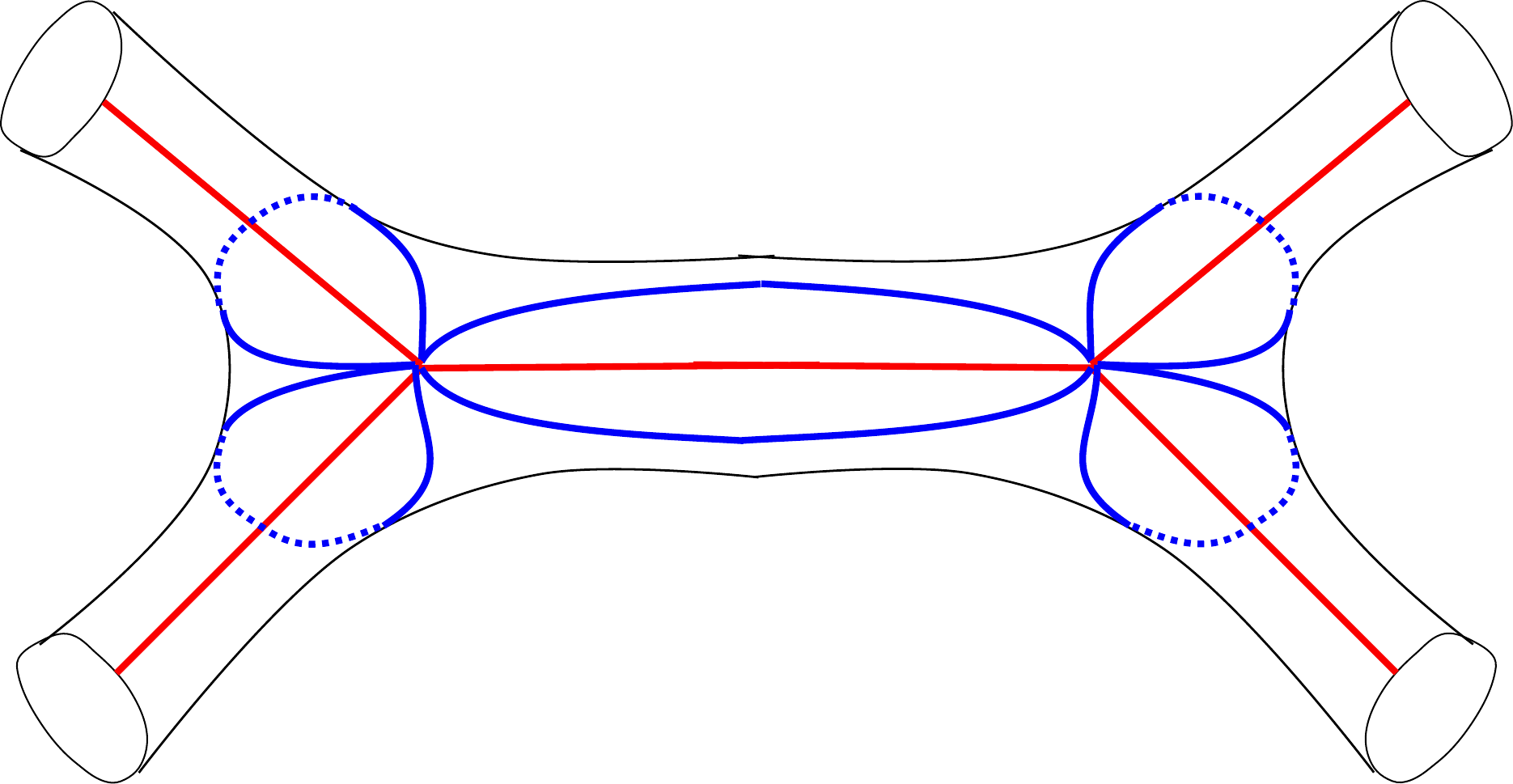}
$$
and consider corresponding cycles of the form
\beq
\Gamma_e^\perp = \sum_{i=1}^6 \gamma_i \otimes E_i
\eeq
The dimension of the space spanned by these cycles is apparently $6\dim\Lieg$. However, the condition $\widehat\partial\, \Gamma_e^\perp=0$ reduces it to $4\dim\Lieg$ as there are $\dim\Lieg$ equations at each of the two vertices. Only four of these six arcs are independent such that moreover, only $2\dim \Lieg$ of these cycles are homologically independent. 
Out of those, we already counted the cycles associated to the two vertices, and the cycles associated to the five edges. 
The number of newly introduced independent cycles associated to the edge is therefore 
\beq
2 \dim\Lieg - 2(\dim\Lieg-3\dim\Lieh) -5\dim\Lieh 
= \dim\Lieh\ ,
\eeq
and we denote them as $\bcycle_{e\cdot E}=e\otimes E + \cdots $, where the dots are contributions of arcs that are not homotopic to the considered edge $e$. There are therefore a total of $2\dim\Lieh$ cycles generically denoted $\acycle_{e\cdot E}$ and $\bcycle_{e\cdot E}$ for each edge. 

The $2\genus$ cycles that we have just constructed being homologically independent, this implies
\beq
\dim \widehat\Ho_1(\mathfrak m)\geq 2\genus.
\eeq
Theorem \ref{thdimmaxH1} then yields the equality
\beq
\dim \widehat \Ho_1(\mathfrak m)= 2\genus.
\eeq
Adding the $M\dim\Lieh$ trivial cycles yields the wanted dimension for $\widehat\Ho{}'_1(\mathfrak m)$. The dimension of $\widehat\Ho{}'''_1(\mathfrak m)$ is then obtained by adding independent generators for the $M$ free boundary components at the punctures. Let us start from the following observation:

Let $\gamma_j$ be an arc from $o$ to $p_j$, and $\gamma_{p_j}$ be a small circle around $p_j$ like before.
Consider $\gamma'_j$ a closed arc starting and ending at $o$, surrounding  $\gamma_j$, i.e. $\gamma'_j$ is a "doubling" of $\gamma_j$.
We then have for any Lie algebra elements $E'_j\in \Lieg$, and $H'_j\in \Lieh_j$
\beq
\gamma_j\otimes \big(H'_j+(1-\Ad_{S_{\gamma_{p_j}}} )\cdot E'_j\big) + \gamma_{p_j}\otimes H_j=\gamma'_j\otimes E'_j + \gamma_j\otimes H'_j 
\eeq
in $\widehat\Ho{}'''_1(\mathfrak m)$, where we denoted by $H_j$ the projection of $E'_j$ on $\Lieh_j$.
This is illustrated by
$$
\includegraphics[scale=0.3]{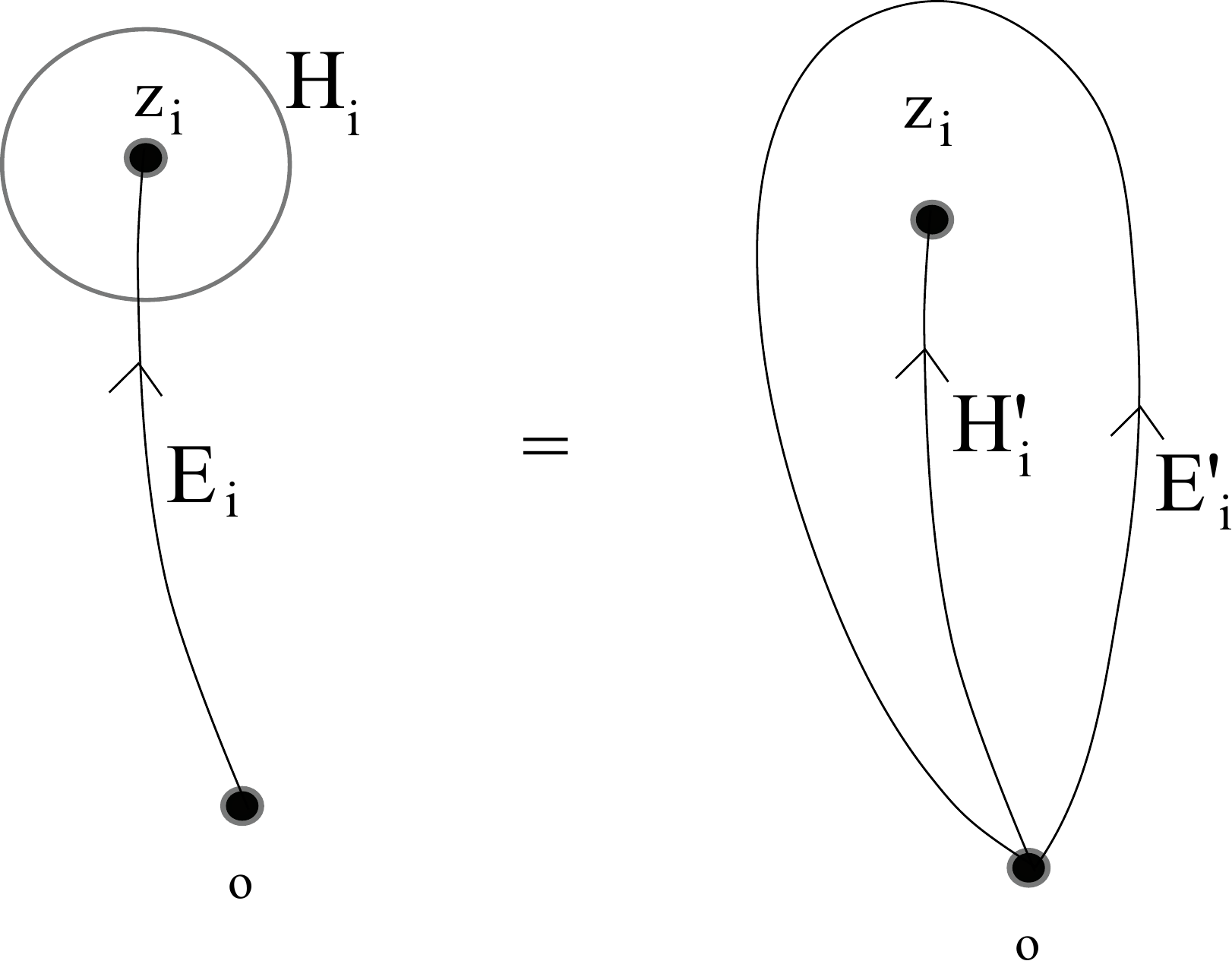}
$$
\label{cyclesqueeze}
In particular this means that a closed contour that surrounds $p_j$ can generically be squeezed into an arc $\gamma_j$ ending at $p_j$, weighted by $E_j = H'_j+(1-\Ad_{S_{\gamma_{p_j}}})\cdot E'_j$, except for the Cartan component $H_j$ of $E'_j$ that remains as a trivial cycle $\gamma_{p_j}\otimes H_j$.

Vice--versa, an arc $\gamma_j\otimes E_j$ can generically be unsqueezed (except for its  Cartan component $\gamma_j\otimes H'_j$) into $\gamma'_j\otimes E'_j$ such that $E'_j=H_j + (1-\Ad_{S_{\gamma_{p_j}}})^{-1}\cdot E_j$ with 
\beq
(1-\Ad_{S_j})^{-1} u  \underset{def}{=} \left\{
\begin{array}{ll}
\frac{1}{1-e^{2\pi\mathbf i\, \mathfrak r(\alpha_j)}}u \quad & \text{if} \quad u \in \Lieg_{\mathfrak r} \\
0 \quad & \text{if} \quad u \in \Lieh_j \\
\end{array}\right.
.
\eeq
This implies that from the $M\dim\Lieg$ free boundary dimensions, only $M\dim\Lieh$ of them are not already counted in $\widehat\Ho{}'_1(\mathfrak m)$. We therefore have $\dim\widehat\Ho{}'''_1(\mathfrak m)=\dim\widehat\Ho{}'_1(\mathfrak m)+M\dim\Lieh$ and this concludes the proof.
\eproof

\paragraph{Lagrangian decomposition of $\widehat\Ho{}'''_1$.}
The total space $\big(\widehat\Ho{}'''_1,\bigcap\big)$ is symplectic and since $\bigcap_{\big|\Ker\widehat B}=0$, we must have
\beq
\dim\Ker\widehat B \leq \genus'''
\eeq
$\Ker\widehat B$ is Lagrangian when this inequality is in fact an equality. This geometry is interpreted (only morally for now) as a quantization of that of the Hitchin integrable system. We postpone this precise discussion to section \ref{wkb} but let us mention that in this classical limit the operator $\widehat B$ is replaced by an equivariant version of the Bergman kernel on the so--called cameral cover and that in particular, its kernel defines a Lagrangian sub--variety of the deformation space.

To show that $\Ker\widehat B$ is Lagrangian, a strategy would be to show that the intersection product vanishes identically on $\operatorname{Im}\widehat C$ as well. As we shall see in section \ref{MM}, $\operatorname{Im}\widehat C$ is Lagrangian in the case of random matrix models. It is moreover also the case of any Fuchsian system on the Riemann sphere. When it comes to geometry of (classical or quantum) integrable systems, random matrix models are always of good advice, we therefore make the working assumption that
\beq
\bigcap{}_{\big|\operatorname{Im}\widehat C}=0
\eeq
an immediate consequence of which is
\bc
$\widehat\Ho{}'''_1=\operatorname{Im}\widehat C\oplus\Ker\widehat B$ is a Lagrangian decomposition of $\big(\widehat\Ho{}'''_1,\bigcap\big)$.
\ec

\paragraph{Trivialization  by Chevalley basis.}
\label{secintegerbasis}

We recall that there is a canonical Chevalley basis of $\Lieg$, consisting of a basis $\{H_{\mathfrak r}\}_{\mathfrak r \in \mathfrak R_0}$ of $\Lieh$ indexed by the set $\mathfrak R_0\subset\mathfrak R$ of simple positive roots, and $\{E_{\mathfrak r}\}_{\mathfrak r \in \mathfrak R}$ forming a basis of a complement of $\Lieh$ in $\Lieg$ indexed by the set of all roots. It generically satisfies
\beq
[H_{\mathfrak r},H_{\mathfrak r'}]=0
\quad , \quad
[H_{\mathfrak r},E_{\mathfrak r'}] 
=  \mathfrak r'(H_{\mathfrak r}) \ E_{\mathfrak r'}
\underset{def}=  \kappa_{\mathfrak r,\mathfrak r'} \ E_{\mathfrak r'},
\eeq
\beq
[E_{\mathfrak r},E_{-\mathfrak r}]= H_{\mathfrak r} \quad \text{if} \ \mathfrak r\in \mathfrak R_0
\quad , \quad
[E_{\mathfrak r},E_{\mathfrak r'}]=0 \quad \text{otherwise}
\eeq
where $\kappa$ is the Cartan matrix.

Assuming that $M\geq 1$ and  $M+2\genusrond-2\geq 0$, our goal is to construct a basis of $\widehat\Ho_1(\mathfrak m)$ from this Chevalley basis of $\Lieg$ and involving only integer coefficients. It will in turn be rigid under infinitesimal deformations.

Still denoting by $o\in\curverond_p$ a generic reference base--point in $\curverond_p$, let
$\gamma_1,\dots,\gamma_M$ be $M$ simple Jordan arcs, $\gamma_j$, $j\in\{1,\dots M\}$, from $o$ to $p_j$ and such that they intersect at $o$ and nowhere else. Consider moreover $\gamma_{M+1},\dots,\gamma_{M+2\genusrond}$ to be $2\genusrond$ additional Jordan arcs from $o$ to $p_1$, that don't intersect any of the other newly introduced arcs, except at $o$ and at $p_1$. Put together, $\gamma_1,\dots,\gamma_{M+2\genusrond}$ generate an $M+2\genusrond $--dimensional $\mathbb Z$--module.

$$
\includegraphics[scale=0.4]{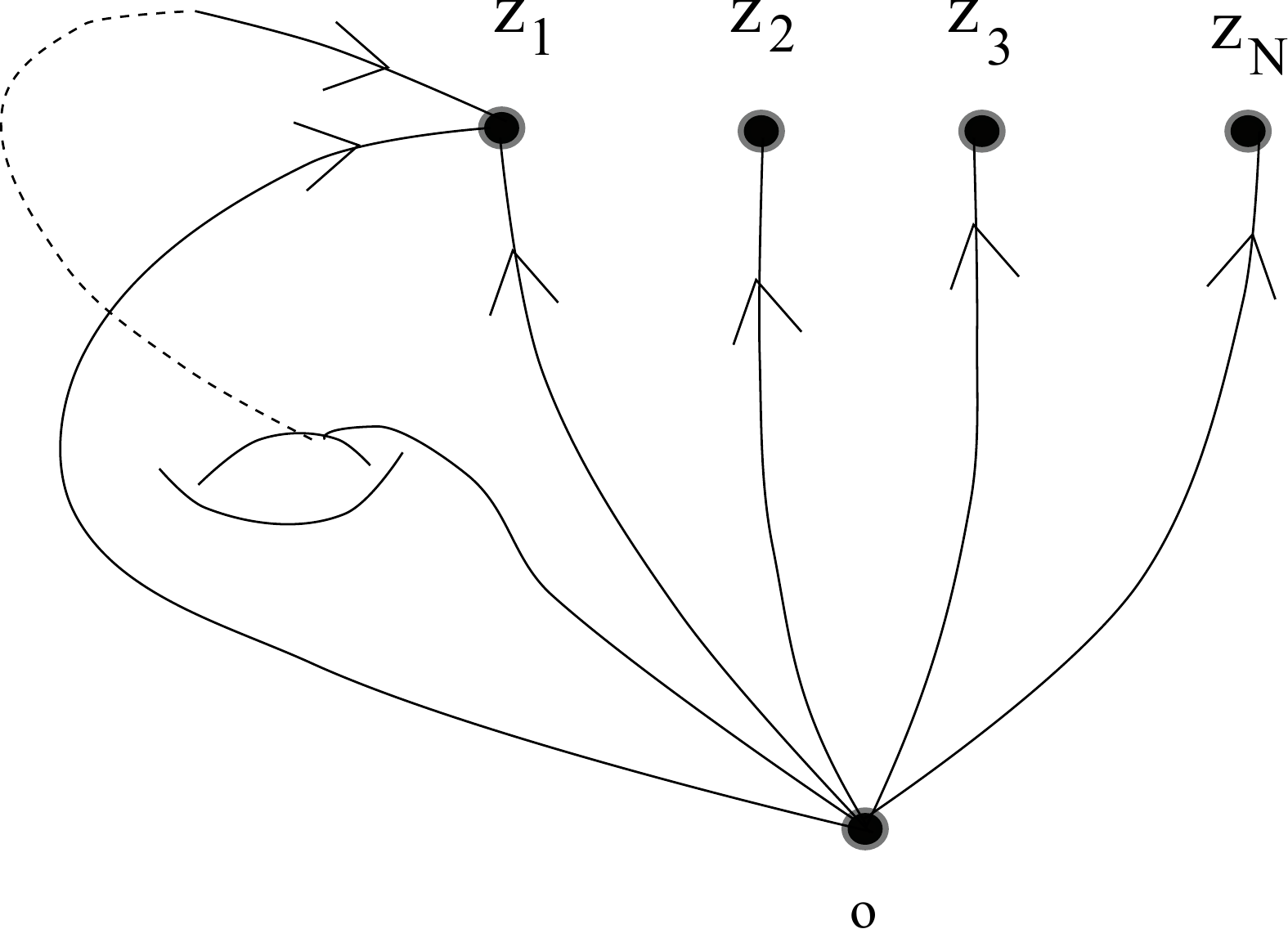}
$$

We also denote as before $\gamma_{p_i}$ a small counterclockwise circle around $p_i$.
We choose moreover the reference Cartan sub--algebra to be $\Lieh=\Lieh_M$, the normalizer of the monodromy $S_{\gamma_{p_M}} = \exp\left(2\pi\mathbf i\, \alpha_M\right)$ around $p_M$.

For any generic cycle written as $\Gamma=\sum_{i=1}^{M+2\genusrond} \gamma_i\otimes \sigma_i$, the vanishing boundary--condition $\widehat\partial\,\Gamma=0$ is non--trivial nowhere else than $o$ and there it takes the simple form
\beq
\sum_{i=1}^{M+2\genusrond} \sigma_i=0
\eeq
A basis of $\widehat\Ho{}'''_1(\mathfrak m)$ is then obtained from the Chevalley basis by defining
\bea
\Gamma_{j,\mathfrak r} \underset{def}{=}& \gamma_j\otimes E_{\mathfrak r} - \gamma_1\otimes E_{\mathfrak r} 
& \qquad j=2,\dots,M-1, \, \mathfrak r\in \mathfrak R \\
\Gamma_{i,\mathfrak r} \underset{def}{=}& \gamma_i\otimes E_{\mathfrak r} - \gamma_1\otimes E_{\mathfrak r} 
& \qquad i=M+1,\dots,M+2\genusrond, \, \mathfrak r\in \mathfrak R \\
\widetilde\Gamma_{j,\mathfrak r} \underset{def}{=} & \gamma_j\otimes H_{\mathfrak r}  - \gamma_1\otimes H_{\mathfrak r}  
 &\qquad j=2,\dots,M, \quad \mathfrak r \in \mathfrak R_0\\
\widetilde\Gamma_{i,\mathfrak r} \underset{def}{=} & \gamma_i\otimes H_{\mathfrak r} - \gamma_1\otimes H_{\mathfrak r} 
& \qquad i=M+1,\dots,M+2\genusrond, \, \mathfrak r \in \mathfrak R_0  \\
\acycle_{j,\mathfrak r} \underset{def}{=} & \gamma_{p_j}\otimes H^{j,\mathfrak r} \qquad \quad & \qquad j=1,\dots,M-1, \quad \mathfrak r \in\mathfrak R_0^{(j)}
\eea
$\mathfrak R_0^{(j)}$ being the set of simple positive roots of $\Lieh_j$ and $\{H_{j,\mathfrak r}\}_{\mathfrak r\in\mathfrak R_0^{(j)}}$ denoting the corresponding Chevalley basis and $\{H^{j,\mathfrak r}\}_{\mathfrak r\in\mathfrak R_0^{(j)}}$. Note in particular that $\gamma_{p_M}$ is expressed in terms of $\gamma_{p_1},\dots,\gamma_{p_{M-1}},\gamma_{M+1},\dots,\gamma_{M+2\genusrond}$ and that the squeezing argument allows a similar expression of generalized cycles supported on $\gamma_M$. Indeed, for any given root $\mathfrak r\in\mathfrak R$, we can express
\bea
\Gamma_{M,\mathfrak r}&\underset{def}=& \gamma_M\otimes E_{\mathfrak r}-\gamma_1\otimes E_{\mathfrak r}\\
&=& \gamma_{p_M}\otimes \frac 1{1-e^{2\pi\mathbf i\,\mathfrak r(\Ad_{C_M}^{-1}\alpha_M)}}E_{\mathfrak r}-\gamma_1\otimes E_{\mathfrak r}\\
&=&-\sum_{j=1}^{M-1} \gamma_{p_j}\otimes \frac 1{1-e^{2\pi\mathbf i\,\mathfrak r(\Ad_{C_M}^{-1}\alpha_M)}}E_{\mathfrak r}-\gamma_1\otimes E_{\mathfrak r}\nonumber\\
&&-\sum_{i=M+1}^{M+2\genusrond}\big(\gamma_i-\gamma_1\big)\otimes  \frac 1{1-e^{2\pi\mathbf i\,\mathfrak r(\Ad_{C_M}^{-1}\alpha_M)}}\big(1-\Ad_{S_{\gamma_i-\gamma_1}}\big)\cdot E_{\mathfrak r}\\
&=&-\sum_{j=1}^{M-1} \big(\gamma_j-\gamma_1\big)\otimes \frac 1{1-e^{2\pi\mathbf i\,\mathfrak r(\Ad_{C_M}^{-1}\alpha_M)}}\big(1-\Ad_{S_{\gamma_{p_j}}}\big)\cdot E_{\mathfrak r}\nonumber\\
&&-\sum_{i=M+1}^{M+2\genusrond}\big(\gamma_i-\gamma_1\big)\otimes  \frac 1{1-e^{2\pi\mathbf i\,\mathfrak r(\Ad_{C_M}^{-1}\alpha_M)}}\big(1-\Ad_{S_{\gamma_i-\gamma_1}}\big)\cdot E_{\mathfrak r}\nonumber\\
&&-\sum_{j=1}^{M-1}\gamma_{p_j}\otimes \frac 1{1-e^{2\pi\mathbf i\,\mathfrak r(\Ad_{C_M}^{-1}\alpha_M)}}\Pi_{\Lieh_j}E_{\mathfrak r}\nonumber\\
&&-\, \gamma_1\otimes\Big( \sum_{j=1}^{M-1}\frac 1{1-e^{2\pi\mathbf i\,\mathfrak r(\Ad_{C_M}^{-1}\alpha_M)}}\big(1-\Ad_{S_{\gamma_{p_j}}}\big)E_{\mathfrak r} + E_{\mathfrak r}\Big)
\eea
where all the terms are expressed in terms of the previously introduced basis elements except the last one that actually vanishes since all the others have vanishing generalized boundaries.

The cardinality of this integer family recovers the dimension  
\bea
\dim \widehat\Ho{}'''_1(\mathfrak m) 
&=& (M-2)(\dim\Lieg-\dim\Lieh) + 2\genusrond (\dim\Lieg-\dim\Lieh) +2\genusrond \dim \Lieh \\
&& + \,\,(M-1)\dim\Lieh+(M-1)\dim\Lieh \\
&=& (M+2\genusrond-2)\dim\Lieg+M\dim\Lieh=2\genus'''.
\eea 

Let us furthermore define $\acycle_{1,\mathfrak r}\underset{def}=\gamma_{p_1}\otimes H^{1,\mathfrak r}$ for any simple root $\mathfrak r\in\mathfrak R_0^{(1)}$..

%
%

This construction relies on discrete data only, namely the fundamental group $\pi_1(\curverond_p,o)$ and the chosen root systems of $\Lieg$. On one hand, it provides a rigid basis of $\widehat\Ho_1(\mathfrak m)$ that can be defined on a whole neighborhood of any $\mathfrak m=[(\mathcal P,\nabla)]$. As such, trivializes locally the bundle $\widehat\Ho_1\longrightarrow\modsp_p$ of generalized first--kind cycles. On the other hand, we see that the generators corresponding to generalized cycles of the third--kind depend on the monodromy data of the connection and are therefore not rigid under infinitesimal deformations. As we shall see in the following, their deformations can however be computed explicitly.

\br
$\acycle_{j,\mathfrak r}\in \Ker\widehat B$.
\er

\section{Tau--function and conformal field theory}

\subsection{Special geometry of theta--series coefficients}
\label{thetacoeff}

Let $\mathcal F\longrightarrow  \modsp_p$ be the bundle whose fiber $\mathcal F_{\mathfrak m}$ over $\mathfrak m\in\modsp_p$ is the Lagrangian Grassmanian of $\widehat\Ho{}'''_1(\mathfrak m)$ consisting of its Lagrangian subspaces. It is equipped with a flat connection defined from the basis introduced in the previous section. For any $\mathfrak m\in\modsp_p$, $\operatorname{dim}\mathcal F_{\mathfrak m}=\frac 12\genus'''(\genus'''+1)$. Let us denote by $\mathcal L$ a flat section of $\mathcal F$ with the property that it is transverse to $\Ker\widehat B$ at any generic point. We then define our tau--function by its theta--series decomposition with respect to internal charges (away from the singularities).

\bd[Theta--series coefficients of the tau--function]
\beq
\log\widehat\Tau(\mathfrak m;\mathcal L) \underset{def}{=} \frac 1{4\pi\mathbf i}\ \widehat \Gamma_{\mathfrak m} \bigcap \widehat \Pi^{\parallel\mathcal L_{\mathfrak m}}_{\Ker\widehat B} \widehat \Gamma_{\mathfrak m}
\eeq
where $\widehat\Gamma=\widehat C(W_1)$ is the universal cycle.
\ed

\br
We assumed $\mathcal L$ to be generically transverse to $\Ker\widehat B$ but observe that
\beq
\log\widehat{\mathfrak T}(\mathfrak m;\Ker\widehat B) = \log\mathfrak T(\mathfrak m;\operatorname{Im}\widehat C)=0
\eeq
\er

\bp[Darboux basis]
Let $\{\acycle_k,\bcycle_k\}_{k=1}^{\genus'''}$ be a Darboux basis of $\widehat \Ho{}'''_1(\mathfrak m)$ such that the Lagrangian $\mathcal L_{\mathfrak m}$ is generated by $\{\bcycle_k\}_{k=1}^{\genus}\bigsqcup\{\bcycle_{j,\mathfrak r}\}_{1\leq j\leq M}^{\mathfrak r \in\mathfrak R_0^{(j)}}$ and furthermore $\{\mathcal A_k\}_{k=\genus+1}^{\genus'''}\underset{def}=\{\mathcal A_{j,\mathfrak r}\}_{1\leq j\leq M}^{\mathfrak r\in\mathfrak R_0^{(j)}}$ and $\{\bcycle_k\}_{k=\genus+1}^{\genus'''}\underset{def}=\{\bcycle_{j,\mathfrak r}\}_{1\leq j\leq M}^{\mathfrak r\in\mathfrak R_0^{(j)}}$. We then have
\bea
4\pi\mathbf i\, \log\widehat\Tau(\mathfrak m;\mathcal L) &=& 
 \sum_{k=1}^{\genus'''} \big(\underset{\acycle'_k}\oint W_1\big)\,\big( \underset{\bcycle_k}\oint W_1\big)\\
&=& 
 \sum_{k=1}^{\genus} \big(\underset{\acycle'_k}\oint W_1\big)\,\big( \underset{\bcycle_k}\oint W_1\big) + 2\pi\mathbf i\,\sum_{j=1}^M\sum_{\mathfrak r\in\mathfrak R_0^{(j)}}\mathfrak r(\Ad_{C_j}^{-1}\alpha_j) \underset{\mathcal B_{j,\mathfrak r}}\oint W_1\nonumber\\
\eea
where $\acycle'_k \underset{def}= \widehat\Pi^{\parallel\mathcal L}_{\Ker\widehat B} \acycle_k$ for any  $1\leq k\leq\genus'''$.
\ep

\proof
By definition we have $\mathcal A'_k-\mathcal A_k\in\mathcal L_{\mathfrak m}$ for any value of the index $k\in\{1,\dots,\genus'''\}$ and as a consequence, the intersection matrix in the basis $\{\mathcal A'_k,\mathcal B_k\}_{i=1}^{\genus'''}$ is identical to that in the basis $\{\mathcal A_k,\mathcal B_k\}_{i=1}^{\genus'''}$. Decomposing the universal cycle as
\beq
\widehat\Gamma_{\mathfrak m} = \sum_{k=1}^{\genus'''}\Big(\mathcal B_k\underset{\mathcal A'_k}\oint W_1-\mathcal A'_k\underset{\mathcal B_k}\oint W_1\Big)
\eeq
and replacing in the expression of $\log\widehat{\mathfrak T}(\mathfrak m;\mathcal L)$ then yields the wanted result together with its expression using the explicit definition of the cycles $\mathcal A_{j,\mathfrak r}$.
\eproof

In the proof of the next theorem, we use the result of corollary of Lemma \ref{deltakerB} (the proof of which we postpone to section \ref{amplitudes}), namely that for any indices $k\in\{1,\dots,\genus'''\}$, $\partial_\Gamma\acycle'_k=0$ for all $\Gamma\in\mathcal L_{\mathfrak m}$.

\bt[Special geometry]
For every $\delta\in T_{\mathfrak m}\modsp_p$ such that $\Gamma_\delta\in\mathcal L$, 
\beq
\delta \log\widehat\Tau(\mathfrak m;\mathcal L) = \underset{\Gamma_\delta}\oint W_1.
\eeq
\et

\proof
From last proposition we have
\bea
4\pi\mathbf i\,\delta\log\widehat{\mathfrak T}(\mathfrak m;\mathcal L) &=&  \sum_{k=1}^{\genus}\Big(\underset{\mathcal A'_k}\oint \delta W_1 \underset{\mathcal B_k}\oint W_1 + \underset{\mathcal A'_k}\oint W_1 \underset{\mathcal B_k}\oint \delta W_1\Big)\nonumber\\
&&+\ 2\pi\mathbf i\,\sum_{j=1}^M \sum_{\mathfrak r\in\mathfrak R_0^{(j)}} \mathfrak r\big(\Ad_{C_j}^{-1}\delta\alpha_j\big)\underset{\mathcal B_{j,\mathfrak r}}\oint W_1\nonumber\\
&&+\  2\pi\mathbf i\,\sum_{j=1}^M\sum_{\mathfrak r\in\mathfrak R_0^{(j)}}\mathfrak r(\Ad_{C_j}^{-1}\alpha_j)\underset{\Gamma_\delta}\oint \widehat B\big(\mathcal B_{j,\mathfrak r}\big)
\eea
where we used $\mathcal A_{j,\mathfrak r}\in\Ker\widehat B$ that $\Gamma_\delta\in\mathcal L_{\mathfrak m}$ implies $\oint_{\mathcal B_{j,\mathfrak r}}\widehat B(\Gamma_\delta)=\oint_{\Gamma_\delta}\widehat B(\mathcal B_{j,\mathfrak r})$ to get the last term. We therefore get
\bea
4\pi\mathbf i\,\delta\log\widehat\Tau(\mathfrak m;\mathcal L) &=&  \sum_{k=1}^{\genus}\Big(\underset{\mathcal A'_k}\oint \widehat B(\Gamma_\delta) \underset{\mathcal B_k}\oint W_1 + \underset{\mathcal A'_k}\oint W_1 \underset{\mathcal B_k}\oint \widehat B(\Gamma_\delta)\Big)\nonumber\\
&&+\ 2\pi\mathbf i\,\sum_{j=1}^M \sum_{\mathfrak r\in\mathfrak R_0^{(j)}} \mathfrak r\big(\Ad_{C_j}^{-1}\delta\alpha_j\big)\underset{\mathcal B_{j,\mathfrak r}}\oint W_1\nonumber\\
&&+\ 2\pi\mathbf i\,\underset{\Gamma_\delta}\oint\,  \widehat B\Big( \sum_{j=1}^M\sum_{\mathfrak r\in\mathfrak R_0^{(j)}}\mathfrak r(\Ad_{C_j}^{-1}\alpha_j)\mathcal B_{j,\mathfrak r}\Big)
\eea
in which we treat the different terms separately. Remark \ref{kerbergint} yields that the first and third terms sum up to
\bea
\sum_{k=1}^{\genus}\underset{\mathcal A'_k}\oint \widehat B(\Gamma_\delta) \underset{\mathcal B_k}\oint W_1&+& 2\pi\mathbf i\, \sum_{j=1}^M \sum_{\mathfrak r\in\mathfrak R_0^{(j)}} \mathfrak r\big(\Ad_{C_j}^{-1}\delta\alpha_j\big)\underset{\mathcal B_{j,\mathfrak r}}\oint W_1\nonumber\\
&&\qquad\qquad\qquad=\quad 2\pi\mathbf i\, \sum_{k=1}^{\genus'''}\mathcal A'_k\bigcap\Gamma_\delta \underset{\mathcal B_k}\oint W_1\\
&&\qquad\qquad\qquad = \quad2\pi\mathbf i\,\underset{\Gamma_\delta}\oint W_1
\eea
where we used the following facts
\begin{itemize}
\item for all $j\in\{1,\dots,M\}$ and $\mathfrak r\in\mathfrak R_0^{(j)}$, $\mathcal A_{j,\mathfrak r}$ belongs to $\Ker\widehat B$ and intersects $\Gamma_\delta$ on $\gamma_j$ and nowhere else with 
\beq
\mathcal A_{j,\mathfrak r}\bigcap \Gamma_\delta = \frac 1{2\pi\mathbf i}\mathfrak r\big( \delta S_{\gamma_j}\cdot S_{\gamma_j}^{-1}\big)= \mathfrak r\big(\Ad_{C_j}^{-1}\delta\alpha_j\big)
\eeq
\item $\Gamma_\delta\in\mathcal L_{\mathfrak m}$
\item $\{\mathcal A'_k,\mathcal B_k\}_{k=1}^{\genus'''}$ forms a Darboux basis of $\widehat\Ho{}_1'''(\mathfrak m)$ with $\operatorname{Span}\{\mathcal B_k\}_{k=1}^{\genus'''}=\mathcal L_{\mathfrak m}$.
\end{itemize}
Again since $\Gamma_\delta\in\mathcal L_{\mathfrak m}$, theorem \ref{berginter} yields the relation $\oint_{\mathcal B_k}\widehat B(\Gamma_\delta)=\oint_{\Gamma_\delta}\widehat B(\mathcal B_k)$ for all $k\in\{1,\dots,\genus'''\}$ and the second and fourth terms therefore sum up to
\bea
\sum_{k=1}^{\genus'''}\underset{\mathcal A'_k}\oint W_1\underset{\mathcal B_k}\oint \widehat B(\Gamma_\delta) &=& \sum_{k=1}^{\genus'''}\underset{\mathcal A'_k}\oint W_1\underset{\Gamma_\delta}\oint \widehat B(\mathcal B_k)\\
&=&\underset{\Gamma_\delta}\oint \widehat B\Big( \sum_{k=1}^{\genus'''}\mathcal B_k\underset{\mathcal A'_k}\oint W_1\Big)
\eea
Observe that the integrand in this last term is equal to $\widehat B\big( \widehat C(W_1)\big)=2\pi\mathbf i\, W_1$. Indeed the sum appearing there differs from $\widehat C(W_1)$ only by a term belonging to $\Ker\widehat B$. Gathering these results then implies the wanted identity $\delta\log\widehat\Tau(\mathfrak m;\mathcal L)=\oint_{\Gamma_\delta}W_1$.
\eproof

This theorem in fact recovers the deformation properties of the logarithm of the tau--function of Jimbo--Miwa--Ueno \cite{JMU1981,GIL2018,BK2019} when restricted to subspaces of the monodromy data on which the curvature of the Malgrange--Bertola form vanishes. This provides an explicit non--perturbative completion of the Seiberg--Witten relations between the pre--potential and the Seiberg--Witten differential in the context of super--symmetric quantum field theories \cite{SW1994}. We end this section by the main definitions of the paper, the quantum matrix of periods and corresponding quantum theta--functions, namely the tau--function.

\bd
Define the matrix $\tau\underset{def}=(\tau_{k,l})_{k,l=1}^{\genus'''}$ by its coefficients
\beq
\bold\tau_{k,l}\underset{def}=\frac1{2\pi\mathbf i}\underset{\bcycle_k}\oint \widehat B(\bcycle_l)
\eeq
for all $k,l\in\{1,\dots,\genus'''\}$.
\ed

\bp[Quantum period matrix]
$\tau$ is a symmetric matrix with positive definite imaginary part, namely
\beq
\operatorname{Im}\tau>0\quad \text{and for any } k,l\in\{1,\dots,\genus'''\}, \quad \tau_{l,k}=\tau_{k,l}
\eeq
\ep

\proof
Symmetry follows from $\mathcal L$ being Lagrangian and positive definiteness is equivalent to Riemann's bilinear inequality. More precisely, with the notations of \ref{RBIn},
\bea
0<\frac{\mathbf i}2\underset{\widehat C(\overline\omega)}\oint\omega &=& \frac{\mathbf i}2\sum_{k=1}^{\genus'''}\big( \underset{\bcycle_k}\oint\omega\ \underset{\overline{\acycle'_k}}\oint\overline\omega-\underset{\acycle'_k}\oint\omega \ \underset{\overline{\bcycle_k}}\oint\overline\omega\big)\\
&=& \operatorname{Im}\big( \overline\omega\cdot\tau\cdot\omega\big)
\eea
since $\underset{\bcycle_k}\oint\omega=\sum_{l=1}^{\genus'''}\tau_{k,l}\underset{\acycle'_l}\oint\omega$.
\eproof

\bd[Tau--function]
\label{tau}
The tau--function is the function on $\modsp_p$ defined for any $\mathfrak m\in\modsp'_p$ by the gauge--invariant 
theta--series expansion
\beq
\Tau(\mathfrak m)\underset{def}=\sum_{\operatorname n \in\mathbb Z^\genus} \exp\Big(-\frac 1{2\pi\mathbf i}\sum_{k=1}^\genus \operatorname n_k\underset{\bcycle_k}\oint W_1\Big)\,\,\widehat\Tau(\mathfrak m+\operatorname n_\acycle;\mathcal L)
\eeq
where for any $\operatorname n\in\mathbb Z^\genus$, $\mathfrak m+\operatorname n_\acycle$ is defined from $\mathfrak m=(\mathcal P,\nabla)$ by the integer shift
\beq
\underset{\acycle'_k}\oint W_1\longrightarrow \underset{\acycle'_k}\oint W_1 + \operatorname n_k\quad\text{namely}\quad W_1\longrightarrow W^{\operatorname n_\acycle}_1\underset{def}=W_1+\frac 1{2\pi\mathbf i}\sum_{k=1}^\genus \operatorname n_k\widehat B(\bcycle_k)
\eeq 
of the corresponding $\acycle$--period  coordinates for any value of the index $k\in\{1,\dots,\genus\}$.
\ed


\br
Since the plane--wave pre--factor in the definition of the tau--function depends only on integers and periods along cycles in the flat Lagrangian, all linear differential properties of $\widehat\Tau$ easily translate to properties of $\Tau$.
\er

Let us now further explore the relationship between our construction and the notion of isomonodromic tau--function before defining higher order analogues of the quantum Liouville form $W_1$ and give a natural interpretation of the whole setup in the terms of conformal field theory on the base Riemann surface $\curverond$. 

\subsection{Second--kind cycles and isomonodromic deformations}
\label{schlesinger}

In this subsection we study how the tau--function defined in the previous section depends on the choice of positions of the simple poles of the connection $\nabla$. To do so, we first restrict to the case where $\curverond=\mathbb P^1$ is the Riemann sphere and consider the previous construction as being fibered over the space $\big(\curverond{}^M-\Delta_M\big)\big/\mathfrak S_M$ of configurations of $M$ distinct points on $\curverond$ up to permutation. The fiber over a choice $p=(p_1,\dots,p_M)$ of distinct points on $\curverond$ (up to permutation) is $\modsp_p$ and we denote the total space of this fibration by $\modsp^{(M)}\subset\modsp$ which is the subspace of the moduli space $\modsp$ consisting of pairs of holomorphic principal $G$--bundles endowed with a meromorphic connection with exactly $M$ simple poles (and no other singularities) up to holomorphic gauge transformations. The construction of the tau--function through the definition of its values $\mathfrak T(\mathfrak m)$ then depends implicitly on the choice of $p_1,\dots,p_M$. Let us now explicit how.

For any $j\in\{1,\dots,M\}$, denote by $\delta_{p_j}$ the horizontal vector field on $\modsp^{(M)}$ obtained from the canonical vector field $\partial_{p_j}$ on the configuration space $\big(\curverond{}^M-\Delta_M\big)\big/\mathfrak S_M$. Consider a section $\mathfrak m:\big(\curverond{}^M-\Delta_M\big)\big/\mathfrak S_M\longrightarrow \modsp^{(M)}$ and given a rigid fundamental domain, decompose the  corresponding connection potential as
\beq
\Phi(x) = \sum_{j=1}^M\Phi_j\, \omega'''_{p_j,o}(x)
\eeq
with no holomorphic part since the Riemann sphere has genus $\genusrond=0$. We obtain the deformation of the quantum Liouville form as
\label{isomdefLiouville}
\beq
\delta_{p_j}W_1 (X)= \partial_{p_j}\omega'''_{p_j,o}(x)\left\langle\Phi_j,\sigma_\Psi(X)\right\rangle\,+\,\sum_{k=1}^M\omega_{p_k,o}(x)\big\langle\delta_{p_j}\Phi_k,\sigma_\Psi(X)\big\rangle
\eeq

\bd[Second--kind generalized cycles]
The space of generalized cycles of the second--kind is defined as being spanned by the localized cycles 
\beq
\mathcal B''_j\underset{def}= \underset{p_j\cdot\Ad_{C_j}^{-1}\alpha_j}{\operatorname{ev}}
\eeq
defined for any $j\in\{1,\dots,M\}$ as the linear map $\mathcal B''_j:\big(\widehat\Ho{}_{\mathfrak m}^1\big){}'''\longrightarrow \mathcal O_{(\curverond{}^M-\Delta_M)/\mathfrak S_M}$ evaluating a given generalized differential form at $p_1\cdot\Ad_{C_1}^{-1}\alpha_1,\dots,p_M\cdot\Ad_{C_M}^{-1}\alpha_M$ respectively, where the group elements $C_1,\dots,C_M$ first appeared in \ref{Psinearzj}. 
\ed

\br
Note that the resulting objects obtained by applying the second--kind generalized cycles $\mathcal B''_1,\dots,\mathcal B''_M$ to generalized differential forms are functions on the configuration space of the points $p_1,\dots,p_M\in\curverond$.
\er

\bt[Schlesinger's isomonodromic flows as special geometry]
\beq
\text{For any}\quad j\in\{1,\dots,M\},\quad
\delta_{p_j}W_1 = \widehat B(\mathcal B''_j)
\eeq
if and only if $\Phi$ is a solution of Schlesinger's equations, namely if and only if it satisfies
\bea
\diff_{p_j}\Phi_j &=& \sum_{k\neq j}\,\omega'''_{p_k,o}(p_j)\,[\Phi_k,\Phi_j],\\
\text{and if}\quad j\neq k,\quad \diff_{p_j}\Phi_k &=& \omega'''_{p_k,o}(p_j)\,[\Phi_j,\Phi_k]\eea
for any $j,k\in\{1,\dots,M\}$. In this case, the one--forms on $\big(\curverond{}^M-\Delta_M\big)\big/\mathfrak S_M$ defined by
\bea
\diff_{p_j}\log\mathfrak T(\mathfrak m) &=& \big(\underset{\mathcal B''_j}\oint W_1\big)\diff p_j = \Big(\underset{p_j\cdot\Ad_{C_j}^{-1}\alpha_j}{\operatorname{ev}} W_1\Big)\diff p_j\\
 &=& \sum_{\underset{k\neq j}{k=1}}^M \omega'''_{p_k,o}(p_j)\left\langle\Phi_k,\Phi_j\right\rangle
\eea
for $j\in\{1,\dots M\}$, yield the Hamiltonians of the Schlesinger integrable system.

In other words, $\mathfrak T$ is an isomonodromic tau--function.
\et

\proof
The explicit expression of the map $\widehat B$ gives
\bea
&\widehat B(\mathcal B''_j)(X)&\nonumber\\
&=\underset{X'=p_j\cdot\Ad_{C_j}^{-1}\alpha_j}{ \operatorname{ev}}&\Big( \diff_x\omega'''_{x,o}(x')\big\langle \sigma_\Psi(X'),\sigma_\Psi(X)\big\rangle-\omega'''_{x,o}(x')\big\langle\big[\Phi(x),\sigma_\Psi(X')\big],\sigma_\Psi(X)\big\rangle\Big)\nonumber\\
\eea
We subsequently use the definition $\underset{x'=p_j}{\operatorname{ev}}\big(\diff_x\omega'''_{x,o}(x')\big)=\partial_{p_j}\omega'''_{p_j,o}(x)$ of the evaluation linear forms and the relationship $\underset{x'\rightarrow p_j}\lim\sigma_\Psi(x'\cdot\Ad_{C_j}^{-1}\alpha_j)=\Phi_j$ implied by \ref{Psinearzj}. The equivalence between the special geometry relation and Schlesinger's equations is then obtained by comparing $\widehat B(\mathcal B_j'')(X)$ with the deformation of the quantum Liouville form \ref{isomdefLiouville}. The theta--series coefficients of the tau--function then deform as 
\bea
4\pi\mathbf i\,\delta_{p_j}\log\widehat\Tau(\mathfrak m;\mathcal L) &=& \sum_{k=1}^{\genus}\Big(\underset{\delta_{p_j}\mathcal A'_k}\oint W_1\underset{\mathcal B_k}\oint W_1+\underset{\mathcal A'_k}\oint \delta_{p_j} W_1\underset{\mathcal B_k}\oint W_1+\underset{\mathcal A'_k}\oint W_1\underset{\mathcal B_k}\oint \delta_{p_j}W_1\Big)\nonumber\\
&+& 2\pi\mathbf i\,\sum_{l=1}^M\sum_{\mathfrak r\in\mathfrak R_0^{(l)}}\mathfrak r\big(\Ad_{C_l}^{-1}\delta_{p_j}\alpha_l\big)\underset{\mathcal B_{l,\mathfrak r}}\oint W_1 \nonumber\\
&+& 2\pi\mathbf i\,\sum_{l=1}^M\sum_{\mathfrak r\in\mathfrak R_0^{(l)}}\mathfrak r\big(\Ad_{C_l}^{-1}\alpha_l\big)\delta_{p_j}\big(\underset{\bcycle_{l,\mathfrak r}}\oint W_1\big) 
\eea
where we used the rigidity of $\{\mathcal B_k\}_{k=1}^{\genus}$ to discard one term in the first line. We furthermore have $\delta_{p_j}\big(\mathcal A'_k-\mathcal A_k\big)=0$ for all $j\in\{1,\dots,M\}$ and $k\in\{1,\dots,\genus\}$ since $\mathcal A'_k-\mathcal A_k\in\mathcal L_{\mathfrak m}$ and by rigidity of $\mathcal L_{\mathfrak m}$. $\mathcal A_k$ being rigid as well then implies that $\delta_{p_j}\mathcal A'_k=0$. We also have $\delta_{p_j}\alpha_l=0$ for all $l\in\{1,\dots,M\}$ since we assumed that $\mathfrak m$ satisfies an isomonodromic flow. The first summand of the first line and the second line itself therefore have vanishing contributions. The theta--series coefficient therefore satisfies
\bea
4\pi\mathbf i\,\delta_{p_j}\log\widehat\Tau(\mathfrak m;\mathcal L)&=& \sum_{k=1}^{\genus}\Big(\underset{\mathcal A'_k}\oint \widehat B(\mathcal B''_j)\underset{\mathcal B_k}\oint W_1+\underset{\mathcal A'_k}\oint W_1\underset{\mathcal B_k}\oint \widehat B(\mathcal B''_j)\Big)\nonumber\\
&+& 2\pi\mathbf i\,\sum_{l=1}^M\sum_{\mathfrak r\in\mathfrak R_0^{(l)}}\mathfrak r\big(\Ad_{C_l}^{-1}\alpha_l\big)\delta_{l,j}\underset{p_j\cdot H_{l,\mathfrak r}}{\operatorname{ev}} W_1\nonumber\\
&+& 2\pi\mathbf i\,\sum_{l=1}^M\sum_{\mathfrak r\in\mathfrak R_0^{(l)}}\mathfrak r\big(\Ad_{C_l}^{-1}\alpha_l\big)\underset{\mathcal B_{l,\mathfrak r}}\oint \widehat B(\mathcal B''_j)
\eea
expression in which we used the special geometry relation $\delta_{p_j}W_1=\widehat B(\mathcal B''_j)$. The generalized cycle $\mathcal B''_j$ is located at the point $p_j$ and as such has no intersection with first--kind cycles. \ref{kerbergint} and $\mathcal A'_k\in\Ker\widehat B$ then imply that
\beq
\underset{\mathcal A'_k}\oint\widehat B(\mathcal B''_j) = 2\pi\mathbf i\,\mathcal A'_k\bigcap \mathcal B''_j = 0
\eeq
and that the first summand has vanishing contribution to the deformation. Regrouping the first and third lines into a single term and using the vanishing intersections $\mathcal B_k\bigcap\mathcal B''_j=0$ and $\mathcal B_{l,\mathfrak r}\bigcap\mathcal B''_j=0$ then yield
\bea
4\pi\mathbf i\,\delta_{p_j}\log\widehat\Tau(\mathfrak m;\mathcal L) &=& \underset{\mathcal B''_j}\oint\sum_{k=1}^{\genus'''}\,  \widehat B(\mathcal B_k)\underset{\mathcal A'_k}\oint W_1\nonumber\\
&+& 2\pi\mathbf i\,\sum_{l=1}^M\delta_{l,j}\sum_{\mathfrak r\in\mathfrak R_0^{(l)}}\mathfrak r\big(\Ad_{C_l}^{-1}\alpha_l\big)\underset{p_j\cdot H_{l,\mathfrak r}}{\operatorname{ev}} W_1
\eea
Recognizing $\sum_{k=1}^{\genus'''}\widehat B(\mathcal B_k)\oint_{\mathcal A'_k}W_1=\widehat B\big(\widehat C(W_1)\big)=2\pi\mathbf i\,W_1$ in the first line and simplifying the second one as
\bea
2\pi\mathbf i\,\sum_{l=1}^M\delta_{l,j}\sum_{\mathfrak r\in\mathfrak R_0^{(l)}}\mathfrak r\big(\Ad_{C_l}^{-1}\alpha_l\big)\underset{p_j\cdot H_{l,\mathfrak r}}{\operatorname{ev}} W_1&=&2\pi\mathbf i\,\sum_{\mathfrak r\in\mathfrak R_0^{(j)}}\mathfrak r\big(\Ad_{C_j}^{-1}\alpha_j\big)\underset{p_j\cdot H_{j,\mathfrak r}}{\operatorname{ev}} W_1\nonumber\\
&&\\
&=&2\pi\mathbf i\,\underset{p_j\cdot \Ad_{C_j}^{-1}\alpha_j}{\operatorname{ev}} W_1
\eea
then yield $\delta_{p_j}\log\widehat\Tau(\mathfrak m;\mathcal L)=\oint_{\mathcal B''_j}W_1$ when put together. We can now compute the variations of the tau--function $\Tau$ itself by linearity. 
\bea
\delta_{p_j}\Tau(\mathfrak m) &=&\sum_{\operatorname n\in\mathbb Z^\genus}\exp\Big(-\frac 1{2\pi\mathbf i}\sum_{k=1}^\genus \operatorname n_k\underset{\bcycle_k}\oint W_1\Big)\nonumber\\
&&\qquad\qquad\times\,\,\Big(-\frac 1{2\pi\mathbf i}\sum_{k=1}^\genus \operatorname n_k\delta_{p_j}\big(\underset{\bcycle_k}\oint W_1\big)+\underset{\bcycle''_j}\oint W^{\operatorname n_\acycle}_1\Big)\times\,\widehat\Tau(\mathfrak m+\operatorname n_\acycle;\mathcal L)\nonumber\\
\eea
where we observe that $\oint_{\bcycle''_k}W_1$ depends on the integer shift and hence comes with the upper--script $\operatorname n_\acycle$. Using the fact that the Lagrangian $\mathcal L$ is rigid under isomonodromic deformations together with the relations $\oint_{\bcycle_k}\widehat B(\bcycle''_j)=\oint_{\bcycle''_j}\widehat B(\bcycle_k)$, we get
\bea
\delta_{p_j}\Tau(\mathfrak m)&=&\sum_{\operatorname n\in\mathbb Z^\genus}\exp\Big(-\frac 1{2\pi\mathbf i}\sum_{k=1}^\genus \operatorname n_k\underset{\bcycle_k}\oint W_1\Big)\nonumber\\
&&\qquad\qquad\times\,\,\Big(\underset{\bcycle''_j}\oint \big(W^{\operatorname n_\acycle}_1-\frac 1{2\pi\mathbf i}\sum_{k=1}^\genus\operatorname n_k\widehat B(\bcycle_k)\big)\Big)\times\,\widehat\Tau(\mathfrak m+\operatorname n_\acycle;\mathcal L)\nonumber\\
\eea
Now noticing that the middle term between parenthesis is independent of $\operatorname n\in\mathbb Z^\genus$ allows to conclude that
\beq
\delta_{p_j}\Tau(\mathfrak m) = \big(\underset{\bcycle''_j}\oint W_1\big)\,\Tau(\mathfrak m)
\eeq
which is the wanted result. Let us furthermore compute these periods using the regularized pairing with third--kind differentials as
\bea
\big(\underset{\mathcal B''_j}\oint W_1\big)\diff p_j &=& \Big(\underset{X=p_j\cdot\Ad_{C_j}^{-1}\alpha_j}{\operatorname{ev}} W_1\Big)\diff p_j\\
&=& \underset{x\rightarrow p_j}\lim\Big(W_1\big(x\cdot \Ad_{C_j}^{-1}\alpha_j\big)-\omega'''_{p_j,o}(x)\big\langle\Phi_j,\sigma_\Psi\big(x\cdot\Ad_{C_j}^{-1}\alpha_j\big)\big\rangle\Big)\\
&=& \sum_{\underset{k\neq j}{k=1}}^M \omega'''_{p_k,o}(p_j)\left\langle\Phi_k,\Phi_j\right\rangle
\eea
\eproof

\pagebreak 

\br
The usual presentation of Schlesinger's Hamiltonians corresponds to the case where the reference point $o$ is actually the point at infinity, implying that third--kind differentials are written as

\bea
\omega'''_{p,\infty}(x)&=&\frac{\diff x}{x-p}\\
\text{and therefore}\quad \delta_{p_j}\log\mathfrak T(\mathfrak m) &=& \sum_{\underset{k\neq j}{k=1}}^M\frac{\left\langle\Phi_j,\Phi_k\right\rangle}{p_j-p_k}\qquad\qquad\qquad\qquad
\eea
\er

\subsection{Higher amplitudes and loop equations}
\label{amplitudes}

Back to the generic situation for the base curve $\curverond$, we now define some higher order amplitudes satisfying infinitely many relations called loop equations. These relations illustrate the integrability of the problem and bridges between this construction and the theory of the topological recursion \cite{EO2007,BEO2013, BBE2015, BEM2016,KS2017} but we will not get into the details of this relationship here.

\bd[Connected amplitudes]
For any integer $n\in\mathbb N^*$, $n\geq 2$, define the $n$-point amplitude to be the $\mathfrak S_n$ and gauge invariant bundle map 
\beq
W_n\in\underset{\modsp_p}{\operatorname{Bun}}\Big(\widehat\curve^{\otimes n}\big/\mathcal G\, , (\overset{\circ}{\pi})^*\operatorname{K}_{\curverond_p}^{\otimes n}\Big)^{\mathfrak S_n}
\eeq
generically defined by the multivalued formula
\bea
W_n( x_1\cdot\sigma_1\otimes\cdots\otimes x_n\cdot\sigma_n)\underset{def}{=} (-1)^{n-1}\sum_{\nu\in\mathfrak S_n^c}\frac{\big\langle \sigma_1(\widetilde x_1)\cdots \sigma_{\nu^{n-1}(1)}(\widetilde x_{\nu^{n-1}(1)})\big\rangle}{\mathcal E( x_1, x_{\nu(1)})\cdots\mathcal E( x_{\nu^{n-1}(1)}, x_1)}
\eea
Note that the last line is only valid outside of the diagonal divisor in $(\curve_\Upsilon)^n$. It can be equivalently defined and extended to this divisor by a choice of multivalued $\nabla$--flat section and through
\bea
 W_n( x_1\cdot E_1\,\,\otimes\,\,\cdots&\otimes&  x_n\cdot E_n)\underset{def}{=}\Big\langle\underset{1\leq i,j\leq n}{\operatorname{det}^c}\big[E_i\cdot \mathcal K_\Psi( x_i,  x_j)\big]\Big\rangle\
\eea
where we use the isomorphism $\widehat\curve_{\mathfrak m}\underset{\Ad_\Psi^{-1}}{\simeq}\Lieg$ and the symbol $\underset{1\leq i,j\leq n}{\operatorname{det}^c}$ denotes the connected determinant, defined much like the usual determinant but with a sum restricted to $\mathfrak S_n^c$, the subgroup of elements of order $n$ in the group of permutations of $n$ elements. 

We will moreover separate arguments of $W_n$ with comas instead of tensor products when the map is assumed to be evaluated on pure tensors.
\ed

\br
Gauge invariance of the $n$-point amplitude is assured by the holomorphic gauge transformations being constant and the bracket $\left\langle\,\bullet\,\right\rangle$ being invariant under conjugation.
\er

\bp
For all $\Gamma\in\widehat\Ho{}'''_1(\mathfrak m)$,
\beq
\underset{\Gamma}\oint\,W_2 = \widehat B(\Gamma)
\eeq
\ep

\proof
The definitions of $W_2$ and $\widehat B$ imply both
\bea
\underset{y\cdot F\in\Gamma}\oint\,W_2(x\cdot E,y\cdot F) &=& \underset{y\cdot F\in\Gamma}\oint\big\langle \sigma_\Psi(y\cdot F),-\frac{\sigma_\Psi(x\cdot E)}{\mathcal E(y,x)\mathcal E(x,y)}\big\rangle\\
\text{and}\quad \widehat B(\Gamma)(x\cdot E) &=& \underset{y\cdot F\in\Gamma}\oint\big\langle \sigma_\Psi(y\cdot F),\diff_x\big(\omega'''_{x,o}(y)\sigma_\Psi(x\cdot E)\big)\big\rangle
\eea
for generic values of the arguments. $-\frac{\sigma_\Psi(x\cdot E)}{\mathcal E(y,x)\mathcal E(x,y)}$ and $\diff_x\big(\omega'''_{x,o}(y)\sigma_\Psi(x\cdot E)\big)$ can then be checked to differ by a holomorphic expression of $y\in\curverond$ thus implying the wanted result by using lemma \ref{holcor}.
\eproof

\br
In particular, $\oint_\Gamma W_2=0$ for any $\Gamma\in\Ker\widehat B$.
\er

\bl
\label{diagonalresidue}
For any integer $n\in\mathbb N^*$ and any generic $ x_1\cdot E_1,\dots,  x_n\cdot E_n\in\widehat\curve_{\mathfrak m}$,
\beq
\underset{ x_i= x_j}{\operatorname{Res}}\, W_{n+1}( x_1\cdot E_1,\dots, x_n\cdot  E_n) = W_n\big(\dots, x_j \cdot [ E_i, E_j],\dots\big)
\eeq
where in the right-hand side we only  wrote the argument featuring the indices $i$ and $j$.
\el

\proof
This result follows from the explicit definition of $W_n$ by extracting the terms which contribute to the residue.
\eproof

\bd[Non-connected amplitudes]
As is customary with generating forms, the non-connected amplitudes are defined from the connected ones by a sum over set-partitions. Namely, for any given integer $n\in\mathbb N^*$,
\beq
\widehat W_n(X_1,\dots, X_n)\underset{def}{=} \widehat\Tau(\mathfrak m;\mathcal L)\,  \sum_{l=1}^n\sum_{\underset{\quad=\{X_1,\dots,X_n\}}{\mu_1\sqcup\cdots\sqcup\mu_l\quad}}\bigotimes_{i=1}^l W_{|\mu_i|}(\mu_i)
\eeq
where $\sqcup$ denotes the disjoint product and for any finite set $\mu$, $|\mu|$ denotes its cardinality. 
\ed

\bc[Operator product expansion]
\label{OPE}
For any integer $n\geq 2$, any generic $ x_1\cdot E_1,\dots,  x_n\cdot E_n\in\widehat\curve_{\mathfrak m}$ and any $i,j\in\{1,\dots,n\}$,
\bea
\widehat W_n(X_1,\dots,X_n)&\underset{x_i\sim x_j}=& \langle E_i,E_j\rangle\frac{\diff\xi_i\diff\xi_j}{(\xi_i-\xi_j)^2}\widehat W_{n-2}(\dots,\widehat X_i,\dots,\widehat X_j,\dots)\nonumber\\
&&+\ \frac{\diff \xi_i}{\xi_i-\xi_j}\widehat W_{n-1}\big(\dots,x_j\cdot[E_i,E_j],\dots\big)+\mathcal O(1)
\eea
close to the diagonal with local coordinates $\xi_i\sim\xi_j$ for $x_i$ and $x_j$. Furthermore,
\beq
\widehat W_n(X_1,\dots,X_n)\underset{x_i\sim p_j}=  \langle E_i,\Phi_j\rangle\frac{\diff \xi_i}{\xi}\widehat W_{n-1}(\dots,\widehat X_i,\dots)\times\mathcal O(1)\quad \text{if}\quad E_i\in\Lieh_j
\eeq
and for any $\mathfrak r\in\mathfrak R^{(j)}$ and $X_i=x_i\cdot E_{\mathfrak r}$ with $\mathbb C E_{\mathfrak r}=\Lieg_{\mathfrak r}^{(j)}\underset{def}=\Ad_{C_j}^{-1}\Lieg_\mathfrak r$,
\beq
\widehat W_n(X_1,\dots,X_n)\underset{x_i\sim p_j}=  \xi_i^{\mathfrak r(\Phi_j)} \diff\xi_i\ \widehat W_{n-1}(\dots,\widehat X_i,\dots)\times\mathcal O(1)
\eeq
\ec

\proof
The non--connected higher amplitude $\widehat W_n$ has a second order pole on the diagonal coming from the factors proportional to $W_2$ in its expression in terms of the connected higher amplitudes. The corresponding residue is then computed from Lemma \ref{diagonalresidue}.
\eproof

Consider the Casimir elements of $\Lieg$, generators of the center $\mathcal Z\,\mathcal U(\Lieg)$ of its universal enveloping algebra. Given a basis $\{v_1,\dots,v_{\dim\Lieg}\}$ of $\Lieg$ and its dual basis $\{v^1,\dots,v^{\dim\Lieg}\}$ such that $\left\langle v_i,\, v^j\right\rangle=\delta_{i,j}$, the $k^{\text{th}}$ Casimir, for $k\in\{1,\dots,\dim\Lieh\}$, has the form
\beq
 C_{d_k}=\sum_{1\leq i_1,\dots, i_k\leq \dim\Lieg} c_{i_1,\dots,i_k} v^{i_1}\otimes\cdots \otimes v^{i_{d_k}}\in\Lieg^{\otimes k}\subset\mathcal U(\Lieg)
\eeq
If $\Lieg$ is simply--laced, then the degrees $d_1<\cdots<d_{\dim\Lieh}$ are such that $(d_1-1,\dots,d_{\dim\Lieh}-1)$ are its Coxeter exponents. The Casimir elements are independent of the choice of basis of $\Lieg$ and defined in this way up to identification of the skew-symmetrized tensor product with the commutator. Additionally define $d_0\underset{def}{=}0$, $C_0\underset{def}{=} 1\in\mathcal U(\Lieg)$ and note that the quadratic Casimir is given by
\beq
C_2 = \sum_{i=1}^{\dim\Lieg} v_i \otimes v^i
\eeq

\bd[Casimir insertion]
For any $n\in\mathbb N$ and any $k\in\{1,\dots, D\}$ the insertion of the $k^{\operatorname{th}}$ Casimir into $W_n$ at the universal covering point $\widetilde x\in\overset{\circ}{\widetilde\curve}_p$ is defined by the formula
\bea
 C_{d_k}\widehat W_n(\widetilde x;X_1,\dots, X_n)&=\, C_{d_k}(\widetilde x)\cdot \widehat W_n(X_1,\dots,X_n)\qquad\qquad\qquad\\
\underset{def}{=} \sum_{1\leq i_1,\cdots,i_{d_k}\leq \dim\Lieg}&c_{i_1,\dots, i_{d_k}} \widehat W_{d_k+n}(\widetilde x\cdot v^{i_1},\dots,\widetilde x\cdot v^{i_{d_k}},X_1,\dots,X_n)
\eea
for generic choices of the arguments $\widetilde x_i\in\overset{\circ}{\widetilde\curve}_p$ and $E_i\in\Lieg$, denoted $X_i\underset{def}{=}\widetilde x_i\cdot E_i$ for $i\in\{1,\dots,n\}$. It is the contraction of the $d_k$ first indices of $W_{d_k+n}$ by the $k^{\operatorname{th}}$ Casimir, at coinciding universal covering point $\widetilde x$, where the regularized evaluation is used. It is independent of the choice of basis of $\Lieg$.
\ed

Consider a faithful $D_{\geq 2}$--dimensional representation of $\Lieg$ that we denote $\rho:\Lieg\subset \mathfrak{gl}_D$ and choose the multilinear bracket to be $\left\langle\,\bullet\,\right\rangle = \underset{\rho}{\Tr}$.

\bt[Loop equations]
\label{loopequations}
For any integer $n\in\mathbb N$ and any generic elements $X_1,\dots,X_n\in \widehat\curve_{\mathfrak m}$ \cite{BBE2015,BEM2016},
\beq
\sum_{l=0}^D(-1)^l  C_l \widehat W_n(\widetilde x;X_1,\dots,X_n)\, \eta^{D-l} = [\varepsilon_1\cdots\varepsilon_n]\, \underset{\rho}{\det}\left( \eta-\Phi(x)-\mathcal M^{(n)}_\varepsilon(x;X_1,\dots,X_n)\right)
\eeq
where $\eta\in\operatorname K$ is a formal one--form, we set $C_l=0$ if $l\neq d_k$ for all $k\in\{1,\dots, \dim\Lieh\}$ and the right--hand side features
\beq
\mathcal M^{(n)}_\varepsilon(x;X_1,\dots,X_n)\underset{def}{=}\sum_{m=1}^n \sum_{1\leq i_1\neq\cdots\neq i_m\leq n}\varepsilon_{i_1}\cdots\varepsilon_{i_m} \frac{\sigma_\Psi(X_{i_1})\cdots\sigma_\Psi(X_{i_m})}{\mathcal E( x, x_{i_1})\cdots\mathcal E( x_{i_m}, x)}
\eeq
We moreover used the symbol defined for any polynomial of the formal variables $\varepsilon_1,\dots,\varepsilon_n$ by
\beq
[\varepsilon_{i_1}\cdots\varepsilon_{i_n}]\,\sum_{m_1,\dots,m_n} f_{m_1,\dots,m_n} \varepsilon_1^{m_1}\cdots\varepsilon_n^{m_n}\underset{def}{=} f_{i_1,\dots,i_n}
\eeq
\et

We extract from the loop equations that the dependence of $\mathcal C_{d_k} \widehat W_n$ in the insertion point of the Casimir element $\mathcal C_k$ is meromorphic on the base curve $\curverond_p$.

\bt[Special geometry]
For any $n\in\mathbb N^*$ and deformation $\delta\in T_{\mathfrak m}\modsp_p$,
\beq
\delta W_n = \underset{\Gamma_\delta}\oint W_{n+1}
\eeq
\et

\proof
This can be shown recursively and is a direct consequence of the deformation property \ref{defrepker} of the self--reproducing kernel $\mathcal K_\Psi$ applied to the determinant formula defining the amplitude $W_n$.
\eproof

\bl
\label{deltakerB}
For any $k\in\{1,\dots,\genus'''\}$, $\oint_{\acycle'_k}W_3=0$ and equivalently, for any choice of indices $k,l\in\{1,\dots,\genus'''\}$, $\partial_{\bcycle_l}\acycle_k'\in\Ker\widehat B$.
\el

\proof
The explicit expression of $W_2$ and the fact that $\oint_{\acycle'_k}W_2=0$ yield that for any $Z\in\widehat\curve_{\mathfrak m}$,
\beq
\big\langle\sigma_\Psi(Z)\underset{X\in\acycle'_k}\oint\frac{\sigma_\Psi(X)}{\mathcal E(z,x)\mathcal E(x,z)}\big\rangle=0
\eeq
which implies $\underset{X\in\acycle'_k}\oint\frac{\sigma_\Psi(X)}{\mathcal E(z,x)\mathcal E(x,z)}=0$ for all generic $z\in\curve_\Upsilon$. Using lemma \ref{holcor} then shows that $\oint_{X\in\acycle'_k}\sigma_\Psi(X)\omega'''_{y,z}(x)=0$ for any $y,z\in\curve_\Upsilon$ such that the definition of $W_3$ and (again) lemma \ref{holcor} then yield
\bea
\underset{X\in\acycle'_k}\oint\, W_3 (X,Y,Z)&=&\underset{X\in\acycle'_k}\oint \frac{\big\langle\sigma_\Psi(X)\sigma_\Psi(Y)\sigma_\Psi(Z)\big\rangle}{\mathcal E(x,y)\mathcal E(y,z)\mathcal E(z,x)}+\underset{X\in\acycle'_k}\oint \frac{\big\langle\sigma_\Psi(X)\sigma_\Psi(Z)\sigma_\Psi(Y)\big\rangle}{\mathcal E(x,z)\mathcal E(z,y)\mathcal E(y,x)}\nonumber\\
\\
&=& -\,\Big\langle \Big(\underset{X\in\acycle'_k}\oint\sigma_\Psi(X)\omega'''_{y,z}(x)\Big)\frac{[\sigma_\Psi(Y),\sigma_\Psi(Z)]}{\mathcal E(y,z)\mathcal E(z,y)}\Big\rangle=0
\eea
The equivalence with the second statement in then obtained by writing
\beq
\partial_{\bcycle_l}\widehat B(\acycle'_k)=0=\widehat B(\partial_{\bcycle_l}\acycle'_k)+\underset{\bcycle_l}\oint\underset{\acycle'_k}\oint\, W_3
\eeq
which is obtained by using the fact that the periods of $W_2$ are evaluations of $\widehat B$ together with the previous special geometry theorem.
\eproof

\bc
For all $k\in\{1,\dots,\genus'''\}$ and $\Gamma\in\Ker\widehat B$, $\partial_\Gamma\acycle'_k=0$
\ec

\proof
This is straightforward since $\acycle'_k$ takes the form
\beq
\acycle'_k=\acycle_k-\sum_{l=1}^{\genus'''}\beta_{k,l}\bcycle_l
\eeq
for some coefficients $\{\bcycle_{k,l}\}_{l=1}^{\genus'''}$. Now $\partial_\Gamma\acycle'_k=-\sum_{l=1}^{\genus'''}\big(\partial_\Gamma\beta_{k,l}\big)\bcycle_l$ such that we have $\partial_\Gamma\acycle'_k\in\Ker \widehat B\bigcap \mathcal L_{\mathfrak m}=\{0\}$.
\eproof

\bp
For any integer $n\geq 2$ and $\Gamma\in\Ker\widehat B$,
\beq
\underset\Gamma\oint\,W_n=0
\eeq
\ep

\proof
We prove this claim recursively. It is true for $n=2$ since $W_2$ has the same periods as $\widehat B$ and for $n=3$ by the previous lemma. Furthermore, if for some $n\geq 2$ $\oint_\Gamma W_n=0$ for all $\Gamma\in\Ker\widehat B$, then for any $k,l\in\{1,\dots,\genus'''\}$,
\bea
0 &=& \partial_{\bcycle_l}\underset{\acycle'_k}\oint\,W_n\\
&=&\underset{\partial_{\bcycle_l}\acycle'_k}\oint W_n+\underset{\bcycle_l}\oint\underset{\acycle'_k}\oint\,W_{n+1}
\eea
which implies the wanted result by the recursion hypothesis since $\partial_{\bcycle_l}\acycle'_k\in\Ker\widehat B$.
\eproof

\subsection{Universal algebra of cycles and W--conformal blocks}

The goal of this section is to provide a conformal field theory interpretation of the previous construction when the Lie algebra $\Lieg$ is simply--laced, namely of type ADE. More precisely we will show that the theta--series coefficients $\widehat\Tau$ are conformal blocks of an associative extension $\mathcal W(\Lieg)$ of the Virasoro algebra $\operatorname{Vir}$ called Casimir W--algebra associated to $\Lieg$.

Let us mention at this point that this paragraph was substantially different in a previous version of this paper but very fruitful discussions with J. Teschner led to considerable improvements worth incorporating.

The strategy we adopt here is to observe that the space of locally holomorphic sections $\mathfrak S'''\underset{def}=\Ho^0_{\operatorname{loc}}\big(\modsp_p,\widehat\Ho{}'''_1\big/\bigotimes_{j=1}^M\gamma_{p_j}\otimes\Lieh_j\big)$ is a Lie algebra and construct a representation of $\mathcal W(\Lieg)$ in a Lagrangian sub--algebra $\mathfrak A$ of a completion of the associated universal envelopping algebra denoted $\mathfrak A'''\underset{def}=\widetilde{\mathcal U}\big(\mathfrak S'''\big)$. It will play the role of algebra of observables of a chiral quantum theory of flat connections whose classical solutions will follow monodromy preserving flows.

Before introducing these notions, let us recall a few definitions. The Virasoro algebra $\operatorname{Vir}$ is the infinite dimensional Lie algebra that generates the conformal transformations of the complex plane. It is defined as the central extension
\begin{eqnarray}
0\longrightarrow \mathbb C \mathfrak c \longrightarrow \operatorname{Vir} \longrightarrow \text{Der}_{\mathbb C} \longrightarrow 0
\end{eqnarray}
where we introduced the Lie algebra $\text{Der}_{\mathbb C}$ of holomorphic derivations of the field of Laurent series on the complex plane as well as the central charge element $\mathfrak c$. In our context it will act as the scalar $c\underset{def}{=}\dim\Lieh$. $\operatorname{Vir}_c\underset{def}=\operatorname{Vir}\big/(c-\mathfrak c)$ defined in this way is generated by the scalar $c$ together with the elements $\{L_n\}_{n\in\mathbb Z}$ satisfying the celebrated commutation relations
\begin{eqnarray}
[L_n,L_m]=\frac {c}{12}n(n^2-1)\delta_{n+m,0}+(n-m)L_{n+m}\quad \text{for}\quad n,m\in\mathbb Z^2
\end{eqnarray}
where for any $n\in\mathbb Z$, $L_n$ generates the one-parameter family of local  conformal transformations $(z\longmapsto t\, z^{n+1})_{t\in\mathbb C}$ e.g. $L_0$ is the dilation operator. It goes to the Witt algebra in the zero central charge limit $c\longrightarrow 0$. 

The conformal bootstrap \cite{BPZ1984} aims at computing correlation functions of physical observables in quantum field theories with conformal symmetry and based on exploiting the information yielded by the corresponding constraints. In particular, the correlation function of some observables in a quantum field theory whose symmetry algebra $\bold A$ is an associative extension $\operatorname{Vir}_c\subset\bold A$ of the Virasoro algebra will decompose on so--called \textit{blocks} satisfying \textit{extended} conformal Ward identities. We use the following definition.

\bd[Blocks of extended conformal algebras]
For any $N\in\mathbb N^*$, an $N$--point $\bold A$--symmetric conformal block on $\curverond$ is an $\bold A$--invariant element in the dual space $\big(R_1\otimes\cdots\otimes R_N\big)^*$ where $R_1,\dots,R_N\in\operatorname{Rep}\bold A$ is any choice of $N$--tuple of representations of the algebra $\bold A$ together with a set of pairwise distinct points $x_1,\dots,x_N\in\curverond$ to which these representations are respectively attached. Namely, it is a linear form $\langle\,\mathfrak a\,|\in\big(R_1\otimes\cdots\otimes R_N\big)^*$ such that for any vector $|\,\mathfrak v\,\rangle=|\,\mathfrak v_1\,\rangle\otimes\cdots\otimes |\,\mathfrak v_N\,\rangle\in R_1\otimes\cdots\otimes R_N$ and any $A\in\bold A$,
\bea
\langle\,\mathfrak a\,|\, A\,|\,\mathfrak v\,\rangle &\underset{def}=& \langle\,\mathfrak a\,|\Big(\sum_{i=1}^N \bold 1_1\otimes\cdots\otimes A^{(i)}\otimes\cdots\otimes\bold 1_N \Big)|\,\mathfrak v\,\rangle\\
&=& \langle\,\mathfrak a\,|\sum_{i=1}^N \Big(|\,\mathfrak v_1\,\rangle\otimes A^{(i)}|\,\mathfrak v_i\,\rangle\otimes\cdots\otimes |\,\mathfrak v_N\,\rangle\Big)\\
&=&0
\eea
where $\bold 1_k$ denotes how the neutral element of $\bold A$ is represented in $R_k$. We will refer to this infinite set of linear equations as the $\bold A$--symmetric Ward identities associated to the choice $R_1,\dots,R_N$ of representations.
\ed

We will consider the particular associative extension
\beq
\operatorname{Vir}_c\subset\bold A\underset{def}= \mathcal W(\Lieg)
\eeq
called the Casimir W--algebra associated to the Lie algebra $\Lieg$. It can be defined in two ways that are equivalent up to isomorphism from the affine algebra at level $\kappa\in\mathbb C$ denoted $\widehat{\mathfrak g}_\kappa$, itself defined very much like the Virasoro algebra. It is the Lie algebra $\widehat{\mathfrak g}_\kappa \underset{def}{=}\widehat{\mathfrak g}\, \big/(\kappa-K)$, where $\widehat{\mathfrak g}$, called the generic affine algebra associated to $\mathfrak g$, is defined as the central extension of vector spaces
\begin{eqnarray}
0\longrightarrow \mathbb C K\longrightarrow \widehat{\mathfrak g} \longrightarrow \mathcal L(\mathfrak g)\oplus\mathbb C \partial \longrightarrow 0
\end{eqnarray}
where $\mathcal L(\mathfrak g)$ is the \textit{loop algebra} of $\mathfrak g$ denoted $\mathcal L(\mathfrak g)\underset{def}{=}\mathfrak g((t))$ (endowed with the natural Lie algebra structure coming from $\mathfrak g$) and the extra generator $\partial$ is defined to satisfy
\begin{eqnarray}
[\partial, M]=\frac{\text{d}}{\text{dt}}M,\quad (\text{and thus}\,\, [\partial, K]=0)
\end{eqnarray}
for any $M\in\mathcal L(\mathfrak g)=\mathfrak g((t))$. We consider here the level one case $\kappa\underset{def}=1$ and $\widehat\Lieg_1$ is then generated by the scalar $K$ acting as the identity together with all possible evaluations of the $\Lieg^*$--valued elements $\{J_n\}_{n\in\mathbb Z}$ satisfying the affine algebra commutation relations
\beq
[J_n(E),J_m(F)]=n\,\langle E,F\rangle\,\delta_{n+m,0}+J_{n+m}([E,F])
\eeq
for any $n,m\in\mathbb Z$ and $E,F\in\Lieg$. These commutation relations can be encoded in the defining operator product expansion of the state--operator map of the corresponding vertex operator algebra
\bea
\label{affinerelations}
J(\xi\cdot E)\,J(\eta\cdot F)&\underset{\xi\sim \eta}=&\langle E,F\rangle\frac{\diff \xi\,\diff \eta}{(\xi-\eta)^2} + \frac{\diff \xi}{\xi-\eta}\, J(\eta\cdot [E,F])+\mathcal O(1)\\
\text{with}\quad J(\xi\cdot E) &\underset{def}= & \sum_{n\in\mathbb Z} \frac{J_n(E)}{\xi^{n+1}}\diff \xi
\eea
where $\xi$ and $\eta$ are as yet formal expansion coordinates. The Casimir algebra $\mathcal W(\Lieg)$ is then naturally defined as the algebra generated by the modes $\{\mathcal W^{(d_k)}_n\}_{\underset{1\leq k\leq \dim\Lieh}{n\in\mathbb Z}}$ of the higher--spin currents defined by the formula
\bea
\sum_{l=0}^D(-1)^k\mathcal W^{(l)}(\xi)\eta^{D-l}&\underset{def}=& \Big(\underset\rho\det \big( \eta-J(\xi)\big)\Big)\\
\mathcal W^{(d_k)}(\xi)&\underset{def}=&\sum_{n\in\mathbb Z}\frac{\mathcal W^{(d_k)}_n}{\xi^{n+d_k}}(\diff \xi)^{d_k} = \Big( C_{d_k}\big(J(\xi)\big)\Big)
\eea
where $\Big( \bullet \Big)$ denotes the interacting normal ordering (picking the only non--vanishing regular limit of an operator product expansion at coinciding points) and computed by $\Big(AB(\xi)\Big)\underset{def}=\underset{\zeta=\xi}\Res\frac{A(\zeta)B(\xi)}{\zeta-\xi}$. These currents in turn satisfy some operator product expansions encoding commutation relations between generators. Note in particular that even though the standard Lie bracket on this associative algebra defines an action of $\mathcal W(\Lieg)$ on itself, the result of this operation is expressed \textit{algebraically} in terms of the generators. See \cite{BS1993} for a review of W--symmetry and its exotic properties.

\br
$\mathcal W(\Lieg)$ is for instance the symmetry algebra of Toda quantum field theory with Lie algebra $\Lieg$ when the corresponding mass parameter $b$ takes the topological value $b=\mathbf i$.
\er

This set of generators contains in particular the element $T\underset{def}=\frac 1{D+h^{\operatorname v}}\mathcal W^{(2)}$ ($h^{\operatorname v}$ being the dual Coxeter number) whose modes generate a Virasoro algebra with central charge $\mathfrak c = \frac{\dim\Lieg}{1+h^{\operatorname v}}\underset{\operatorname{ADE}}=\dim\Lieh$. This turns both the vertex algebras associated to $\widehat\Lieg_1$ and $\mathcal W(\Lieg)$ into vertex \textit{operator} algebras with the sequence of inclusions
\beq
\operatorname{Vir}_c\subset\mathcal W(\Lieg)\subset\widetilde{\mathcal U}(\,\widehat\Lieg_\kappa)
\eeq
valid for $c=\frac{\dim\Lieg}{1+h^{\operatorname v}}\underset{\operatorname{ADE}}=\dim\Lieh$ and $\kappa=1$ only and where $\widetilde{\mathcal U}$ denotes a completion of the universal enveloping algebra.

\bp
The coset $\mathfrak S\underset{def}=\mathfrak S'''\big/\Ker\widehat B$ is a commutative finite dimensional Lie algebra with dimension $\genus+M\dim\Lieh$ and whose linear dual is identified with the locally holomorphic sections $\mathfrak S^{\operatorname v}\underset{def}=\Ho^0_{\operatorname{loc}}(\modsp_p,\big(\widehat\Ho{}^1\big){}''')$ of the vector bundle of generalized differential forms of the third kind.
\ep

\proof
The Lie algebra structure on $\mathfrak S'''=\Ho^0_{\operatorname{loc}}\big(\modsp_p,\widehat\Ho{}'''_1\big/\bigotimes_{j=1}^M\gamma_{p_j}\otimes\Lieh_j\big)$ is pulled--back from that of the locally holomorphic vector fields over $\modsp_p$ by the cycle deformation duality. More concretely, the Lie bracket of $\Gamma_1,\Gamma_2\in\mathfrak S$ is given by
\beq
[\Gamma_1,\Gamma_2]\underset{def}=\Gamma_{[\partial_{\Gamma_1},\partial_{\Gamma_2}]}
\eeq
which vanishes on $\mathfrak S=\mathfrak S'''\big/\Ker\widehat B$. The pairing with the proposed dual is performed by comparing dimensions and by the generalized integration pairing introduced in previous sections. 
\eproof

The space of states (or equivalently of observables) of the chiral theory is defined as a completion of the symmetric algebra
\beq
\mathfrak A\underset{def}=\Big(\widetilde{\operatorname S}(\mathfrak S),\big\langle\bullet|\bullet\big\rangle\Big)
\eeq
together with the bilinear form $\big\langle\bullet\big|\bullet\big\rangle$ defined by
\beq
\big\langle\, \bcycle_1^{n_1}\cdots\bcycle_{\genus'''}^{n_{\genus'''}}\,\big|\,\bcycle_1^{m_1}\cdots\bcycle_{\genus'''}^{m_{\genus'''}}\, \big\rangle\underset{def}=\prod_{k=1}^{\genus'''}\delta_{n_k,m_k}
\eeq
 where we denoted $\big|\,\Gamma_1\cdots\Gamma_n\,\big\rangle\underset{def}=\Gamma_1\cdots\Gamma_n$ the pure tensors of the symmetric algebra. $\mathfrak A$ is generated with help of ladder operators $Q$ and its adjoint ${}^tQ$ defined by
\bea
Q,{}^tQ: & \mathfrak S & \longrightarrow\, \operatorname{End}(\mathfrak A)\nonumber\\
Q(\bcycle_k)\,\big|\,\bcycle_1^{n_1}\cdots\bcycle_{\genus'''}^{n_{\genus'''}}\,\big\rangle & \underset{def}=& \sqrt{n_k+1}\,\big|\,\bcycle_1^{n_1}\cdots\bcycle_k^{n_k+1}\cdots\bcycle_{\genus'''}^{n_{\genus'''}}\,\big\rangle\\
{}^tQ(\bcycle_k)\,\big|\,\bcycle_1^{n_1}\cdots\bcycle_{\genus'''}^{n_{\genus'''}}\,\big\rangle & \underset{def}=& \sqrt{n_k}\,\big|\,\bcycle_1^{n_1}\cdots\bcycle_k^{n_k-1}\cdots\bcycle_{\genus'''}^{n_{\genus'''}}\,\big\rangle
\eea
 where $n_k$ generically denotes the number of occurrences of $\bcycle_k$ in $\bcycle_1^{n_1}\cdots\bcycle_{\genus'''}^{n_{\genus'''}}$. They satisfy $[{}^tQ(\widetilde\Gamma),Q(\Gamma)]=0$ if $\widetilde\Gamma\neq\Gamma$ and $[Q(\widetilde\Gamma),Q(\Gamma)]=0$, $[{}^tQ(\widetilde\Gamma),{}^tQ(\Gamma)]=0$ and $[{}^tQ(\Gamma),Q(\Gamma)]=\bold 1_{\operatorname{End}(\mathfrak A)}$ for any $\widetilde\Gamma,\Gamma\in\mathfrak S$. Note that the operator $\bold N_k\underset{def}=Q{}^tQ(\bcycle_k)$ yields the $k^{\operatorname{th}}$ occupation number and $\bold N\underset{def}=\sum_{k=1}^{\genus'''}\bold N_k$ the total one
\bea
\bold N_k\,\big|\,\bcycle_1^{n_1}\cdots\bcycle_{\genus'''}^{n_{\genus'''}}\,\big\rangle & = & n_k\,\big|\,\bcycle_1^{n_1}\cdots\bcycle_{\genus'''}^{n_{\genus'''}}\,\big\rangle\\
\bold N\,\big|\,\bcycle_1^{n_1}\cdots\bcycle_{\genus'''}^{n_{\genus'''}}\,\big\rangle & = & \Big(\sum_{k=1}^{\genus'''}n_k\Big)\,\big|\,\bcycle_1^{n_1}\cdots\bcycle_{\genus'''}^{n_{\genus'''}}\,\big\rangle
\eea
\beq
\text{We then write}\quad\big|\,\bcycle_1^{n_1}\cdots\bcycle_{\genus'''}^{n_{\genus'''}}\,\big\rangle = \Big(\prod_{k=1}^{\genus'''}\frac{Q(\bcycle_k)^{n_k}}{\sqrt{(n_k!)}}\Big)\big|\,\bold 1_{\mathfrak A}\,\big\rangle
\eeq

Define a linear form $\widehat\varphi_{\mathfrak m}\in\mathfrak A^*$ for any $\mathfrak m\in\modsp_p$ by $\widehat\varphi_{\mathfrak m}(\bold 1_{\mathfrak A})\underset{def}=\widehat\Tau(\mathfrak m;\mathcal L)$ and for any $n\in\mathbb N^*$ and $\Gamma_1,\dots,\Gamma_n\in\mathfrak S$, 
\beq
\widehat\varphi_{\mathfrak m}(\Gamma_1\cdots\Gamma_n)\underset{def}=\underset{\Gamma_1}\oint\cdots\underset{\Gamma_n}\oint \widehat W_n(\mathfrak m)
\eeq
It allows to define the $S$--matrix operator $\bold S(\mathfrak m)\in\operatorname{End}(\mathfrak A)$ characterizing the model by
\bea
\big\langle\, \bcycle_1^{n_1}\cdots\bcycle_{\genus'''}^{n_{\genus'''}}\,\big|\,\bold S(\mathfrak m)\,\big|\,\bcycle_1^{m_1}\cdots\bcycle_{\genus'''}^{m_{\genus'''}}\, \big\rangle &\underset{def}= & \bold S^n_m(\mathfrak m)\\
&\underset{def}=&\frac{\widehat\varphi_{\mathfrak m}\big(\bcycle_1^{n_1+m_1}\cdots\bcycle_{\genus'''}^{n_{\genus'''}+m_{\genus'''}}\big)}{\prod_{k=1}^{\genus'''}\sqrt{(n_k!)}\sqrt{(m_k!)}}\nonumber
\eea
defined by its symmetric matrix elements. The normalization factors are chosen such that $\bold S$ intertwines between $Q$ and ${}^tQ$ in the sense that for all $\Gamma\in\mathfrak S$,
\beq
{}^tQ(\Gamma)\,\bold S(\mathfrak m) = \bold S(\mathfrak m)\,Q(\Gamma)
\eeq
Note furthermore that the unit element $\bold 1_{\mathfrak A}\in\widetilde{\operatorname S}(\mathfrak S)$ satisfies by convention 
\beq
\big\langle\,\bcycle_1^{n_1}\cdots\bcycle_{\genus'''}^{n_{\genus'''}}\,\big|\,\bold S(\mathfrak m)\,\big|\,\bold 1_{\mathfrak A}\,\big\rangle = \bold S^{ n}_0(\mathfrak m)=\big\langle\,\bold 1_{\mathfrak A}\,\big|\,\bold S(\mathfrak m)\,\big|\,\bcycle_1^{n_1}\cdots\bcycle_{\genus'''}^{n_{\genus'''}}\,\big\rangle
\eeq
for any values of the indices, thus defining the co--vector $\big\langle\,\bold S(\mathfrak m)\,\big|\underset{def}=\big\langle\,\bold 1_{\mathfrak A}\,\big|\bold S(\mathfrak m)\in\mathfrak A^*$.

The periods of the higher amplitudes $\{\widehat W_n\}_{n\in\mathbb N^*}$ therefore compute matrix elements of observables acting on $\mathfrak A$ in the general form
\bea
\label{periodsascorrelators}
\underset{\Gamma}\oint\cdots\underset{\Gamma_n}\oint\,\widehat W_n (\mathfrak m) &=& \big\langle\,\bold S(\mathfrak m)\,\big|\,Q(\Gamma_1)\cdots Q(\Gamma_n)\,\big|\,\bold 1_{\mathfrak A}\,\big\rangle\quad\text{for}\quad n>0\nonumber\\
\text{and}\quad  \widehat\Tau(\mathfrak m;\mathcal L)& = & \big\langle\,\bold S(\mathfrak m)\,\big|\,\bold 1_{\mathfrak A}\,\big\rangle\quad\text{for}\quad n=0
\eea


These commuting charges are the conserved quantities of the time--evolution of the model. They are encoded in the conserved current
\beq
J\underset{def}=\sum_{k,l=1}^{\genus'''} Q(\bcycle_k)\,\tau^{-1}_{k,l}\,\widehat B(\bcycle_l)\in \operatorname{End}\big(\mathfrak A\big)\otimes \big(\widehat\Ho{}^1\big){}'''
\eeq
satisfying $\underset{\Gamma}\oint J = Q(\Gamma)$ for any $\Gamma\in\mathfrak S$ and whose correlation functions exactly compute the regularized definitions of the higher amplitudes. Namely, for all $n\in\mathbb N^*$, 
\beq
\underset{z_1=x_1}\Res\cdots\underset{z_n=x_n}\Res\ \frac{\widehat W_n(z_1\cdot E_1,\cdots,z_n\cdot E_n)(\mathfrak m)}{\mathcal E(z_1,x_1)\cdots\mathcal E(z_n,x_n)} \underset{def}= \big\langle\, \bold S(\mathfrak m)\,\big|\,\prod_{i=1}^n J(x_i\cdot E_i)\,\big|\,\bold 1_{\mathfrak A}\,\big\rangle
\eeq
for all values of the arguments. We (loosely) still denote by $\{\widehat W_n\}_{n\in\mathbb N^*}$ the regularized  amplitudes and drop the explicit dependence in $\mathfrak m\in\modsp_p$ such that for any $n\in\mathbb N^*$, 
\beq
\widehat W_n(X_1,\cdots,X_n)= \big\langle\, \bold S\,\big|\,\prod_{i=1}^n J(X_i)\,\big|\,\bold 1_{\mathfrak A}\,\big\rangle
\eeq

Recall the first expansion of corollary \ref{OPE} of the non--connected  amplitude $\widehat W_n$, 
\bea
\widehat W_n(X_1,\dots,X_n)&\underset{x_i\sim x_j}=& \langle E_i,E_j\rangle\frac{\diff\xi_i\diff\xi_j}{(\xi_i-\xi_j)^2}\widehat W_{n-2}(\dots,\widehat X_i,\dots,\widehat X_j,\dots)\nonumber\\
&&+\ \frac{\diff \xi_i}{\xi_i-\xi_j}\widehat W_{n-1}(\dots,x_j\cdot[E_i,E_j],\dots)+\mathcal O(1)
\eea
for any integer $n\geq 2$, close to the diagonal with local coordinates $\xi_i\sim\xi_j$ for $x_i,x_j$ and we denoted $X_k=x_k\cdot E_k\in\widehat\curve_{\mathfrak m}$. Translated in terms of correlators of the current, this yields the operator product expansion
\beq
J(\xi\cdot E) J(\eta\cdot F) \underset{\xi\sim\eta}=\langle E,F\rangle \frac{\diff\xi\diff\eta}{(\xi-\eta)^2}+\frac{\diff\xi}{\xi-\eta}J(\eta\cdot[E,F])+\mathcal O(1)
\eeq
corresponding to the commuting relations \ref{affinerelations} of the generators of the affine Lie algebra $\widehat\Lieg_1$ for expansion modes in a local coordinate $\xi$ around a generic point $x_0\in\curve_\Upsilon$
\beq
J(x\cdot E)\underset{def}=\sum_{n\in\mathbb Z}\frac{J_n(E)_{x_0}}{\xi^{n+1}}\diff\xi
\eeq

 The second asymptotic expansion of corollary \ref{OPE} reads
\beq
\widehat W_n(X_1,\dots,X_n)\underset{x_i\sim p_j}=  \xi_i^{\mathfrak r(\Phi_j)} \diff\xi_i\ \widehat W_{n-1}(\dots,\widehat X_i,\dots)\times\mathcal O(1)
\eeq
for all $j\in\{1,\dots,M\}$, where $X_i=x_i\cdot E_{\mathfrak r}$ with $\mathfrak r\in\mathfrak R^{(j)}$. In general, $\mathfrak r(\Phi_j)\notin\mathbb Z$ such that $J$ is only locally holomorphic with monodromy generically given by
\beq
J(x+\gamma\cdot E) = J(x\cdot\Ad_{S_\gamma}^{-1}E)
\eeq
The asymptotic behavior of the coefficient
\beq
\big\langle\,\bold S\,\big|\,J(X)\,Q(\bcycle_{j_1})\cdots Q(\bcycle_{j_m})\,\big|\,\bold 1_{\mathfrak A}\,\big\rangle  =  \underset{X_1\in\bcycle_{j_1}}\oint\cdots\underset{X_m\in\bcycle_{j_m}}\oint\widehat W_{m+1}(X,X_1,\dots,X_m)
\eeq
near a puncture $p_j$, for some index $j\in\{1,\dots,M\}$, is given by
\beq
\big\langle\,\bold S\,\big|\,J(X)\,Q(\bcycle_{j_1})\cdots Q(\bcycle_{j_m})\,\big|\,\bold 1_{\mathfrak A}\,\big\rangle \underset{x\sim p_j}=  \langle H,\Phi_j\rangle\frac{\diff\xi}{\xi}\widehat\varphi(\bcycle_{j_1}\cdots\bcycle_{j_m})\,+\,\mathcal O(1)
\eeq
when $X=x\cdot H$ with $H\in\Lieh_j$. This implies that for any $H\in\Lieh_j$ and any $n\in \mathbb N$,
\beq
\big\langle\,\bold S\,\big|J_n(H)_{p_j} = \big\langle\,\bold S\,\big|\,\langle H,\Phi_j\rangle\,\delta_{n,0}\eeq

In turn, $\big\langle\,\bold S\,\big|$ is simultaneously a highest--weight co--vector for copies $\widehat\Lieg{}^{(j)}_1\simeq\widehat\Lieg_1$ of the affine Lie algebra respectively generated by $\{J_n(H)\}_{\underset{H\in\Lieh_j}{n\in\mathbb Z}}$, $j\in\{1,\dots,M\}$. This defines a right--action of $\widehat\Lieg_1$ through $\widehat\Lieg{}^{(1)}_1\otimes\cdots\otimes\widehat\Lieg{}^{(M)}_1$ on the dual Fock space $\mathcal F^*\underset{def}=\big\langle\,\bold S\,\big|\,\widetilde{\operatorname S}(\mathfrak S)$ defined from $\big\langle\,\bold S\,\big|$, in which the one representing $\widehat\Lieg{}^{(j)}_1$ has corresponding highest--weight dual to $\alpha_j\in\Lieh$ (and not to $\Phi_j\in\Lieh_j$).

From this last point follows the existence of a right--action of the Casimir W--algebra $\mathcal W(\Lieg)$ on $\mathcal F^*$ that is generated by the asymptotic expansion modes of differential--valued operators $\mathcal W^{(d_k)}\underset{def}=\Big( C_{d_k}(J)\Big)\in\operatorname{End}(\mathfrak A)\, \otimes \, \Ho^0(\curverond_p, \operatorname K^{\otimes d_k}_{\curverond_p})$ near the punctures $p_1,\dots,p_M\in\curverond$ and for $k\in\{1,\dots,\dim\Lieh\}$. The definition of $\mathcal W(\Lieg)$ and the loop equations \ref{loopequations} together imply that these generating  higher--spin currents satisfy
\beq
\big\langle\,\bold S\,\big|\,\mathcal W^{(d_k)}(x) \, Q(\bcycle_{j_1})\cdots Q(\bcycle_{j_m})\,\big|\,\bold 1_{\mathfrak A}\,\big\rangle = \underset{X_1\in\bcycle_{j_1}}\oint\cdots\underset{X_m\in\bcycle_{j_m}}\oint  C_{d_k}\widehat W_m (x;X_1,\dots, X_m)
\eeq
as meromorphic expressions of $x\in\curverond_p$ and with asymptotic expansions of the form
\bea
\big\langle\,\bold S\,\big|\,\mathcal W^{(d_k)}(x) \, Q(\bcycle_{j_1})\cdots Q(\bcycle_{j_m})\,\big|\,\bold 1_{\mathfrak A}\,\big\rangle &\underset{x\sim p_j}=& C_{d_k}(\alpha_j)\frac{(\diff\xi)^{d_k}}{\xi^{d_k}}\widehat\varphi(\bcycle_{j_1}\cdots\bcycle_{j_m})\times\mathcal O(1)\nonumber\\
\\
&\underset{x\sim p_j}=& C_{d_k}(\alpha_j)\frac{(\diff\xi)^{d_k}}{\xi^{d_k}}\Big(\underset{\bcycle_{j_1}}\oint\cdots\underset{\bcycle_{j_m}}\oint \widehat W_m\Big)\times\mathcal O(1)\nonumber\\
\eea
near the puncture $p_j$, where we used 
\beq
\big\langle\,\bold S\,\big|\,Q(\bcycle_{j_1})\cdots Q(\bcycle_{j_m})\,\big|\,\bold 1_{\mathfrak A}\,\big\rangle=\widehat\varphi(\bcycle_{j_1}\cdots\bcycle_{j_m})=\underset{\bcycle_{j_1}}\oint\cdots\underset{\bcycle_{j_m}}\oint \widehat W_m
\eeq
The corresponding residues are expressed through commutators with the Casimir element $C_{d_k}$ and therefore vanish. As a direct consequence of the last facts and re--expliciting the dependence on $\mathfrak m\in\modsp_p$, we obtain the following

\bt[W--symmetric conformal blocks]
Let $k\in\{1,\dots,\dim\Lieh\}$ and $\gamma_t$ be a small loop surrounding $t\in\curverond_p$. Then for any $\mathfrak m\in\modsp_p$ and any homogeneous meromorphic differential operator $D^{(d_k-1)}$ of degree $d_k-1$ with poles located at $p_1,\dots,p_M$, 
\bea
-\frac1{2\pi\mathbf i}\big\langle\,\bold S(\mathfrak m)\,\big|\oint_{\gamma _t}\mathcal W^{(d_k)} D^{(d_k-1)}\big|\,\bcycle_1^{n_1}\cdots\bcycle_{\genus'''}^{n_{\genus'''}}\,\big\rangle &= &0\\
=\big\langle\,\bold S(\mathfrak m)\,\big|\sum_{j=1}^M&\underset{p_j}\Res &\big(\mathcal W^{(d_k)} D^{(d_k-1)}\big)\big|\,\bcycle_1^{n_1}\cdots\bcycle_{\genus'''}^{n_{\genus'''}}\,\big\rangle\nonumber\\
\eea
thus identifying $\big\langle\,\bold S(\mathfrak m)\,\big|\in\mathfrak A^*$ and furthermore its coefficient $\widehat\Tau(\mathfrak m;\mathcal L)=\big\langle\,\bold S(\mathfrak m)\,\big|\,\bold 1_{\mathfrak A}\,\big\rangle$ as an $M$--point conformal block of the algebra $\mathcal W(\Lieg)$ on $\curverond$, corresponding to representations $R^{(j)}\in\operatorname{Rep}\mathcal W(\Lieg)$, with highest--weight associated to $p_j$ dual to $\alpha_j\in\Lieh$, $j\in\{1,\dots,M\}$.
\et

\proof
We are left to prove the second equality which is true by deforming the contour $\gamma_t$, since $\mathcal W^{(d_k)}$ has no monodromy around closed contours of $\curverond$.
\eproof

\section{Semi--classical analysis}
\label{wkb}

\subsection{Cameral cover and WKB--asymptotics}

In this section we extend the construction from flat--connections to $\varepsilon$--connections, for some non--zero complex parameter $\varepsilon\in\mathbb C^*$, and explain how, in a subsequent work, the authors together with J. Hurtubise reconstruct the corresponding asymptotic expansions, in the WKB--limit $\varepsilon\longrightarrow 0$. Indeed, the introduction of this parameter $\varepsilon\in\mathbb C^*$ is interpreted as a hyper--K\"ahler rotation $\modsp_p\longrightarrow\modsp_p(\varepsilon)$ from the moduli--space of flat meromorphic $G$--connections in principal $G$--bundles over $\curverond$ to that of flat $\varepsilon$--connections $\modsp_p(\varepsilon)$. These spaces are analytically isomorphic to $\modsp_{\operatorname{Betti}}$ via the Riemann-Hilbert correspondence. They are however only diffeomorphic as real manifolds to the moduli--space $\modsp_{\operatorname{Dol}}$ of Higgs bundles, consisting of holomorphic principal $G$-bundles on $\curverond$ together with a meromorphic section of the adjoint bundle $\Phi\in\Ho^0(\curverond_p,\Ad\mathcal P\otimes\operatorname K_{\curverond_p})$ called a Higgs field. We therefore repeat the previous construction starting from $\mathfrak m_\varepsilon\underset{def}{=}[(\mathcal P,\nabla_\varepsilon)]\in\modsp_p(\varepsilon)$, with a $G$--bundle $\mathcal P\longrightarrow\curverond$ equipped with an $\varepsilon$--connection $\nabla_\varepsilon$ with simple poles located at $p_1,\dots,p_M$ and no other singularities. On the fundamental domain $\curve_\Upsilon=\curverond-\Upsilon$ obtained by removing a one--face graph $\Upsilon\subset\curverond$ with $\partial\Upsilon=\{p_1,\dots,p_M\}$ and such that $\pi_1(\curve_\Upsilon,o)=0$, the connection is written
\beq
\nabla_\varepsilon = \varepsilon \diff-\Phi_\varepsilon,\qquad 
\Phi_\varepsilon \underset{def}{=}\sum_{k=0}^\infty \varepsilon^k\Phi^{(k)}
\eeq
which amounts to the simultaneous rescaling $\{\alpha_j\}_{j=1}^M \longrightarrow \{\varepsilon^{-1}\alpha_j\}_{j=1}^M$ of all the charges, also called the heavy limit. We moreover assume that $\Phi_\varepsilon$ admits a power--series expansion in the limit $\varepsilon\longrightarrow 0$ and this in turn defines a Higgs bundle $\underset{\varepsilon\rightarrow 0}{\lim}\ \mathfrak m_\varepsilon = \mathfrak m_0\in\modsp_{\operatorname{Dol}}$, with $\mathfrak m_0=[(\mathcal P,\Phi^{(0)})]$. Indeed the local gauge transformation $g\in\Ho^0(\curve_\Upsilon,\Ad\mathcal P_{|\curve_\Upsilon}\otimes\operatorname K_{\curve_\Upsilon})$ acts on the $\varepsilon$--connection potentials as
\beq
g\cdot \Phi_\varepsilon = \Ad_g\Phi_\varepsilon\, +\, \varepsilon\diff g\cdot g^{-1}
\eeq
and therefore transforms $\Phi^{(0)}$ as a Higgs field
\beq
g\cdot\Phi^{(0)}=\Ad_g\Phi^{(0)}
\eeq
Let us diagonalize it over generic base--points as
\beq
\Phi^{(0)} \underset{def}{=} \Ad_V Y
\eeq
where $V$ and $Y$ are defined up to the action of a Weyl group element $w\in\Weyl$
\beq
V\longrightarrow V\cdot w\qquad\text{and}\qquad Y\longrightarrow \Ad_{w^{-1}} Y
\eeq
More precisely, they are sections $V\in\Ho^0(\curve_\Weyl, (\pi_\Weyl)^*\mathcal P_{|\curve_\Upsilon})$, which is holomorphic at $p_1,\dots,p_M$, and $Y\in\Ho^0(\curve_\Weyl, \mathfrak h\otimes  (\pi_\Weyl)^*\operatorname K_{\curve_\Upsilon})$, a Cartan--valued meromorphic one--form with asymptotic behavior at the poles given by
\beq
Y(z)\underset{\pi_\Weyl(z)\sim p_j}{=} \frac{\alpha_j}{\pi_\Weyl(z)-z_j}\diff\pi_\Weyl(z)\, +\, \mathcal O(1),
\eeq
where $\pi_\Weyl:\curve_\Weyl\longrightarrow \curve_\Upsilon$ is the cameral cover associated to the Higgs field $\Phi^{(0)}$.  It is defined as
\bea
&\curve_\Weyl\ \ \underset{def}{=} & (\pi^\Weyl)^{-1}\mathcal H(\Phi^{(0)})\subset \mathfrak h\otimes \operatorname K_{\curve_\Upsilon} \\
\text{with}\qquad\mathcal H:&\modsp_{\operatorname{Dol}} & \longrightarrow\quad \mathcal B \underset{def}{=} \bigoplus_{k=1}^{\dim\Lieh}\Ho^0(\curve_\Upsilon,\operatorname K_{\curve_\Upsilon}^{\otimes d_k})\simeq [\mathfrak h\otimes \operatorname K_{\curve_\Upsilon}]/\Weyl\\
&[(\mathcal P,\Phi^{(0)})] &\longmapsto\quad \left(C_{d_1}(\Phi^{(0)}),\dots,C_{d_{\dim\Lieh}}(\Phi^{(0)})\right)
\eea
defining the Hitchin integrable system and $\pi^\Weyl:\mathfrak h\otimes \operatorname K_{\curve_\Upsilon}\longrightarrow  [\mathfrak h\otimes \operatorname K_{\curve_\Upsilon}]/\Weyl$ is the canonical projection map.

$\curve_\Weyl$ is a Galois $\Weyl$--covering of $\curve_\Upsilon$ and for any choice of multivalued $\nabla_\varepsilon$--flat section $\Psi_\varepsilon$, satisfying
\beq
\nabla_\varepsilon\Psi_\varepsilon = 0
\eeq
there exists a piecewise constant section $C\in\Ho^0(\curve_\Weyl,G)$ such that the WKB--asymptotics leading order
\beq
\Psi_\varepsilon(\widetilde x)\underset{\varepsilon\rightarrow 0}{=} V(z)\cdot\left(\bold 1_G+\mathcal O(\varepsilon)\right)\cdot \exp\left(\frac 1\varepsilon \int^{\widetilde x} Y\right)\cdot C_{z}
\eeq
hold with projection sequence
\bea
\overset{\circ}{\widetilde\curve}_\Upsilon & \ \ \longrightarrow\quad \curve_\Weyl & \longrightarrow\quad \curve_\Upsilon\\
\widetilde x & \longmapsto\quad z & \longmapsto\quad x
\eea

Recall that for any Lie algebra element $E\in\Lieg$, $\sigma_{\Psi_\varepsilon}(*\cdot E)$ defines a multivalued $\nabla_\varepsilon$--flat section of $\Ad\mathcal P$. We are interested, for a given $\widetilde x\in\overset\circ{\widetilde\curve}_p$, to the subspace $\mathfrak B(\widetilde x)\subset\Lieg$ consisting of Lie algebra elements $E\in\Lieg$ such that $\sigma_{\Psi_\varepsilon}(\widetilde x\cdot E)$ admits a finite limit when $\varepsilon\longrightarrow 0$. This defines a parabolic sub--bundle that is independent of $\varepsilon$ and in turn corresponds to a sub--space of the quantum spectral curve at $\varepsilon=1$ denoted $\widehat{\mathfrak B}\subset\widehat\curve_{\mathfrak m_1}$.

\subsection{Perturbative reconstruction}

The amplitudes $\{W_n\}_{n\in\mathbb N^*}$ are defined as before and we are interested in the expansions of their restrictions to $\widehat{\mathfrak B}$. In the regime $\varepsilon\longrightarrow 0$, their dependence localizes to a curve in $\widehat{\mathfrak B}$ identified with the cameral cover $\curve_\Weyl\subset\widehat{\mathfrak B}$.

These expansions are said to be of topological type if they take the form
\beq
W_n=\sum_{g=0}^\infty \varepsilon^{2g-2+n} \omega_{g,n}
\eeq
where $\omega_{g,n}\in\Ho^0\left(\curve_\Weyl^n,(\mathfrak h\otimes\operatorname K_{\curve_\Weyl})^{\boxtimes n}\right)$ is a $\Weyl$--equivariant symmetric $n$--differential on $\curve_\Weyl$ regular at the ramification points of $\pi_\Weyl:\curve_\Weyl\longrightarrow\curve_\Upsilon$. If so, these differentials can be computed recursively from an equivariant version of the spectral curve topological recursion very similar to that of \cite{DNOPS2015}.

When a highest--weight $h\in\mathfrak h^*$ is chosen, the cameral cover can be projected to a spectral curve $h(\curve_\Weyl)\subset T^*\curve_\Upsilon$ such that the expansions of the amplitudes computed from the cameral curve topolgical recursion and then projected are equal to those computed from the original topological recursion applied to the spectral curve obtained by applying the highest--weight to the cameral curve.

Special geometry then allows for the computation of the corresponding expansion of the tau--function
\beq
\log\widehat\Tau(\mathfrak m_\varepsilon;\mathcal L) \underset{def}{=} \exp\left(\sum_{g=0}^\infty \varepsilon^{2g-2}  F_g(\mathfrak m_0,\mathcal L)\right)
\eeq
where we introduced the so--called free energies $\{ F_g\}_{g=0}^\infty$.

\section{Extensions: higher order poles and boundaries}

Our geometric construction for Fuchsian connections extends straightforwardly to  wild connections, or to surfaces with boundaries,
by using generalized cycles in a way similar to \cite{Eyn2017}. Recall that $\widehat{\mathfrak M}^1$ is the space of all possible meromorphic one forms in the sense of section \ref{formsandcycles}.

\newcommand \remBE[1]{\par {\color{red} \textbf{Remarque BE:}} {\color{red} #1}\par }
\newcommand \remRB[1]{\par {\color{blue} \textbf{Remarque RB:}} {\color{blue} #1}\par }

In this context, define generalized cycles, as elements of the dual of meromorphic one--forms, whose image by $\widehat B$ is a meromorphic one--form
\beq
\widehat{\mathfrak M}_1(\mathfrak m) \underset{def}= \{ \Gamma\in (\widehat{\mathfrak M}{}^1_{\mathfrak m})^* \ | \ \widehat B(\Gamma)\in \widehat{\mathfrak M}{}^1_{\mathfrak m} \}.
\eeq
We endow it with an intersection form by integrating $\widehat B$, i.e.
\beq
2\pi\mathbf i\, \Gamma_1\bigcap \Gamma_2 \underset{def}=  \underset{\Gamma_1}\int \widehat B(\Gamma_2) - \underset{\Gamma_2}\int \widehat B(\Gamma_1)
\eeq
which is a non--degenerate symplectic form on the infinite--dimensional space $\widehat{\mathfrak M}_1(\mathfrak m)$.

\subsection{Higher order poles}
\label{sec:wildconnection}

Assume that the connection $\nabla$ has a pole of order $d_j\geq 1$ at $p_j$, and let $S_j$ be the monodromy of a flat section $\Psi$ along a small circle $\gamma_j$ around $p_j$, and denote as before $\alpha_j\in\Lieh$ such that $S_j = \Ad_{C_j}^{-1} \exp\left(2\pi\mathbf i\, \alpha_j\right) $.

Define the following family of cycles in $\widehat{\Ho}{}''_1(\mathfrak m)$
\bea
\acycle_{p_j,\mathfrak r,k} &\underset{def}=& \gamma_j\cdot(x-z_j)^{k-\mathfrak r(\alpha_j)} \otimes E_{\mathfrak r},  \quad \mathfrak r\in \mathfrak R \\
\acycle_{p_j,\mathfrak r_0,k} &\underset{def}=& \gamma_j\cdot(x-z_j)^{k} \otimes H_{\mathfrak r_0},  \quad \mathfrak r_0\in \mathfrak R_0  \\
\bcycle_{p_j,\mathfrak r,k} &\underset{def}=& \frac{1}{2\pi\mathbf i\, k}\, \gamma_j\cdot (x-z_j)^{\mathfrak r(\alpha_j)-k} \otimes E^{\mathfrak r},  \quad \mathfrak r\in \mathfrak R  \\
\bcycle_{p_j,\mathfrak r_0,k} &\underset{def}=& \frac{1}{2\pi\mathbf i\, k}\, \gamma_j\cdot(x-z_j)^{-k} \otimes H^{\mathfrak r_0},  \quad \mathfrak r_0\in \mathfrak R_0
\eea
Their intersections are given by
\beq
\acycle_{p_j,\mathfrak r,k}\bigcap \acycle_{p_{j'},\mathfrak r',k'} = 0
\qquad ,\qquad
\bcycle_{p_j,\mathfrak r,k}\bigcap \bcycle_{p_{j'},\mathfrak r',k'} = 0,
\eeq
\beq
\acycle_{p_j,\mathfrak r,k}\bigcap \bcycle_{p_{j'},\mathfrak r',k'} = \delta_{j,j'} \delta_{\mathfrak r,\mathfrak r'} \delta_{k,k'}.
\eeq
They also have zero intersection with $\widehat{\Ho}{}'''_1(\mathfrak m)$.
They do not form a basis of $\widehat{\Ho}{}''_1(\mathfrak m)$, but together with $\widehat{\Ho}{}'''_1(\mathfrak m)$, they form a generating family of $\widehat{\Ho}{}''_1(\mathfrak m)$.

\br
$\acycle_{p_j,\mathfrak r,k}\in \Ker \widehat B$
\er

\bd[KP times]
\beq
t_{p_j,\mathfrak r,k} \underset{def}= \frac{1}{2\pi\mathbf i} \underset{\acycle_{p_j,\mathfrak r,k}}\oint W_1
= \Res_{x= p_j} (x-z_j)^{k-\mathfrak r(\alpha_j)}\left< \Psi_{U_j}(z)^{-1}\diff\Psi_{U_j}(z) , E_{\mathfrak r} \right>
\eeq
\ed

These times satisfy 
\beq
\frac{\partial}{\partial{t_{p_j,\mathfrak r,k}}} = \partial_{\bcycle_{p_j,\mathfrak r,k}}
\eeq
such that there is a map
\bea
\widehat{\mathfrak M}_1(\mathfrak m)  & \longmapsto & T_{\mathfrak m}\modsp_{\text{wild}}      \cr
\bcycle_{p_j,\mathfrak r,k} & \longmapsto & \frac{\partial}{\partial{t_{p_j,\mathfrak r,k}}} 
\eea
which pushes the intersection symplectic form to the Goldman form for wild connections.

The corresponding tau--function is then
\beq
4\pi\mathbf i\log\widehat\Tau(\mathfrak m;\mathcal L) \underset{def}= 
 \sum_{k=1}^{\genus'''} \big(\underset{\acycle'_k}\oint W_1\big)\,\big( \underset{\bcycle_k}\oint W_1\big)
+ \sum_{j=1}^M \underset{\mathfrak r\in\mathfrak R_0^{(j)}}\sum \sum_{k=1}^{\deg_{p_j} W_1} t_{p_j,\mathfrak r,k} \underset{\bcycle_{p_j,\mathfrak r,k}}\oint W_1,
\eeq
and satisfies the higher order Jimbo-Miwa relation
\bea\label{TauMiwaJimbo}
\frac{\partial}{\partial{t_{p_j,\mathfrak r,k}}} \log\widehat\Tau(\mathfrak m;\mathcal L) &=& 
\underset{\bcycle_{p_j,\mathfrak r,k}}\oint W_1\\
&=&\frac 1k\Res_{z= p_j} (x-z_j)^{\mathfrak r(\alpha_j)-k} \left< \Psi_{U_j}(z)^{-1}\diff\Psi_{U_j}(z) , E_{-\mathfrak r} \right>
\eea
All the previous sections of this article extend straightforwardly to the case of higher order poles.

\br
The $k=1$ case recovers the Schlesinger equations of section \ref{schlesinger}.
\er

\subsection{Boundaries}
\label{boundaries}

The same discussion applies to bounded Riemann surfaces.
Let us assume that $\curverond$ has $L$ boundaries $b_j= \partial_j\curverond$, $j=1,\dots,L$, each with the topology of a circle, and for each boundary, choose a chart $\widetilde U_j\subset \CC^*$ as an outer neighborhood of the circle $|z|=1$.
The local coordinate is denoted $\xi_j=\exp(\theta_j\mathbf i)$, with a complex angle $\theta_j\in \RR$ along the boundary and $\operatorname{Im}\theta_j<0$ on the surface.
The monodromy of the flat section $\Psi$ along the boundary is denoted as before $S_j=\Ad_{C_j}^{-1} \exp\left(2\pi\mathbf i\, \alpha_j\right)$.

The following cycles then define a rigid symplectic family in $\widehat{\mathfrak M}_1(\mathfrak m)$
\bea
\acycle_{b_j,\mathfrak r,k} \underset{def}=& \partial_i\curverond\cdot\xi_j^{k-\mathfrak r(\alpha_j)} \otimes E_{\mathfrak r}, \qquad \quad &  k\geq 0, \mathfrak r\in \mathfrak R  \\
 \acycle_{b_j,\mathfrak r_0,k} \underset{def}=& \partial_i\curverond\cdot\xi_j^{k} \otimes H_{\mathfrak r_0}, \qquad\qquad\quad & k\geq 0,\mathfrak r_0\in \mathfrak R_0  \\
\bcycle_{b_j,\mathfrak r,k} \underset{def}=& \frac{1}{2\pi\mathbf i\, k} \gamma_j\cdot\xi_j^{\mathfrak r(\alpha_j)-k} \otimes E^{\mathfrak r}, \quad &    k>0, \mathfrak r\in \mathfrak R  \\
\bcycle_{b_j,\mathfrak r_0,k} \underset{def}=& \frac{1}{2\pi\mathbf i\, k} \gamma_j\cdot\xi_j^{-k} \otimes H^{\mathfrak r_0},   \qquad & k>0,\mathfrak r_0\in \mathfrak R_0  \\
\bcycle_{b_j,\mathfrak r,0} \underset{def}=& \frac{1}{2\pi\mathbf i} \gamma_j\cdot \log{\xi_j}  \otimes E^{\mathfrak r}, \qquad & \mathfrak r\in \mathfrak R  \\
\bcycle_{b_j,\mathfrak r_0,0} \underset{def}=& \frac{1}{2\pi\mathbf i } \gamma_j\cdot \log{\xi_j}  \otimes H^{\mathfrak r_0}, \qquad & \mathfrak r_0\in \mathfrak R_0
\eea
Their intersections are
\beq
\acycle_{b_j,\mathfrak r,k}\bigcap \acycle_{b_{j'},\mathfrak r',k'} = 0
\qquad ,\qquad
\bcycle_{b_j,\mathfrak r,k}\bigcap \bcycle_{b_{j'},\mathfrak r',k'} = 0,
\eeq
\beq
\acycle_{b_j,\mathfrak r,k}\bigcap \bcycle_{b_{j'},\mathfrak r',k'} = \delta_{j,j'} \delta_{\mathfrak r,\mathfrak r'} \delta_{k,k'}.
\eeq
They also have zero intersection with all the previously defined cycles in $\widehat{\Ho}{}'''_1(\mathfrak m)$ and for higher order poles at points in the interior of the surface as well.

Let us introduce the Fourier coefficients of $W_1$ as
\bea
t_{b_j,\mathfrak r,k} 
&\underset{def}=& \frac{1}{2\pi\mathbf i} \underset{\acycle_{b_j,\mathfrak r,k}} \oint W_1\\
&=&  \int_0^{2\pi} \diff\theta_j\, e^{ (k-\mathfrak r(\alpha_j)) \theta_j\mathbf i} \ \left<\Psi_{\widetilde U_j}(e^{\theta_j\mathbf i})^{-1} \diff\Psi_{\widetilde U_j}(e^{\theta_j\mathbf i}), E_{\mathfrak r}  \right> \\
t_{b_j,\mathfrak r_0,k} &\underset{def}=& \frac{1}{2\pi\mathbf i} \underset{\acycle_{b_j,\mathfrak r_0,k}}\oint W_1\\
&=&  \int_0^{2\pi} \diff\theta_j\, e^{k \theta_j\mathbf i} \ \left<\Psi_{\widetilde U_j}(e^{\theta_j\mathbf i})^{-1} \diff\Psi_{\widetilde U_j}(e^{\theta_j\mathbf i}), H_{\mathfrak r_0}  \right>
\eea
together with the dual coefficients
\bea
\widetilde t_{b_j,\mathfrak r,k} &
\underset{def}=& \underset{\bcycle_{b_j,\mathfrak r,k}}\oint W_1\\
&=& \frac 1k \int_0^{2\pi} \diff\theta_i e^{-(k-\mathfrak r(\alpha_j)) \theta_j\mathbf i} \  \left<\Psi_{\widetilde U_j}(e^{\theta_j\mathbf i})^{-1} \diff\Psi_{\widetilde U_j}(e^{\theta_j\mathbf i}) , E^{\mathfrak r}  \right>\\
\widetilde t_{b_j,\mathfrak r_0,k} &\underset{def}=& \underset{\bcycle_{b_j,\mathfrak r_0,k}}\oint W_1\\
&=& \frac 1k \int_0^{2\pi} \diff\theta_i e^{-k \theta_j\mathbf i} \  \left<\Psi_{\widetilde U_j}(e^{\theta_j\mathbf i})^{-1} \diff\Psi_{\widetilde U_j}(e^{\theta_j\mathbf i}) , H^{\mathfrak r_0}  \right>
\eea
such that 
\beq
\frac{\partial}{\partial{t_{b_j,\mathfrak r,k}}} = \partial_{\bcycle_{b_j,\mathfrak r,k}}.
\eeq
If it would be possible to glue a disc to the boundary $b_j$ in such a way that $W_2$ could be analytically continued to that disc, then we would have $\acycle_{b_j,\mathfrak r,k}\in \Ker \widehat B$ as wanted. This analytic continuation property is not fulfilled in general. If it were, the tau--function would be defined such that it additionally contains a summation over all Fourier modes
\bea
4\pi\mathbf i\log\widehat\Tau(\mathfrak m;\mathcal L)& \underset{def}=& 
 \sum_{k=1}^{\genus'''} \big(\underset{\acycle'_k}\oint W_1\big)\,\big( \underset{\bcycle_k}\oint W_1\big) \nonumber\\
&+& \sum_{j=1}^M \underset{\mathfrak r\in\mathfrak R_0^{(j)}}\sum \sum_{k=1}^{\deg_{p_j} W_1} t_{p_j,\mathfrak r,k} \underset{\bcycle_{p_j,\mathfrak r,k}}\oint W_1 + \sum_{j=1}^L \underset{\mathfrak r\in\mathfrak R_0^{(j)}}\sum \sum_{k=0}^{\infty} t_{b_j,\mathfrak r,k} \underset{\bcycle_{b_j,\mathfrak r,k}}\oint W_1\nonumber\\
\eea
The last sum over $k\in\mathbb N$ is absolutely convergent as its coefficients are the Fourier modes of a periodic function. All the previous sections again extend straightforwardly to the case with boundaries.

\br
Poles of order higher than one can also be described in the same way boundaries are, namely by removing a disc around $p_j$. The corresponding Fourier modes would then appear as Taylor-Laurent expansion coefficients around $p_j$. In other words, a pole corresponds to a boundary with only a finite number of non--zero Fourier modes.
\er

\section{Examples}

\subsection{Liouville theory at $c=1$}
\label{DOZZ}

This is the case described in \cite{ER2013}: the group $G=\operatorname{SL}_2(\CC)$ on the sphere $\curverond=\mathbb P^1$ with three marked points located at $0$, $1$ and  $\infty$. The connection takes the form
\beq
\nabla=\diff-\Phi(x)
\quad , \quad
\Phi(x) = \Phi_0\frac{\diff x}{x}+\Phi_1\frac{\diff x}{x-1}
\qquad , \quad
\Phi_\infty \underset{def}= -\Phi_0-\Phi_1.
\eeq
The 3 charges, i.e. the eigenvalues $\operatorname{diag}(\alpha_j,-\alpha_j)$ of $\Phi_j$ are assumed to be given and their evaluations by the second Casimir are denoted
\beq
\Delta_j\underset{def}=\mathfrak C_2(\alpha_j) = \alpha_j^2 = -\det\operatorname{diag}(\alpha_j,-\alpha_j) = -\det \Phi_j = \frac12 \Tr \Phi_j^2.
\eeq
Choosing a gauge in which $\Phi_0$ is diagonal we obtain $\Phi_0=\operatorname{diag}(\alpha_0,-\alpha_0)$ and 
\beq
\Phi_1=\begin{pmatrix}
-\frac{\Delta_0+\Delta_1-\Delta_\infty}{2\alpha_0} & -B \cr -C & \frac{\Delta_0+\Delta_1-\Delta_\infty}{2\alpha_0}
\end{pmatrix}
,
\Phi_\infty=\begin{pmatrix}
-\frac{\Delta_0-\Delta_1-\Delta_\infty}{2\alpha_0} & B \cr C & \frac{\Delta_0-\Delta_1-\Delta_\infty}{2\alpha_0}
\end{pmatrix}
\eeq
where $B$ and $C$ can be chosen arbitrarily in $\CC^*$ provided that their product reads
\beq
BC=\frac{2\Delta_0\Delta_1+2\Delta_1\Delta_\infty+2\Delta_\infty\Delta_0-\Delta_0^2-\Delta_1^2-\Delta_\infty^2}{4\Delta_0}
\eeq

A flat section of $\nabla$ can be constructed from the hypergeometric function $F={}_{2}F_1 $, as
\bea
\Psi(x) &\underset{def}=& \begin{pmatrix}
\psi_+(x) & \psi_-(x) \cr
\psi'_+(x) & \psi'_-(x) \cr
\end{pmatrix}\\
\text{where}\quad \psi_+(x) &\underset{def}=& x^{\alpha_0}(1-x)^{\alpha_1} F(\alpha_0+\alpha_1+\alpha_\infty,\alpha_0+\alpha_1-\alpha_\infty,2\alpha_0,x)\\
\text{and}\quad \psi_-(x) &\underset{def}=& \frac{B}{1-2\alpha_0}\  x^{1-\alpha_0}(1-x)^{\alpha_1} F(\alpha_\infty+\alpha_1-\alpha_0,\alpha_1-\alpha_0-\alpha_\infty,2-2\alpha_0,x)\nonumber\\
\eea
In \cite{ER2013} it was moreover shown that the Liouville 3-point function \cite{DO1994,ZZ1996} at $c=1$ satisfies our formalism, namely
\beq
\left(\frac\partial{\partial \alpha_0}-\frac\partial{\partial \alpha_1}\right)\log Z_3 = \int_0^1 \Tr \Phi(x) \ \Psi(x) E \Psi(x)^{-1} = \int_{]0,1[\otimes E} W_1
\eeq
in the notation of this paper, where $]0,1[\, \otimes\, E\in \widehat{\Ho}{}'''_1$ is the third--kind cycle with
\beq
E=\begin{pmatrix}
-1 & -\frac{f_{+-}}{f_{++}}\cr \frac{f_{-+}}{f_{--}} & 1
\end{pmatrix}
\eeq
where the coefficients $f_{++}, f_{+-}, f_{-+}$ and $f_{--}$ are given by
\beq
\begin{pmatrix}
f_{++} & f_{+-} \cr
f_{-+} & f_{--}
\end{pmatrix}
=
\begin{pmatrix}
\frac{\Gamma(2\alpha_0)\Gamma(-2\alpha_1)}{\Gamma(\alpha_0-\alpha_1-\alpha_\infty)\Gamma(\alpha_0-\alpha_1+\alpha_\infty)} &
B \frac{\Gamma(1-2\alpha_0)\Gamma(-2\alpha_1)}{\Gamma(1-\alpha_0-\alpha_1-\alpha_\infty)\Gamma(1-\alpha_0-\alpha_1+\alpha_\infty)} \cr
B^{-1} \frac{\Gamma(2\alpha_0)\Gamma(1+2\alpha_1)}{\Gamma(\alpha_0+\alpha_1+\alpha_\infty)\Gamma(\alpha_0+\alpha_1-\alpha_\infty)} 
&
\frac{\Gamma(1-2\alpha_0)\Gamma(1+2\alpha_1)}{\Gamma(1-\alpha_0+\alpha_1+\alpha_\infty)\Gamma(1-\alpha_0+\alpha_1-\alpha_\infty)} 
\end{pmatrix}.
\eeq

\subsection{Painlev\'e VI}
\label{P6}

In this example the Lie group is still given by $G=\operatorname{SL}_2(\CC)$ on the sphere $\curverond=\mathbb P^1$ but now with a number $M\geq 4$ of distinct singularities $z_1,\dots,z_M$ for the connection
\beq
\nabla=\diff-\Phi(x)
\quad , \quad
\Phi(x) = \sum_{j=1}^M \Phi_j \frac{ \diff x}{x-z_j}
\qquad , \quad
\sum_{j=1}^M \Phi_j=0.
\eeq
The charges, again the eigenvalues $\operatorname{diag}(\alpha_j,-\alpha_j)$ of $\Phi_j$, are assumed to be given and their evaluation on the second Casimir are again denoted
\beq
\Delta_j\underset{def}=\mathfrak C_2(\alpha_j) = \alpha_j^2 = -\det\operatorname{diag}(\alpha_j,-\alpha_j) = -\det \Phi_j = \frac12 \Tr \Phi_j^2.
\eeq
The classical spectral curve would be given by
\beq
y^2 = \frac12 \Tr \Phi(x)^2 = \left( \sum_{j=1}^M \frac{\Delta_j}{(x-z_j)^2} + \frac{\beta_j}{x-z_j} \right)(\diff x)^2
\quad , \quad
\beta_j \underset{def}= \sum_{j\neq k} \frac{\Tr \Phi_j\Phi_k}{z_j-z_k},
\eeq
which is the classical stress energy tensor of Liouville theory.
The genus of that algebraic curve is
\beq
\genus = M-3
\eeq
and coincides with \eqref{genusbasis} when $\dim\Lieg=3$, $\dim\Lieh=1$ and $\genusrond=0$. In this case, our tau--function coincides with the notion of isomonodromic tau--function for Schlesinger's integrable system (see section \ref{schlesinger}) and provides for instance a non-perturbative completion to the 2-parameter tau--function of \cite{I}.

\subsection{Random matrices}
\label{MM}

This is the again a case of the sphere $\curverond=\mathbb P^1$ with $G=\operatorname{SL}_2(\mathbb C)$, but this time with a wild connection. This allows to recover the tau-function of the Toda Lattice integrable system. We use here the orthogonal polynomials method \cite{MehtaBook}.

Let $V'\in \CC(x)$ a rational function with poles located at $\{p_j\}_{j=1}^M$ and choose a primitive $V$ of $V'$. 
Let $\gamma\in \Ho_1(\mathbb P^1,\{p_1,\dots,p_M\},e^{-V(x)}\diff x)$ be a relative homology class of Jordan arcs on which the one--form $e^{-V(x)}\diff x$ is integrable and such that the expression $e^{-V(x)}$ vanishes at $p_1,\dots,p_M$.

In the case where $\gamma=\RR$, consider the probability measure 
\beq
\DD M \underset{def}= \frac1{\Tau_n}\frac 1{ \operatorname{Vol}(U(n)/U(1)^n)} e^{-\Tr V(M)} \prod_{i,j} \diff M_{i,j}.
\eeq
on the space of Hermitian matrices $\mathcal H_n$ of size $n\times n$, where $\operatorname{Vol}(U(n)/U(1)^n)$ is the Haar volume of the group $U(n)/U(1)^n$.
The normalization factor $\Tau_n$ is called the partition function and will be identified with our tau--function. It is given by
\beq
\Tau_n = \frac 1{\operatorname{Vol}(U(n)/U(1)^n)} \int_{\mathcal H_n(\gamma)} e^{-\Tr V(M)} \prod_{i,j} \diff M_{i,j}.
\eeq
where for $\gamma\neq \RR$, we consider the set of corresponding normal matrices $\mathcal H_n(\gamma)$ (namely the set of diagonalizable matrices with unitary diagonalizing matrix and  eigenvalues located on $\gamma$) with the probability measure
\beq
\DD M \underset{def}= \frac{1}{\Tau_n}\frac 1{\operatorname{Vol}(U(n)/U(1)^n)} e^{-\Tr V(M)} \mathcal D_0 M
\eeq
where $\mathcal D_0 M$ is the standard measure on $\mathcal H_n(\gamma)$ \cite{EKR2015}.

We shall define the "Baker-Akhiezer (\textit{resp.} dual) function", also called the (\textit{resp.} Hilbert transform of the) orthogonal polynomial, as the expectation value of the characteristic polynomial (\textit{resp.} inverse of the characteristic polynomial) of the random matrix
\beq
\psi_n(x) = e^{-\frac12 V(x)}\ \mathbb E_{\DD M} \left( \det(x-M)\right)_{\mathcal H_n(\gamma)}
\quad , \quad
\phi_n(x) = e^{\frac12 V(x)}\ \mathbb E_{\DD M} \left( \frac{1}{\det(x-M)}\right)_{\mathcal H_{n}(\gamma)}
\eeq
and let the matrix
\beq
\Psi_n(x) = \begin{pmatrix}
\psi_{n-1}(x) & \phi_{n}(x) \cr
\psi_n(x) & \phi_{n+1}(x)
\end{pmatrix}.
\eeq
 $\det\Psi_n(x)$ is a constant \cite{EKR2015} such that up to normalisation, we have
\beq
\Psi_n(x) \in \operatorname{SL}_2(\CC).
\eeq
In other words, $\Psi_n$ is can be viewed as a section of a trivial $\operatorname{SL}_2(\CC)$--bundle over $\mathbb P^1-\{p_1,\dots,p_M\}$. It is moreover a flat section of the connection $\nabla=\diff-\Phi_n(x)$ with
\beq
\Phi_n \underset{def}= \diff\Psi_n\cdot \Psi_n^{-1},
\eeq
meromorphic with poles at the singularities of $V'$ of degrees at most equal to the degree of $V'$.
For instance
\begin{itemize}
\item if $V'$ only exhibits simple poles (i.e. $V=\sum_j \alpha_j \log{(x-z_j)}$), then $\nabla$ is Fuchsian, and $\Tau_n$ is the Dotsenko-Fateev integral of conformal field theory (see for instance \cite{MMS2010} and references therein)
\item if $V$ is a polynomial of some degree $d+1\geq 2$, then $\Phi_n(x)$ is a polynomial of degree $\leq d$
\end{itemize}
{}
$\Tau_n$ satsfies
\bea
\partial_t \log\Tau_n &=& - \int_\gamma \frac{\partial V(x)}{\partial t} \,\rho_n(x)  \diff x \\
&=& - \int_\gamma \frac{\partial V(x)}{\partial t} \  (\psi'_n(x)\psi_{n-1}(x)-\psi_n(x)\psi'_{n-1}(x))  \diff x \\
&=&  \frac 1{2\pi\mathbf i}\int_\gamma \underset x{\operatorname{Disc}}\, \Big(\frac{\partial V}{\partial t} \  (\psi'_n\phi_{n}-\phi_{n+1}\psi'_{n-1})\Big)  \diff x \\
&=&  -\underset{x\rightarrow\infty}{\operatorname{Res}}\, \frac{\partial V(x)}{\partial t} \  \Tr \begin{pmatrix}
1 & 0 \cr 0 & 0\end{pmatrix}
 \Psi_n(x)^{-1} \ \diff \Psi_n(x) \ \diff x \\
&=&  -\underset{x\rightarrow\infty}{\operatorname{Res}}\, \frac{\partial V(x)}{\partial t} \  \Tr\, \Psi_n(x) \begin{pmatrix}
1 & 0 \cr 0 & 0\end{pmatrix} \Psi_n(x)^{-1} 
\ \  \Phi_n(x)   \diff x 
\eea
where $t$ is any parameter on which the potential depends and  $\rho_n$ is the eigenvalue density. We set $W_1(x\cdot E) = \Tr \Phi_n(x) \Ad_{\Psi_n(x)}E$ such that 
\bea
\partial_t\log\mathfrak T_n=-\underset{x\rightarrow\infty}\Res \frac{\partial V(x)}{\partial t}\,W_1\Big(x\cdot\begin{pmatrix}
1 & 0 \cr 0 & 0\end{pmatrix}\Big)=\underset{\Gamma_t}\oint W_1 \quad \text{with} \quad \Gamma_t\in \widehat{\mathfrak M}_1
\eea
a generalized cycle in the sense of section \ref{sec:wildconnection}. The matrix integral under consideration is in other words a Jimbo--Miwa--Ueno tau--function \cite{JMU1981,GIL2018,BK2019}.

The functions $\psi_n(x)$ are moreover orthogonal \cite{MehtaBook} in the sense that
\beq
\int_\gamma \psi_n(x)\psi_m(x)\diff x = h_n\delta_{n,m},
\eeq
for some constants $h_n$, which is equivalent to the property that $\Tau_n$ satisfies Hirota equations \cite{AK1996}. The generalized relsolvents and their cumulants are defined as the correlation functions 
\bea
\widehat W_n(x_1,\dots,x_n) &\underset{def}= &\mathbb E_{\DD M} \left( \prod_{i=1}^n \Tr \frac{1}{x_i-M} \right)_{\mathcal H_n(\gamma)}\\
W_n(x_1,\dots,x_n) &\underset{def}=& \mathbb E_{\DD M}^{\text{cumulant}} \left( \prod_{i=1}^n \Tr \frac{1}{x_i-M} \right)_{\mathcal H_n(\gamma)}
\eea
They coincide with the determinantal formulas \cite{BE2009}
\beq
W_n(x_1,\dots,x_n) = (-1)^{n-1}\sum_{\sigma\in\mathfrak S_n^c} \Tr \prod_{i=1}^n 
\begin{pmatrix} 1 & 0 \cr 0 & 0 \end{pmatrix}
\mathcal K_{\Psi_n}(\widetilde x_i,\widetilde x_{\sigma(i)})
\eeq
with the self--reproducing kernel given by
\beq
\mathcal K_{\Psi_n}(\widetilde x,\widetilde y) = \frac{\sqrt{\diff x\, \diff y}}{x-y}\  \Psi_n(\widetilde y)^{-1} \Psi_n(\widetilde x)  .
\eeq
Our formalism therefore applies to study matrix integrals where it was born.

\br
Let us mention without going into the details that the two--matrix model \cite{BM2008} can also be described with our formalism with group $G=\operatorname{GL}_r(\CC)$ for some $r\geq 2$.
\er

\section*{Aknowledgements}

This work was supported by the ANR grant Quantact : ANR-16-CE40-0017.
B.E. thanks the CRM for its hospitality during the Aisenstadt chair in 2015 during which part of this this work was conducted.
We thank S. Ribault who played a key role in developing these ideas, as well as J. Hurtubise.
We thank M. Bertola, G. Borot, D. Korotkin, P. Gavrylenko, J. Harnad, O. Lisovyy, J. Teschner and D. Yang for fruitful discussions.

\appendix{}

\section*{Appendices}

\section{Pairing with with 3$^{\operatorname{rd}}$ kind cycles}
\label{appintegral3rdkind}

Given a generalized meromorphic one--form $\omega\in\widehat{\mathfrak M}{}^1_{\mathfrak m}$, consider a cycle $\Gamma\in \widehat\Ho{}'''_1(\mathfrak m)$ with a boundary component at some pole $p_j$ of $\nabla$ such that

\beq
\Gamma = (o,p_j)\otimes \sigma + 3^{\operatorname{rd}}-\text{kind component with no boundary at }p_j
\eeq
where $o$ still denotes a smooth generic point (at which $\omega$ is holomorphic), $(o,p_j)$ is a chain on $\curverond$ with boundary $p_j-o$ and $\sigma$ is a $\nabla$--flat section, that we write

\beq
\sigma_U(\widetilde x) \underset{x\sim p_j}= \Ad_{\Psi_U(\widetilde x)} E
\eeq
in a chart $U$ containing $p_j$. There, the behavior \eqref{Psinearzj} and the decomposition

\beq
\Ad_{C_j}(E) = E_0 + \sum_{\mathfrak r\in \mathfrak R} E_{\mathfrak r}
\eeq
of $E$ on root spaces yields

\beq
\sigma_U(\widetilde x) \underset{x\sim p_j}= \Ad_{V_j} E_0 +O(x-z_j) + \sum_{\mathfrak r\in \mathfrak R} (x-z_j)^{\mathfrak r(\alpha_j)} \ ( \Ad_{V_j} E_{\mathfrak r} +O(x-z_j))
\eeq

We shall here give a precise meaning to the integral

\beq
\int_o^{p_j} \left\langle\omega,\sigma\right\rangle
\eeq

For that purpose, we replace $\Gamma$ by an equivalent representative of the same generalized homology class

\beq
\Gamma = (o,p_j)\otimes \sigma_0 + \sum_{\mathfrak r\in \mathfrak R}  \ \gamma_{p_j}\otimes \sigma_{\mathfrak r}
\eeq
where $\gamma_{p_j}$ is a closed loop from $o$ to $o$ that surrounds $p_j$ and moreover the $\nabla$--flat sections $\sigma_0$ and $\sigma_{\mathfrak r}$ -- for $\mathfrak r\in\mathfrak R$-- have the local behaviors 

\beq
\sigma_{0,U}(\widetilde x) \underset{x\sim p_j}= \Ad_{\Psi_U(\widetilde x)} \Ad_{C_j}^{-1}E_0
\eeq
\beq
\sigma_{\mathfrak r,U}(\widetilde x) \underset{x\sim p_j}=\frac 1{1-e^{2\pi\mathbf i \mathfrak r(\alpha_j)}} \Ad_{\Psi_U(\widetilde x)} \Ad_{C_j}^{-1} E_{\mathfrak r}
\eeq
around $p_j$. Integrals of the form $\oint_{\gamma_{p_j}}\left\langle\omega,\sigma_{\mathfrak r}\right\rangle$ are convergent such that we are left to define symbols $\int_o^{p_j} \left\langle \omega,\sigma_0\right\rangle$.
A problem occurs when the $\Lieh^*$--component of $\omega$ has a pole at $p_j$. Let us therefore write the Laurent expansion of the restriction $\omega|_{\Lieh^*}$ at $p_j$

\beq
\omega(x)|_{\Lieh^*} \underset{x\sim p_j}= \sum_{k=0}^d  \frac{\Ad_{V_j}(r_{j,k})\diff x}{(x-z_j)^{k+1}} + O(1) 
\eeq
where $r_{j,k}\in \Lieh^*$.
We define

\bea
\int_o^{p_j} \left\langle \omega,\sigma_0\right\rangle
& \underset{def}= &  \int_o^{p_j} \big\langle \omega(x)-\sum_{k=0}^d \frac{\Ad_{V_j}(r_{j,k}) \diff x}{(x-z_j)^{k+1}},\sigma_0(\widetilde x)\big\rangle \cr
&& \qquad-\quad r_{j,0}(E_0)\log(o-z_j) + \sum_{k=1}^{d} \frac{1}{k (o-z_j)^k} r_{j,k}(E_0)   \cr
\eea

\section{Fundamental domain}
\label{appfunddom}

It is useful to consider charts given by fundamental domain, i.e. a $(4\genusrond+2M)$--gon, with pairwise--glued edges

\beq
\curve_\Upsilon\underset{def}{=}\curverond-\Upsilon
\eeq
where $\Upsilon$ is a one--face graph drawn on the compact surface $\curverond$, with $2\genusrond+M$ edges, and $M$ one--valent vertices located at the poles of $\nabla$ and one vertex $o$ of valence $4\genusrond+M$.

Let us choose an orientation for the edges of $\Upsilon$.
The boundary of the chart $\curve_\Upsilon$ is then given by

\beq
\partial \curve_\Upsilon = \sum_{e\, :\, \text{edge of }\Upsilon} e_+ - e_-
\eeq
where $e_+$ (resp. $e_-$) is the left (resp. right) side of $e$.
We shall call $e^\perp$ a loop crossing edge $e$ and no other edge of $\Upsilon$, oriented from $e_-$ towards $e_+$ in the fundamental domain.

\begin{equation}
\includegraphics[scale=0.6]{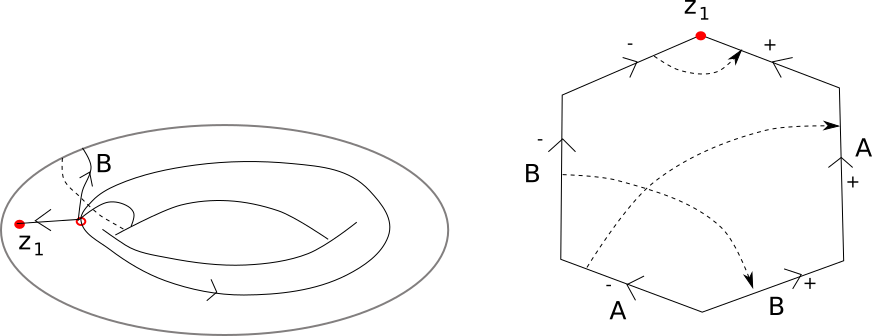}
\end{equation}

\section{Prime form and Klein form}
\label{primeform}

Following Fay's lectures \cite{Fay}, we will define the prime form (twisted or not) and Klein form of a compact Riemann surface $\curverond$. 
All one-forms on $\curverond$ can be built from the prime form, the amplitudes defined in sections\ref{qLiouville} and \ref{amplitudes} are moreover built using a twisted prime form. 

The prime form $\Erond$ and twisted prime form $\Erond_\zeta$ are spinor $(-\frac12, -\frac12)$-forms that respectively live on $\curverond$ and on its universal cover $\curveuniv$. In any local coordinate $x$, both forms behave as 
\beq
\Erond(\widetilde x,\widetilde x') \underset{x\sim x'}\sim \Erond_\zeta(x,x') \underset{x\sim x'}= \frac{x-x'}{\sqrt{\diff x \diff x'}}\big(1+O(x-x')\big) \ .
\label{ezd}
\eeq 
If our Riemann surface has genus zero, both forms are the same, and have a simple expression in terms of the global coordinate $x$ on $\curverond = \bar{\mathbb{C}}$, 
\beq
\Erond(x,x') \overset{\genusrond = 0}{=} \Erond_\zeta(x,x') \overset{\genusrond = 0}{=} \frac{x-x'}{\sqrt{\diff x\diff x'}}\ .
\eeq
If $\genusrond >0$, let $\{\acyclerond_i,\bcyclerond_i\}_{i=1}^{\genusrond}$ be  a symplectic basis\footnote{Symplectic basis always exist and are not unique. The prime form depends on this choice, but the twisted prime form and Klein form will not.} of cycles of $\curverond$ -- also known as a Torelli marking. Let then $(\diff u_i)_{i=1}^{\genusrond}$ be holomorphic one-forms such that $\oint_{\acyclerond_i} \diff u_j = \delta_{i,j}$, their primitives $(u_i)_{i=1}^{\genusrond}$ constitute the Abel--Jacobi map, and their integrals $\tau_{i,j}=\oint_{\bcyclerond_i} \diff u_j$ define the matrix of periods $\tau$. We also introduce the Riemann theta function $\theta(u,\tau)$ with modulus $\tau$. 

\bd[Prime form if $\genusrond > 0$]

Let $c=\frac12 (\vec n+\tau \vec m)\in \mathbb C^{\genusrond}$ be and odd regular half-integer characteristic, that is $\vec n,\vec m \in \mathbb{Z}^{\genusrond}$ and $\vec n\cdot \vec m \in 2\mathbb Z+1$.
Using the $c$-dependent holomorphic one-form
\beq
h_c(x) = \sum_{i=1}^{\genusrond} \left(\partial_i \theta\right)(c)\, \diff u_i(x)\ ,
\eeq
we write the prime form as 
\beq
\Erond(\widetilde x,\widetilde x') = \frac{\theta\left(u(x)-u(x')+c,\tau\right)}{\sqrt{h_c(x)h_c(x')}}\ .
\eeq

\ed
The prime form $\Erond$ has no monodromies around $\acyclerond$-cycles, but it has monodromies around $\bcyclerond$-cycles,
\beq
\Erond\big(\widetilde x+\bcyclerond_i,\widetilde x'\big) = \Erond(\widetilde x,\widetilde x')\,e^{-2\pi\mathbf i (u_i(x)-u_i(y)+c_i)}\,e^{-\pi\mathbf i \tau_{i,i}}\ .
\eeq
This is why it is defined on $\curveuniv$. But we need a form on $\curverond$, which is why we will now introduce the twisted prime form. The idea is to correct the monodromies around $\bcyclerond$-cycle with the help of a meromorphic one-form $f$ on $\curverond$ such that $\oint_{\acyclerond_i} f = 0$ and 
\beq
\frac{1}{2\pi\mathbf i}\oint_{\bcyclerond_i} f = \zeta_i \qquad \text{with} \qquad 
\zeta=(\zeta_1,\dots,\zeta_{\genusrond})\in \mathbb C^{\genusrond} - \big(\mathbb Z^{\genusrond}+\tau\cdot\mathbb Z^{\genusrond} \big)\ .
\eeq
Our condition on the polarization $\zeta$ implies that $f$ has poles. 

\bd[Twisted prime form if $\genusrond > 0$]

Given an odd regular half-integer characteristic $c$ and a meromorphic one-form $f$ with polarization $\zeta$, let 
\beq
\Erond_\zeta(x,x') = \Erond(\widetilde x,\widetilde x')\frac{\theta(\zeta+c,\tau)}{\theta(u(x)-u(x')+\zeta+c,\tau)}e^{-\int_{x'}^x f}\ ,
\eeq
where the integral $\int_{x'}^x f$ is taken around the unique homology chain that does not intersect our $\acyclerond$- and $\bcyclerond$-cycles. 

\ed
This twisted prime form $\Erond_\zeta$ has no monodromies and is a form on $\curverond$ rather than $\curveuniv$.
Moreover, given a value of $\zeta\mod \mathbb Z^{\genusrond}+\tau\cdot\mathbb Z^{\genusrond}$, the twisted prime form depends only weakly on the characteristic $c$ and one-form $f$, in the sense that changing these quantities only changes $\Erond_\zeta$ as
\beq
\Erond_\zeta(x,x') \longmapsto \Erond_\zeta(x,x')\times \frac{F(x)}{F(x')}
\eeq
\label{ete}

where is a meromorphic function on $\curverond$. In our amplitudes that are built using $\Erond_\zeta$, ratios such as $\frac{F(x)}{F(x')}$ cancel, and there is no dependence on $c$ or $f$. This justifies writing $\Erond_\zeta$ as a quantity that depends solely on $\zeta \in \mathbb C^{\genusrond}\Big/\mathbb Z^{\genusrond}+\tau\cdot\mathbb Z^{\genusrond}$. In particular, the essential singularities at the poles of $f$  cancel in our amplitudes. Moreover, the amplitudes involve $\frac{1}{\Erond_\zeta(x,x')}$ and are therefore regular at the poles of $\theta(u(x)-u(x')+\zeta+c,\tau)$. The only singularities of the amplitudes are poles that come from the zero \eqref{ezd} of $\Erond_\zeta(x,x')$ at the diagonal $x=x'$.

These properties of the amplitudes also hold in the case of the Klein form, that we now define.

\bd[Klein form]

The Klein form 
\beq
\Brond_\zeta(x,x') = - \frac{1}{\Erond_\zeta(x,x')\Erond_\zeta(x',x)}\ ,
\eeq
is a meromorphic symmetric $(1,1)$-form on $\curverond$, whose only singularity is a double pole on the diagonal $\Delta$,
\beq
\Brond_\zeta(x,x') \underset{x\to x'}{=} \frac{\diff x \, \boxtimes \diff x'}{(x-x')^2} + \mathrm{analytic} \qquad \text{i.e.} \qquad \Brond_\zeta\in \Ho^0(\curverond, \operatorname K_{\curverond}\boxtimes \operatorname K_{\curverond}(2\Delta))^{\mathrm{sym}}\ .
\eeq
The normalized fundamental second kind differential, or Bergman kernel is then 
\beq
\Brond(x,x')=\Brond_\zeta(x,x')- 2\pi\mathbf i \sum_{i,j=1}^{\genusrond} \left(\partial^2_{i,j}\log\theta\right)(\zeta+c)\, \diff u_i(x) \diff u_j(x')\ ,
\eeq
and is independent of the polarization $\zeta$ although it still depends on the choice of symplectic basis of cycles.

\ed

\bp[Fay identities] 

Let $k>0$ and $x_1,\dots,x_k$ and $y_1,\dots,y_k$ be $2k$ distinct points of $\curverond$, then
\beq
\frac{\theta\big(\sum_i (u(x_i)-u(y_i)) + \zeta+c\big)}{\theta(\zeta+c)} \det \left(\frac{1}{\Erond_\zeta(x_i,y_j)}\right) = \frac{\prod_{i<j} \Erond_\zeta(x_i,x_j) \Erond_\zeta(y_i,y_j)}{\prod_{i,j} \Erond_\zeta(x_i,y_j)}\ .
\eeq

\ep
In genus $\genusrond=0$ the Fay identities reduces to the Cauchy identity,
\beq
\det\left( \frac{1}{x_i-y_j}\right) = \frac{\prod_{i<j} (x_i-x_j) (y_i-y_j)}{\prod_{i,j} (x_i-y_j)}\ .
\eeq

\bd[Klein's third kind form]\label{def:Klein3rd}

Given $x_1,x_2\in \curveuniv$, let Klein's third kind form be 
\beq\label{defomrond3rd}
\omegarond_{x_1,x_2;\zeta}(x) = \int_{x_2}^{x_1} \Brond_\zeta(x,.)\ .
\eeq
This is a meromorphic one-form of $x$, with simple poles at $x=x_1,x_2$ and residues
\beq
\Res_{x_1} \omegarond_{x_1,x_2;\zeta}  = - \Res_{x_2} \omegarond_{x_1,x_2;\zeta} = 1\ .
\eeq
Klein's third kind form moreover satisfies 
\begin{multline}\label{eq:klein3rd}
\omegarond_{x_1,x_2;\zeta}(x) +\frac{\Erond_\zeta(x_1,x_2)}{\Erond_\zeta(x_1,x)\,\Erond_\zeta(x,x_2)} 
\\ = 
 2\pi\mathbf i \sum_{j=1}^{\genusrond} \left(\frac{(\partial_j\theta)\big(u(x_1)-u(x_2)+\zeta+c,\tau\big)}{\theta\big(u(x_1)-u(x_2)+\zeta+c,\tau\big)}-\frac{(\partial_j\theta)(\zeta+c,\tau)}{\theta(\zeta+c,\tau)}\right)\diff u_j(x)
\end{multline}

\ed

In this paper we assume a choice of polarization $\zeta$ made once and for all and drop its explicit dependence, therefore writing $\mathcal E \underset{def}=\Erond_\zeta$ for the twisted prime--form at hand.

Let us note moreover that in order to consider second--kind deformations -- including for instance deformations of the complex structure of the base Riemann surface -- one has to consider

\bd

The space of marked Riemann surfaces equipped with a twisted prime form (modulo appropriate identifications) is the Jacobian bundle over the Torelli space $\mathcal T_{\genusrond,M}$ of marked Riemann surfaces,
\beq
\mathcal T'_{\genusrond,M} \to \mathcal T_{\genusrond,M}\ .
\eeq
The fiber is the Jacobian $\mathrm{Jac}(\curverond) \underset{def}= \mathbb C^{\genusrond}\Big/\mathbb Z^{\genusrond}+\tau\cdot\mathbb Z^{\genusrond}$ of the base curve  that parameterize choices of polarization.
\ed


\section{Tangent space to cycles}
\label{apptangenttocycles}

Proof of theorem \ref{thTangenttocycles}.

Let $\Psi$ be a multivalued $\nabla$--flat section of $\mathcal P$ and for each deformation $\delta\in\operatorname T_{\mathfrak m}\modsp'_p$, define the element
\beq
F_\delta \underset{def}{=} \delta \Psi \cdot \Psi^{-1}
\eeq
which takes values in the adjoint bundle and is almost by definition single--valued on $\curverond_p$. It satisfies the zero--curvature condition
\beq
[\,\delta\,-\,F_\delta\, ,\, \diff\, -\, \Phi\,]\,=\, \delta\Phi\, -\, \diff F_\delta\, +\, [\, F_\delta\, ,\, \Phi\, ]\, =\, 0
\eeq
Recall that we chose a fundamental domain $\curve_\Upsilon=\curverond-\Upsilon$ constructed from the graph $\Upsilon\subset\curverond$. Considering a differential $\omega'''_{x,o}$ on $\curverond$ with a simple pole at $x$ with residue $+1$, a simple pole at $o$ with residue $-1$ and no other singularities, the Cauchy residue formula yields
\beq
F_\delta(x)-F_\delta(o) = \Res_{x,o}\left( \omega'''_{x,o}\, F_\delta\right)
\eeq
By deforming the integration contour we can push it to the boundary of the fundamental domain to obtain
\beq
F_\delta(x)=F_\delta(o) + \frac{1}{2\pi\mathbf i} \sum_{e\, : \, \text{edge of }\Upsilon}\underset{e_+-e_-}\int \omega'''_{x,o}\,F_\delta
\eeq
Let $\gamma_e$ a loop on $\curverond$ that crosses edge $e$ and no other edge, oriented from $e_-$ to $e_+$ in $\curve_\Upsilon$.
For $x\in \curve_\Upsilon$ close to $e_-$, we have
\beq
\Psi(\widetilde x+\gamma_e) = \Psi(\widetilde x)\cdot S_{\gamma_e}
\eeq
and thus
\beq
\delta\Psi(\widetilde x+\gamma_e) \cdot \Psi(\widetilde x+\gamma_e)^{-1} = \delta\Psi(\widetilde x)\cdot\Psi(\widetilde x)^{-1} + \Ad_{\Psi(\widetilde x)} \left(\delta S_{\gamma_e} \cdot S_{\gamma_e}^{-1}\right)
\eeq
in other words
\beq
F_\delta(x+\gamma_e)-F_\delta(x) =\Ad_{\Psi(\widetilde x)} \left(\delta S_{\gamma_e} \cdot S_{\gamma_e}^{-1}\right)
\eeq
from which follows that
\beq
F_\delta(x)=F_\delta(o) + \frac{1}{2\pi\mathbf i} \sum_{e\, : \, \text{edge of }\Upsilon} \underset{e_-}\int  \omega'''_{x,o}\, \Ad_\Psi \left(\delta S_{\gamma_e} \cdot S_{\gamma_e}^{-1}\right)
\eeq
This leads us to define the following linear combination of Jordan arcs valued in $\Ad \mathcal P$
\bea
\Gamma_\delta & \underset{def}{=} &  \frac{1}{2\pi\mathbf i} \sum_{e\, : \, \text{edge of }\Upsilon} \ e_- \otimes \left(\delta S_{\gamma_e} \cdot S_{\gamma_e}^{-1}\right)\\
\text{such that}\qquad F_\delta(x)&=&F_\delta(o)\ +\,\underset{X'\in\Gamma_\delta}\int \omega'''_{x,o}(x') \sigma_\Psi(X')
\eea
Its boundary lies above the vertices of $\Upsilon$ and for each internal vertex $v$, we have $\underset{e\rightarrow v}\prod S_{\gamma_e}=\operatorname{Id}_{G_o}$ and this implies in particular that
\beq
\sum_{e\rightarrow v} \delta S_{\gamma_e} \cdot S_{\gamma_e}^{-1}=0
\eeq
so that there is no boundary component at $v$. The boundary thus lies at the external vertices of $\Upsilon$, poles $p_1,\dots,p_M$ of $\nabla$, which means that
\beq
\Gamma_\delta\in \widehat\Ho{}'''_1(\mathfrak m)
\eeq
If we moreover keep the charges $[\alpha]$ fixed, there is no boundary component at any of the poles of the connection either and therefore  
\beq
\Gamma_\delta\in \widehat \Ho_1(\mathfrak m).
\eeq
Quotienting by the gauge group means identifying $\Psi\equiv g\cdot\Psi$. In turn $F_\delta\equiv F_\delta + \delta g \cdot g^{-1}$ and $F_\delta$ is only defined modulo an additive constant. The term $F_\delta(o)$ is therefore irrelevant. Vice versa, if $\Gamma\in \widehat \Ho{}'''_1(\mathfrak m)$ be a third--kind cycle and define
\bea
\partial_\Gamma \Psi(\widetilde x) &\underset{def}{=}& F_{\Gamma}(o)\cdot \Psi(\widetilde x) + \underset{X'\in\Gamma}\oint \omega'''_{x,o}(x')\ X'\cdot \Psi(\widetilde x)\\
\text{which implies}\qquad \partial_\Gamma \Phi &=& \diff F_\Gamma(x) + [F_\Gamma,\Phi]
\eea
and therefore defines a tangent vector in $\modsp_p$ that is indeed independent of $F_\Gamma(o)$.

\bibliographystyle{morder5}

\end{document}